\documentclass[prl,aps,nofootinbib,superscriptaddress,showpacs,floatfix,preprintnumbers,twocolumn]{revtex4}
\usepackage{hyperref,amssymb,amsmath,graphicx,xcolor,slashed}

\DeclareMathOperator\Tr{Tr}

\begin{document}

\title{Nucleon Tomography and Generalized Parton Distribution at Physical Pion Mass from Lattice QCD}

\author{Huey-Wen Lin}
\email{hwlin@pa.msu.edu}
\affiliation{Department of Physics and Astronomy, Michigan State University, East Lansing, MI 48824}
\affiliation{Department of Computational Mathematics,
  Science and Engineering, Michigan State University, East Lansing, MI 48824}

\preprint{MSUHEP-20-014}

\pacs{12.38.-t, 
      11.15.Ha,  
      12.38.Gc  
}

\begin{abstract}
We present the first lattice calculation of the nucleon isovector unpolarized generalized parton distribution (GPD) at the physical pion mass using a lattice ensemble with 2+1+1 flavors of highly improved staggered quarks (HISQ) generated by MILC Collaboration, with lattice spacing $a\approx 0.09$~fm and volume $64^3\times 96$. We use momentum-smeared sources to improve the signal at nucleon boost momentum $P_z \approx 2.2$~GeV, and report results at nonzero momentum transfers in $[0.2,1.0]\text{ GeV}^2$.
Nonperturbative renormalization in RI/MOM scheme is used to obtain the quasi-distribution before matching to the lightcone GPDs. The three-dimensional distributions $H(x,Q^2)$ and $E(x,Q^2)$ at $\xi=0$ are presented, along with the three-dimensional nucleon tomography and impact-parameter--dependent distribution for selected Bjorken $x$ at $\mu=3$~GeV in $\overline{\text{MS}}$ scheme.
\end{abstract}

\maketitle

Nucleons (that is, protons and neutrons) are the building blocks of all ordinary matter, and the study of nucleon structure is a central goal of many worldwide experimental efforts.
Gluons and quarks are the underlying degrees of freedom that explain the properties of nucleons, and fully understanding how they contribute to the properties of nucleons (such as their mass or spin structure) helps to decode the Standard Model.
In quantum chromodynamics (QCD), gluons strongly interact with themselves and with quarks, binding both nucleons and nuclei. However, due to their confinement within these bound states, we cannot single out individual constituents to study them.
More than half a century since the discovery of nucleon structure, our understanding has improved greatly;
however, there is still a long way to go in unveiling the nucleon's detailed structure, which is characterized by functions such as the generalized parton distributions (GPDs)~\cite{Mueller:1998fv,Ji:1996ek,Radyushkin:1996nd}.
GPDs can be viewed as a hybrid of parton distributions (PDFs), form factors and distribution amplitudes.
They play an important role in providing a three-dimensional spatial picture of the nucleon~\cite{Burkardt:2000za} and in revealing its spin structure~\cite{Ji:1996ek}.
Experimentally, GPDs can be accessed in exclusive processes such as deeply virtual Compton scattering~\cite{Ji:1996nm} or meson production~\cite{Kriesten:2019jep}.
Experimental collaborations and facilities worldwide have been devoted to searching for these last unknowns of the nucleon, 
including HERMES at DESY, COMPASS at CERN, GSI in Europe, BELLE and JPAC in Japan, 
Halls~A, B and C at Jefferson Laboratory, and PHENIX and STAR at RHIC  at Brookhaven National Laboratory in the US.
There are also plans for future facilities: 
a US electron-ion collider (EIC)~\cite{NAP25171} at Brookhaven National Laboratory, 
an EIC in China (EicC)~\cite{Chen:2018wyz,Chen:2020ijn}, 
and the Large Hadron-Electron Collider (LHeC) in Europe~\cite{AbelleiraFernandez:2012cc,Agostini:2020fmq}.

Although interest in GPDs has grown enormously, we still need fresh theoretical and phenomenological ideas, including reliable model-independent techniques.
Most QCD models have issues associated with three-dimensional structure that are not yet fully understood, so a reliable framework for extracting three-dimensional parton distributions and fragmentation functions from experimental observables does not yet exist.
Theoretically, there are factorization issues in hadron production from hadronic reactions, and theoretical efforts are striving to answer key questions that lie along the path to a precise mapping of three-dimensional nucleon structure from experiment.
It has become common understanding that we need to develop a program in both theory and experiment that will allow an accurate flavor decomposition of the nucleon GPDs, including flavor differences in the quark structure, the gluon structure and the nucleon sea-quark GPDs.
Most current theoretical issues are associated with nonperturbative features of QCD, that is, where the strong coupling is too large for analytic perturbative methods to be valid.
Using a nonperturbative theoretical method that starts from the quark and gluon degrees of freedom, lattice QCD (LQCD), allows us to compute these properties on supercomputers.

Probing hadron structure with lattice QCD was for many years limited to the first few moments, due to complications arising from the breaking of rotational symmetry by the discretized Euclidean spacetime.
The breakthrough for LQCD came in 2013, when a technique was proposed to connect
quantities calculable on the lattice to those on the lightcone.
Large-momentum effective theory (LaMET), also known as the ``quasi-PDF method''~\cite{Ji:2013dva,Ji:2014gla,Ji:2017rah}, allows us to calculate the full Bjorken-$x$ dependence of distributions for the first time.
Much progress has been made since the first LaMET paper ~\cite{Lin:2013yra,Lin:2014zya,Chen:2016utp,Lin:2017ani,Alexandrou:2015rja,Alexandrou:2016jqi,Alexandrou:2017huk,Chen:2017mzz,Alexandrou:2018pbm,Chen:2018xof,Chen:2018fwa,Alexandrou:2018eet,Lin:2018qky,Fan:2018dxu,Liu:2018hxv,Wang:2019tgg,Lin:2019ocg,Chen:2019lcm,Xiong:2013bka,Ji:2015jwa,Ji:2015qla,Xiong:2015nua,Ji:2014hxa,Lin:2014yra,Monahan:2017hpu,Ji:2018hvs,Stewart:2017tvs,Constantinou:2017sej,Green:2017xeu,Izubuchi:2018srq,Xiong:2017jtn,Wang:2017qyg,Wang:2017eel,Xu:2018mpf,Zhang:2017bzy,Ishikawa:2016znu,Chen:2016fxx,Ji:2017oey,Ishikawa:2017faj,Ishikawa:2019flg,Li:2016amo,Monahan:2016bvm,Radyushkin:2016hsy,Rossi:2017muf,Carlson:2017gpk,Ji:2017rah,Briceno:2018lfj,Hobbs:2017xtq,Jia:2017uul,Xu:2018eii,Jia:2018qee,Spanoudes:2018zya,Rossi:2018zkn,Liu:2018uuj,Ji:2018waw,Bhattacharya:2018zxi,Radyushkin:2018nbf,Zhang:2018diq,Li:2018tpe,Braun:2018brg,Detmold:2019ghl,Ebert:2019tvc,Ji:2019ewn,Bali:2017gfr,Bali:2018spj,Sufian:2019bol,Bali:2019ecy,Orginos:2017kos,Karpie:2017bzm,Karpie:2018zaz,Karpie:2019eiq,Joo:2019jct,Joo:2019bzr,Radyushkin:2018cvn,Zhang:2018ggy,Lin:2020ssv,Zhang:2020dkn,Fan:2020nzz,Sufian:2020vzb,Shugert:2020tgq,Green:2020xco,Chai:2020nxw,Shanahan:2020zxr,Braun:2020ymy,Bhattacharya:2020cen,Ji:2020ect,Ebert:2020gxr,Lin:2020ijm,Joo:2020spy,Bhat:2020ktg,Fan:2020cpa}.
Most work has been done using only one lattice ensemble, but there has been some progress in determining the size of lattice systematic uncertainties.
For example, finite-volume systematics were studied in Refs.~\cite{Joo:2019bzr,Lin:2019ocg}.
Machine-learning algorithms have been applied to the inverse problem~\cite{Karpie:2019eiq,Zhang:2020gaj} and to making predictions for larger boost momentum and larger Wilson-displacement~\cite{Zhang:2019qiq}.
On the lattice discretization errors, a $N_f=2+1+1$ superfine ($a \approx 0.042$~fm) lattice at 310-MeV pion mass was used to study nucleon PDFs in Ref.~\cite{Fan:2020nzz}, and results using multiple lattice spacings were reported in Refs.~\cite{Zhang:2020gaj,Lin:2020ssv,Sufian:2020vzb}.
The first attempt to obtain strange and charm distributions of the nucleon was recently reported~\cite{Zhang:2020dkn}.
However, beyond one-dimensional hadron structure, there is little work available.
Last spring, the first lattice study of GPDs was made for pions~\cite{Chen:2019lcm}.
During the completion of this work, ETM Collaboration reported their findings on both unpolarized and polarized nucleon GPDs with largest boost momentum 1.67~GeV at pion mass $M_\pi \approx 260$~MeV~\cite{Alexandrou:2019dax,Alexandrou:2020zbe}. 
In this work, we present the first lattice-QCD calculation of the nucleon GPD at the physical pion mass using the LaMET method and study the three-dimensional structure of the unpolarized nucleon GPDs. 

The unpolarized GPDs $H(x,\xi,t)$ and $E(x,\xi,t)$ 
are defined in terms of the matrix elements
\begin{widetext}
\begin{align}
F(x,\xi,t)&=\int\frac{dz^-}{4\pi}e^{ixp^+ z^-}\left\langle p''\left|\bar\psi\left(-\frac{z}{2}\right)\gamma^+ L\left(-\frac{z}{2},\frac{z}{2}\right)\psi\left(\frac{z}{2}\right)\right|p'\right\rangle_{z^+=0,\vec z_\perp=0}\nonumber\\
&=\frac{1}{2p^+}\left[H(x,\xi,t)\bar u(p'')\gamma^+ u(p')+E(x,\xi,t)\bar u(p'')\frac{i\sigma^{+\nu}\Delta_\nu}{2m}u(p')\right],
\end{align}
\end{widetext}
where $L(-z/2,z/2)$ is the gauge link along the lightcone and
\begin{equation}
\Delta^\mu=p''^\mu-p'^\mu, \qquad
t=\Delta^2, \qquad
\xi=\frac{p''^+-p'^+}{p''^+ +p'^+}.
\end{equation}
In the limit $\xi, t\to 0$, $H$ reduces to the usual unpolarized parton distributions
while the information encoded in $E$ cannot be accessed, since they are multiplied by the four-momentum transfer $\Delta$.
Only in exclusive processes with a nonzero momentum transfer can $E$ be probed.
The one-loop matching~\cite{Ji:2015qla,Liu:2019urm} for the GPD $H$ and $E$ turns out to be similar to that for the parton distribution. 

In this work, we focus on the nucleon isovector unpolarized GPDs and their quasi-GPD counterparts defined in terms of spacelike correlations calculated in Breit frame. 
We use clover valence fermions on an ensemble with lattice spacing $a\approx 0.09$~fm, spatial (temporal) extent around 5.8~(8.6)~fm, and with the physical pion mass $M_\pi \approx 135$~MeV and $N_f=2+1+1$ (degenerate up/down, strange and charm) flavors of highly improved staggered dynamical quarks (HISQ)~\cite{Follana:2006rc} generated by MILC Collaboration~\cite{Bazavov:2012xda}.
The gauge links are one-step hypercubic (HYP) smeared~\cite{Hasenfratz:2001hp} to suppress discretization effects.
The clover parameters are tuned to recover the lowest sea pion mass of the HISQ quarks.
The ``mixed-action'' approach is commonly used, and there has been promising agreement between the calculated lattice nucleon charges, moments and form factors and the experimental data when applicable~\cite{Mondal:2020cmt,Jang:2019jkn,Gupta:2018lvp,Lin:2018obj,Gupta:2018qil,Rajan:2017lxk,Yoon:2016jzj,Bhattacharya:2016zcn,Bhattacharya:2015esa,Bhattacharya:2013ehc,Briceno:2012wt,Bhattacharya:2011qm,Lin:2020reh}.
Gaussian momentum smearing~\cite{Bali:2016lva} is used on the quark field to improve the overlap with ground-state nucleons of the desired boost momentum, allowing us to reach higher boost momentum for the nucleon states. 
We calculate the matrix elements of the form 
$\left\langle P_f\left|\bar\psi\left(-\frac{z}{2}\right)\Gamma L\left(-\frac{z}{2},\frac{z}{2}\right)\psi\left(\frac{z}{2}\right)\right|P_i\right\rangle$ with projection operators $\frac{1+\gamma_t}{2}(1+i \gamma_5\gamma_{x,y,z})$. 
We also use high-statistics measurements, 501,760 total over 1960 configurations, to drive down the increased statistical noise at high boost momenta, $P_z=|\frac{\vec{P_i}+\vec{P_f}}{2}| = |\frac{2\pi}{L}\{0,0,10\}a^{-1}|$,
and vary spatial momentum transfer $\vec{q}=\vec{P_f}-\vec{P_i}=\frac{2\pi}{L}\{n_x,n_y,0\}a^{-1}$ with integer $n_{x,y}$ and $n_x^2+n_y^2 \in \{0,4,8,16,20\}$ 
with four-momentum transfer squared { $Q^2=-q_\mu q^\mu=\{{0, 0.19, 0.39, 0.77, 0.97}\} $}~GeV$^2$. 
We solve a set of linear equations to obtain $H$ and $E$ (similar to form-factor extraction) 
with all $|q|$ at fixed $Q^2$. 
Technical details (such as renormalization) and more information on how the matrix elements are extracted can be found in the supplement and our previous work~\cite{Yoon:2016dij,Chen:2018xof,Lin:2018qky,Liu:2018hxv}. 

\begin{figure}[tb]
\centering
\includegraphics[width=0.4\textwidth]{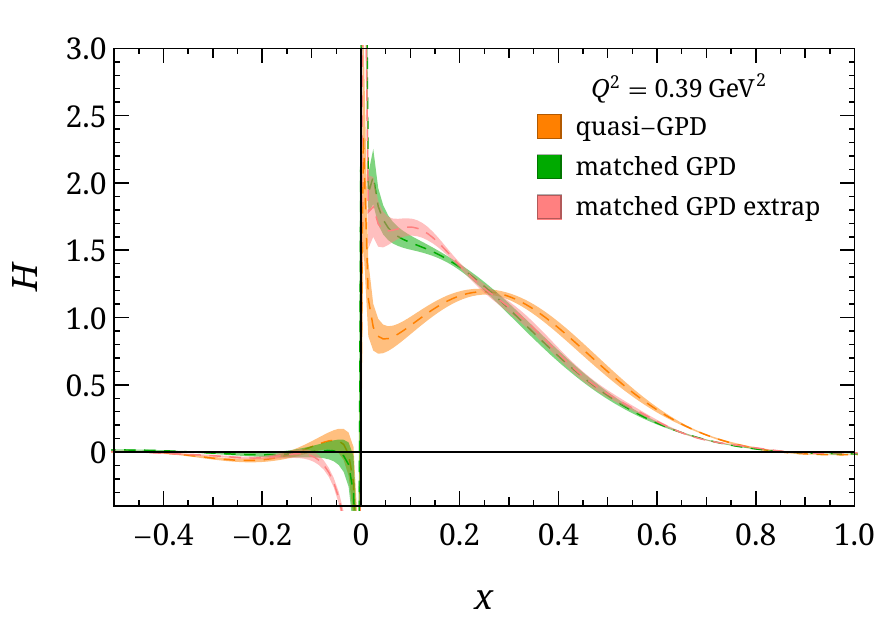}
\includegraphics[width=0.4\textwidth]{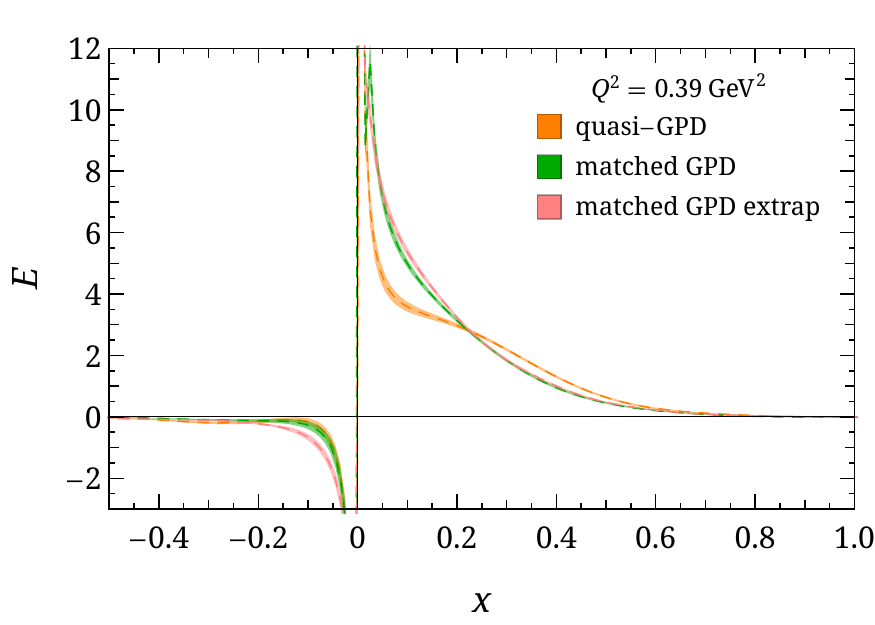}
\caption{
Nucleon isovector $H$ and $E$ quasi-GPDs and matched GPDs at momentum transfer $Q^2=0.39\text{ GeV}^2$.
The orange and green bands are the quasi-GPD and matched GPDs from derivative method~\cite{Lin:2017ani}, while the pink band corresponds to the matched GPD using quasi-GPD from the extrapolation formulation suggested by Ref.~\cite{Ji:2020brr}.
We find both methods give reasonable agreement in the $x$-dependent behavior, except in the small-$x$ region, which is dominated by the large-$z$ matrix elements that rely on the extrapolation.
\label{fig:quasi-matched}}
\end{figure}

\begin{figure}[tb]
\includegraphics[width=0.4\textwidth]{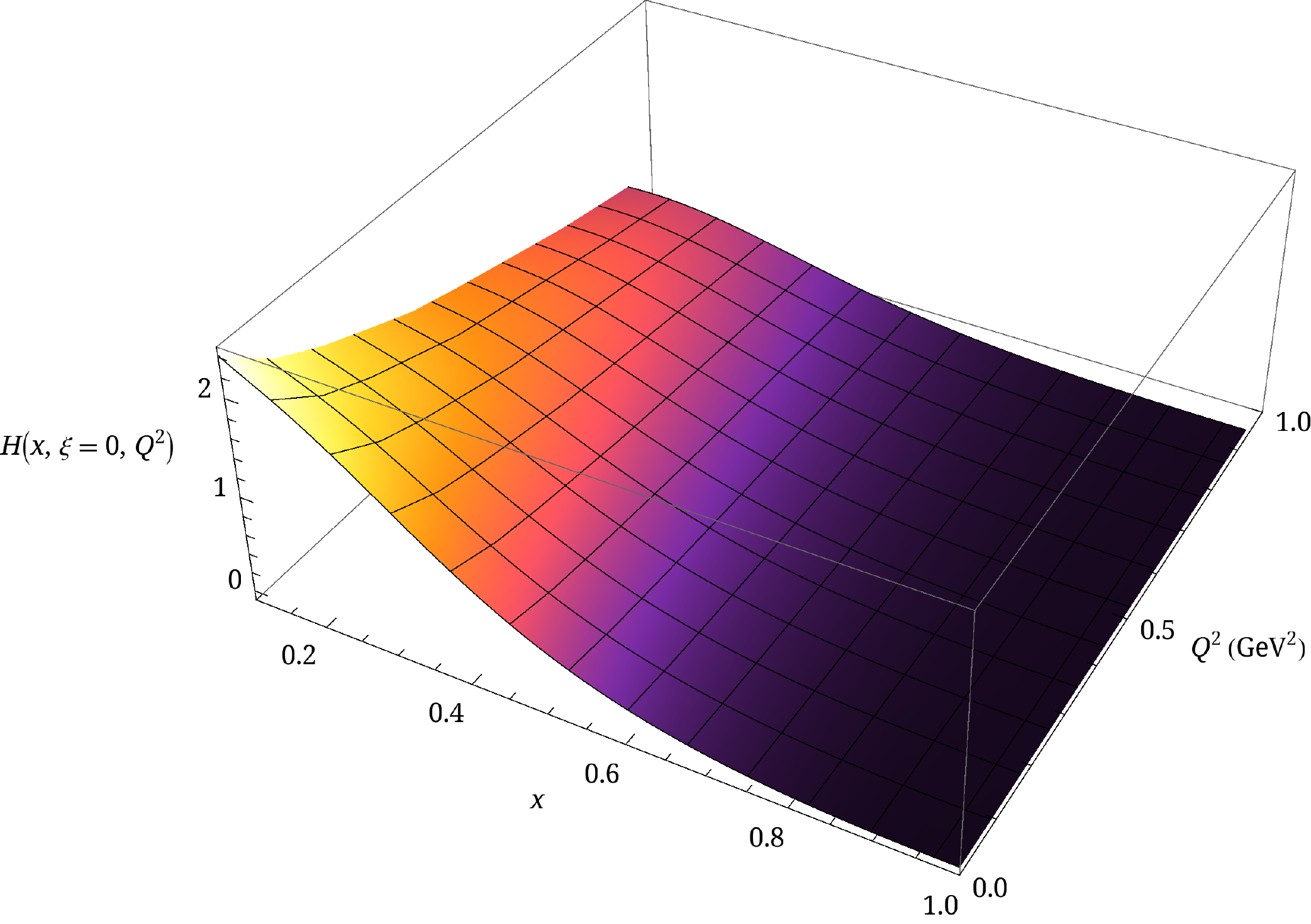}
\includegraphics[width=0.4\textwidth]{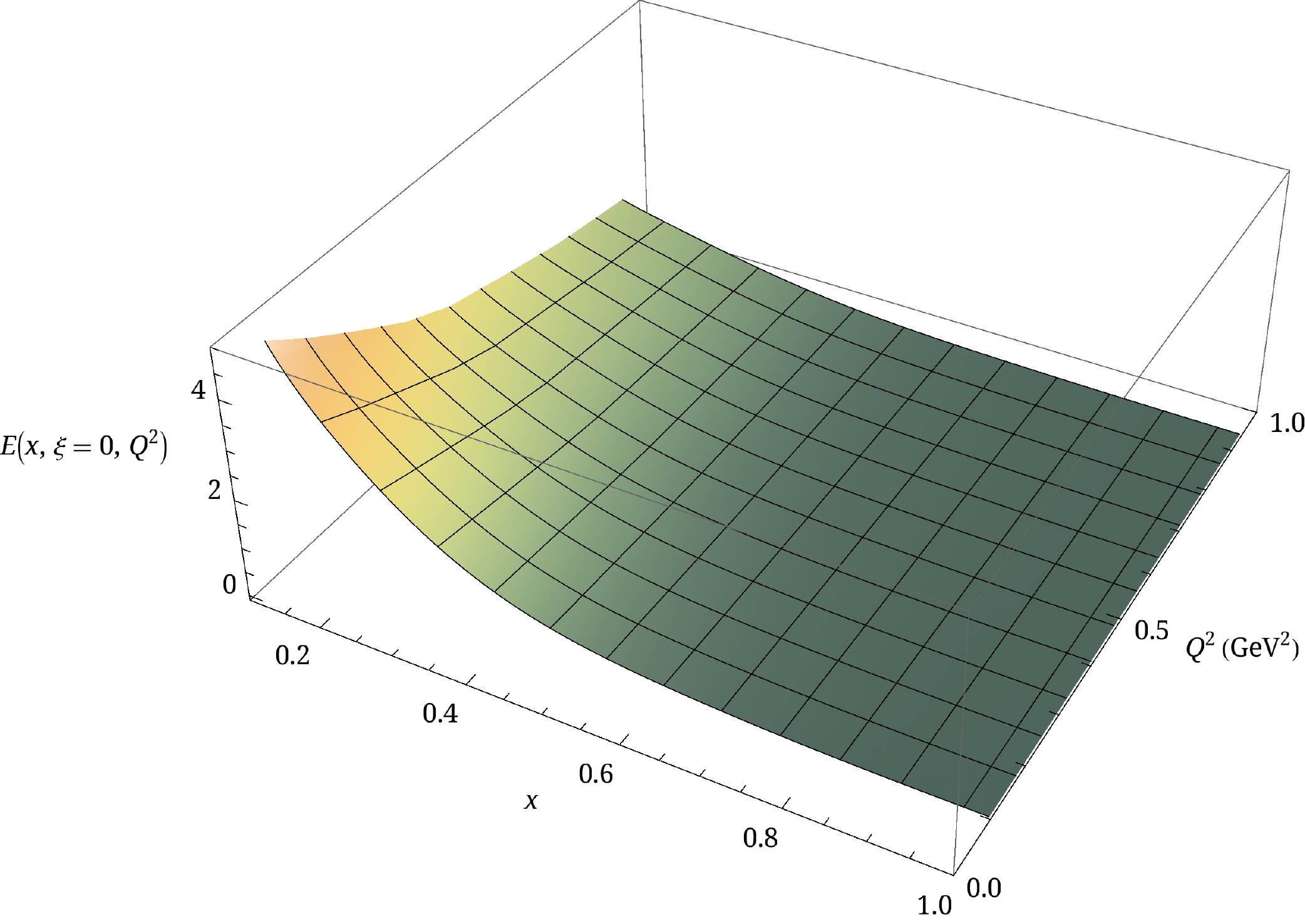}
\caption{
Nucleon isovector $H$ and $E$ GPDs at $\xi=0$ as functions of $x$ and momentum transfer $Q^2$.
\label{fig:3DGPD}}
\end{figure}

The nonperturbatively renormalized matrix elements  are then Fourier transformed into quasi-GPDs through two different approaches. 
Following the recent work~\cite{Chen:2018xof,Lin:2018qky,Liu:2018hxv}, we take the matrix elements $z \in [-12,12]$ and apply the simple but effective ``derivative'' method,
$\tilde{Q} = i\int_{-z_\text{max}}^{+z_\text{max}} \!\! dz\, e^{i x P_z z}  \tilde{h}'_R/x$, to obtain the quasi-GPDs. 
Alternatively, we adopt the extrapolation formulation suggested by Ref.~\cite{Ji:2020brr} by fitting $|z| \in \{10,15\}$ using the formula $c_1(-izP_z)^{-d_1}+c_2 e^{izP_z}(izP_z)^{-d_2}$, inspired by the Regge behavior, to extrapolate the matrix elements into the region beyond the lattice calculation and suppress Fourier-transformation artifacts.
Then, both quasi-GPDs are matched to the physical GPDs by applying the matching condition~\cite{Stewart:2017tvs,Liu:2018uuj,Chen:2018xof,Lin:2018qky}.
Examples of the GPDs at momentum transfer $Q^2\approx 0.4\text{ GeV}^2$ are shown in Fig.~\ref{fig:quasi-matched}.
Figure~\ref{fig:quasi-matched} compares the $H$ and $E$ GPDs at $Q^2\approx 0.4\text{ GeV}^2$ with the quasi-distribution and matched distribution using $P_z \approx 2.2$~GeV.
The matching lowers the positive mid-$x$ to large-$x$ distribution, as expected;
as one approaches lightcone limit, the probability of a parton carrying a larger fraction of its parent nucleon's momentum should become smaller.
We also find that the derivative and Regge-inspired extrapolation agree in the mid- to large-$x$ regions, but their difference grows as $x$ approaches zero in both the quark and antiquark distribution.
This is expected, since they differ mainly in the treatment of the large-$z$ matrix elements in the quasi-GPD Fourier transformation, which contributes more significantly to the small-$x$ distribution. 
By repeating a similar analysis for each available $Q^2$ in this calculation, we can construct the full three-dimensional shape of $H$ and $E$ as functions of $x$ and $Q^2$, as shown in Fig.~\ref{fig:3DGPD}. 
Due to the limited reliable $zP_z$ reach of the lattice calculation, we find that the small-$x$ region is unreliable, due to lack of precision lattice data to constrain it.
Thus, due to charge conservation, the antiquark (negative-$x$) distribution can also be sensitive to $P_z$.
It has been found in past work~\cite{Lin:2017ani,Chen:2018xof,Lin:2018qky,Liu:2018hxv} that higher boost momenta are needed to improve the antiquark region.
Therefore, for the rest of the work, we will mainly focus on the $x>0.05$ region.
For convenience, we will focus on showing the GPD results from derivative method, and use the Regge-inspired extrapolation to estimate the uncertainties in the small-$x$ region by reconstructing GPD moments from our $x$-dependent GPD functions. 

\begin{figure}[tb]
\centering
\includegraphics[width=0.22\textwidth]{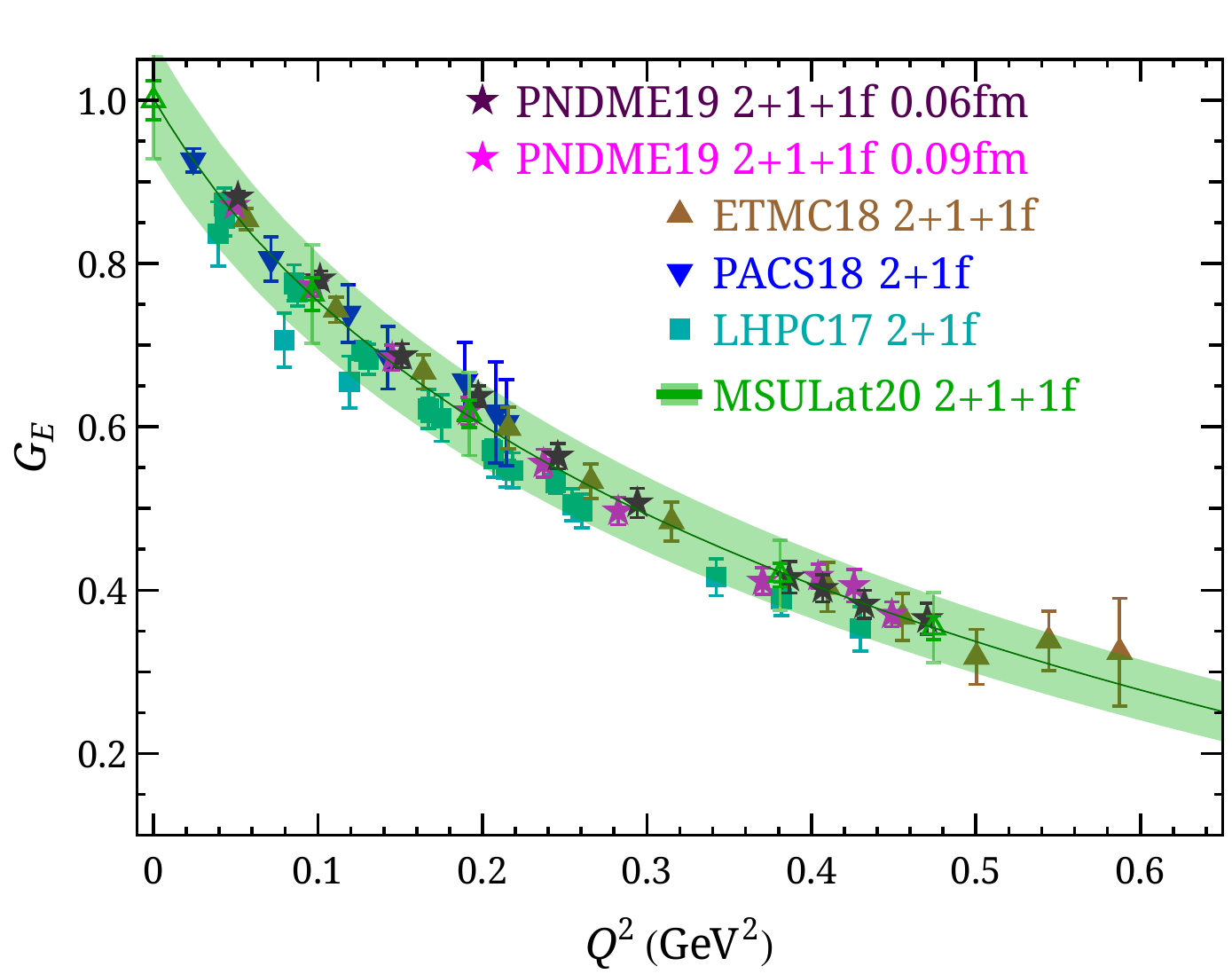} 
\includegraphics[width=0.22\textwidth]{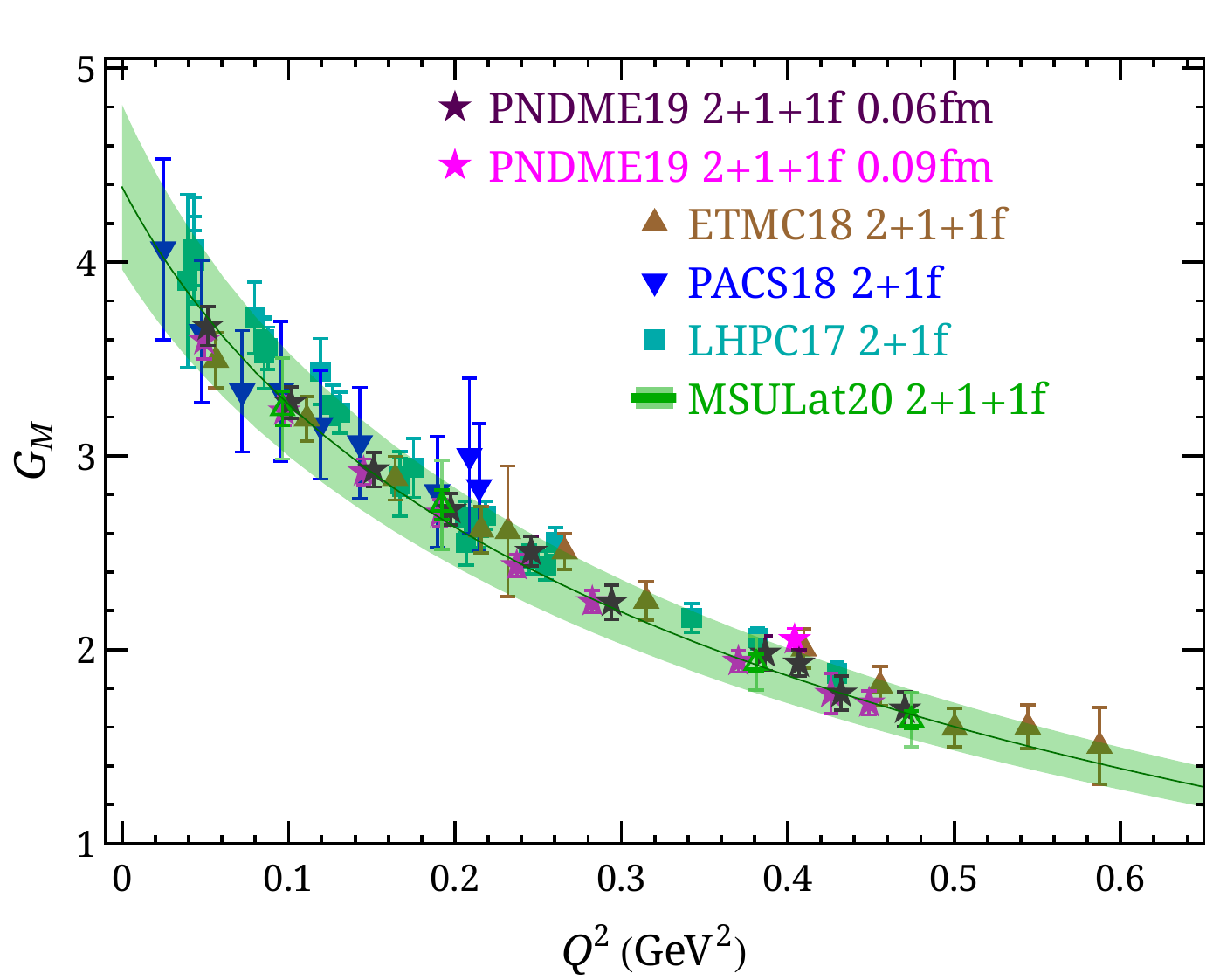}
\includegraphics[width=0.23\textwidth]{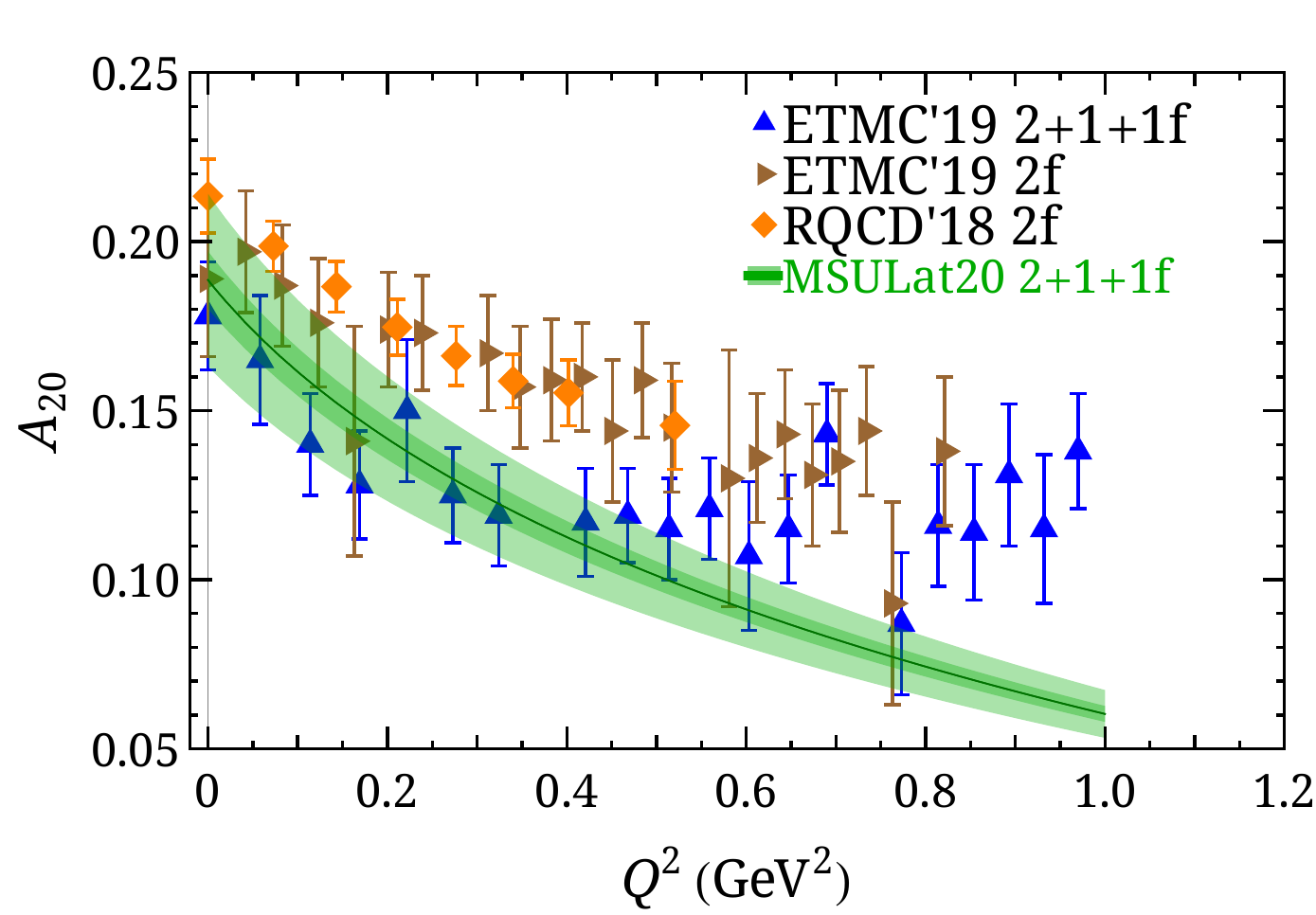} 
\includegraphics[width=0.23\textwidth]{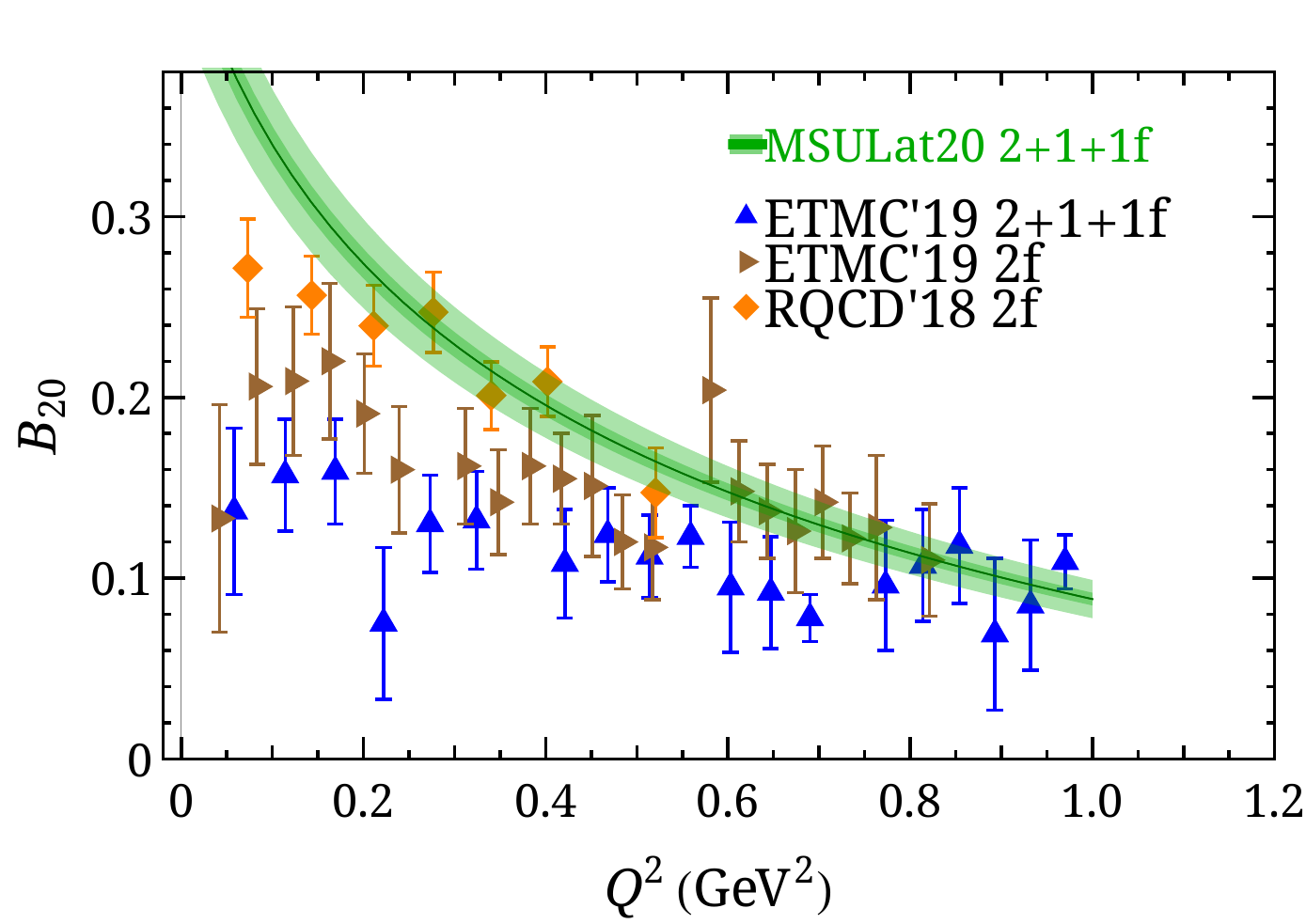}
\caption{
\label{fig:LatGFF}
(top) The nucleon isovector electric and magnetic form factor results obtained from this work (labeled as ``MSULat20 2+1+1'') as functions of transferred momentum $Q^2$, and comparison with other lattice works calculated near physical pion mass: 
$N_f=2$ ETMC18~\cite{Alexandrou:2017ypw} 
$N_f=2+1$ LHPC14~\cite{Green:2014xba}, 
LHPC17~\cite{Hasan:2017wwt}, 
PACS18~\cite{Shintani:2018ozy}; 
 $N_f=2+1+1$ ETMC18~\cite{Alexandrou:2018sjm},
PNDME19~\cite{Jang:2018djx} with 2 lattice spacings of 0.06 and 0.09~fm.
(bottom) The unpolarized nucleon isovector GFFs obtained from this work, compared with other lattice results calculated near physical pion mass as functions of transfer momentum $Q^2$:
ETMC19~\cite{Alexandrou:2019ali},
and RQCD19~\cite{Bali:2018zgl}.
}
\end{figure}

Since this is the first lattice calculation with full three-dimensional $x$ and $Q^2$ dependence of the $H$ and $E$ GPD functions, we would like to check how the new results using LaMET approach compare with the previous moment approaches to the generalized form factors.  
In the $\xi\to 0$ limit, the $H$ and $E$ GPDs decrease monotonically as $x$ or  $Q^2$ increases.
We take Mellin moments of the GPDs to compare with previous lattice calculations done using local matrix elements through the operator product expansion (OPE).
Taking the $x$-moments of $H$ and $E$~\cite{Ji:1998pc,Hagler:2009ni}:
\begin{subequations}
\begin{align}
\label{eq:GFFs}
\int_{-1}^{+1}\!\!dx \, x^{n-1} \, H(x, \xi, Q^2)&=  \nonumber \\
\sum\limits_{i=0,\text{ even}}^{n-1} (-2\xi)^i A_{ni}(Q^2) &+ (-2\xi)^{n} \,  C_{n0}(Q^2)|_{n\text{ even}}, \\
\int_{-1}^{+1}\!\!dx \, x^{n-1} \, E(x, \xi, Q^2)&=  \nonumber \\
\sum    \limits_{i=0,\text{ even}}^{n-1}	(-2\xi)^i B_{ni}(Q^2)	 &- (-2\xi)^{n} \,  C_{n0}(Q^2)|_{n\text{ even}}\,,
\end{align}
\end{subequations}
where the generalized form factors (GFFs) $A_{ni}(Q^2)$, $B_{ni}(Q^2)$ and $C_{ni}(Q^2)$ in the $\xi$-expansion on the right-hand side are real functions.
When $n=1$, we get the Dirac and Pauli electromagnetic form factors $F_1(Q^2) = A_{10}(Q^2)$ and $F_2(Q^2) = B_{10}(Q^2)$.
To compare with other lattice results, we plot the Sachs electric and magnetic form factors using $F_{1,2}$ as $G_E(Q^2)=F_1(Q^2)+q^2F_2(Q^2)/(2 M_N)^2$ and $G_M(Q^2)=F_1(Q^2)+F_2(Q^2)$ 
in Fig.~\ref{fig:LatGFF} 
with selected results from near-physical pion masses.
PACS has the largest volume among these calculations and is able to probe the smallest $Q^2$. 
Overall, our results are not only consistent within errors with the earlier PNDME study using the same ensemble (but which used local operators) but are also in good agreement with other lattice collaborations.
When $n=2$, we obtained GFFs of $A_{20}(Q^2)$ and $B_{20}(Q^2)$ so that we can compare our moment results with past lattice calculations using the OPE, as shown 
in Fig.~\ref{fig:LatGFF}.
We compare our moment results with those obtained from simulations at the physical point by ETMC using three ensembles~\cite{Alexandrou:2019ali} and the near-physical calculation of RQCD~\cite{Bali:2018zgl}. 
We note that even with the same OPE approach by the same collaboration, the two data sets for $A_{20}$ in the ETMC calculation exhibit some tension.
This is an indication that the systematic uncertainties are more complicated for these GFFs.
Given that the blue points correspond to finer lattice spacing, larger volume and larger $m_\pi L$, we expect that the blue points have suppressed systematic uncertainties.
Our moment result $A_{20}(Q^2)$ is in better agreement with those obtained using the OPE approach at small momentum transfer $Q^2$, while $B_{20}(Q^2)$ is in better agreement with OPE approaches at large $Q^2$. 
The comparison between the $N_f=2$ ETMC data and $N_f=2$ RQCD data reveals agreement for $A_{20}$ and $B_{20}$. 
However, the RQCD data have a different slope than the ETMC data, which is attributed to the different analysis methods and systematic uncertainties.
Both our results and ETMC's are done using a single ensemble;
future studies to include other lattice artifacts, such as lattice-spacing dependence, are important to account for the difference in the results.

Note that the error bands in Fig.~\ref{fig:LatGFF} include the systematics from the following:
1) The systematics due to the negative- and small-$x$ regions of the current GPD extraction.
We create pseudo-lattice data using input of CT18NNLO PDF~\cite{Hou:2019efy} with the same lattice parameters (such as $z$ and $P_z$) used in this calculation and apply the same analysis procedure described above.
We take the upper limit of the reconstructed and original CT18 moments as an estimate of the systematics introduced by the analysis procedure (e.g. by Fourier truncation);
2) We vary the maximum Wilson-line length $z$ within 2 lattice units and take half the difference as an estimate of the systematic due to finite $z$;
3) We estimate $1/P_z$ systematics due to higher-twist effects by comparing our $Q^2=0$ PDFs and to those in the previous works with 3 boost momenta~\cite{Chen:2018xof,Lin:2018qky}.
The final errors are summed in quadrature to create the final error bands shown in Fig.~\ref{fig:LatGFF}.

\begin{figure*}[tb]
\includegraphics[width=0.3\textwidth]{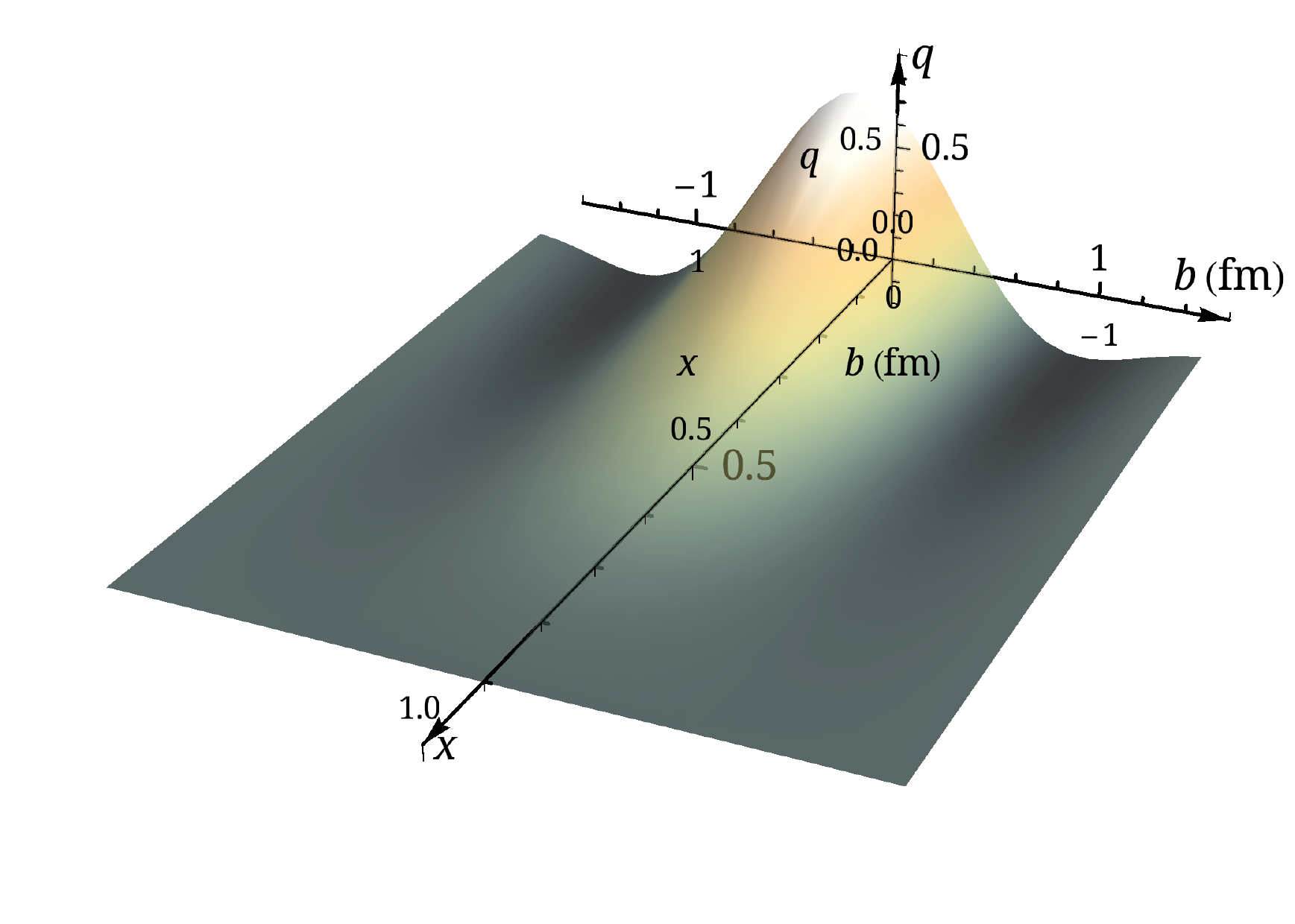}
\includegraphics[width=0.6\textwidth]{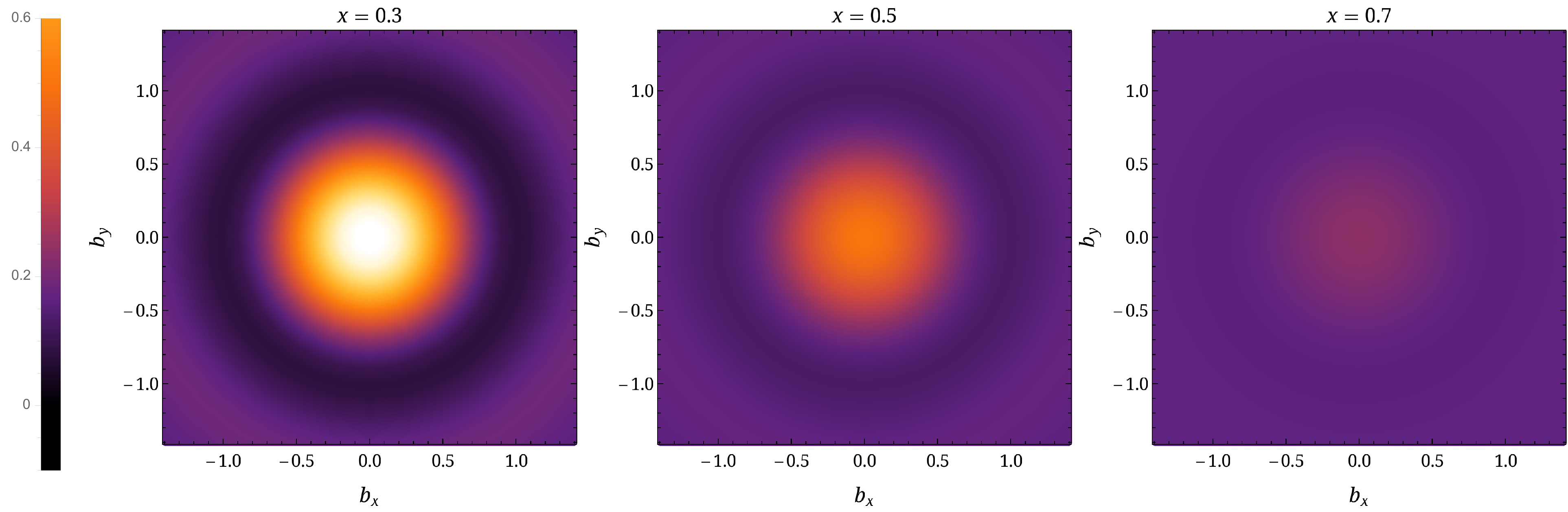}
\caption{
(Left) Nucleon tomography: three-dimensional impact parameter--dependent parton distribution as a function of $x$ and $b$ using lattice $H$ at physical pion mass.
(Right) The two-dimensional impact-parameter--dependent distribution for $x=0.3$, 0.5 and 0.7.
\label{fig:impact-distribution}}
\end{figure*}

The Fourier transform of the non--spin-flip GPD $H(x,\xi=0,Q^2)$ gives the impact-parameter--dependent distribution $\mathsf{q}(x,b)$~\cite{Burkardt:2002hr}
\begin{equation}\label{eq:impact-dist}
\mathsf{q}(x,b) = \int \frac{ d \mathbf{q}}{(2\pi)^2} H(x,\xi=0,t=-\mathbf{q}^2) e^{i\mathbf{q}\,\cdot \, \mathbf{b} },
\end{equation}
where $b$ is the transverse distance from the center of momentum.
Figure~\ref{fig:impact-distribution} shows the first results of impact-parameter--dependent distribution from lattice QCD:
a three-dimensional distribution as function of $x$ and $b$, and two-dimensional distributions at $x=0.3$, 0.5 and 0.7.
The impact-parameter--dependent distribution describes the probability density for a parton with momentum fraction $x$ at distance $b$ in the transverse plane.
Figure~\ref{fig:impact-distribution} shows that the probability decreases quickly as $x$ increases.
Using Eq.~\ref{eq:impact-dist} and the $H(x,\xi=0,t=-\mathbf{q}^2)$ obtained from the lattice calculation at the physical pion mass, we can take a snapshot of the nucleon in the transverse plane to perform $x$-dependent nucleon tomography using lattice QCD for the first time.

In this work, we compute the isovector nucleon unpolarized GPDs at physical pion mass using boost momentum $2.2$~GeV with nonzero momentum transfers in $[0.2,1.0]\text{ GeV}^2$. 
We are able to map out the first three-dimensional GPD structures using lattice QCD in the special limit $\xi=0$.
There are residual lattice systematics are not yet included in the current calculation:
In our past studies, we found the finite-volume effects to be negligible for isovector nucleon quasi-distributions calculated within the range $M_\pi^\text{val} L \in \{3.3, 5.5\}$.
We anticipate such systematics should be small compared to the statistical errors. 
The lattice discretization has been studied by MSULat collaboration in Refs.~\cite{Zhang:2020gaj,Lin:2020ssv} with multiple lattice spacings  in the LaMET study of pion and kaon distribution amplitudes and PDFs;
similarly, a comparison of nucleon isovector PDFs with 0.045 and 0.12~fm lattice spacing is shown in supplementary materials.  
There was mild lattice-spacing dependence for a majority of the Wilson-link displacements studied with similar largest boost momenta with same valence/sea lattice setup. 
EMTC also report LaMET isovector nucleon PDFs in Ref.~\cite{Alexandrou:2020qtt} using  twisted-mass fermion actions and reports different findings. 
Future work will investigate ensembles with smaller lattice spacing to reach even higher boost momentum (either directly or with the aid of machine learning~\cite{Zhang:2019qiq}) so that we can push toward reliable determination of the smaller-$x$ and antiquark regions.

\begin{acknowledgments}
We thank the MILC Collaboration for sharing the lattices used to perform this study. The LQCD calculations were performed using the Chroma software
suite~\cite{Edwards:2004sx}.
This research used resources of the National Energy Research Scientific Computing Center, a DOE Office of Science User Facility supported by the Office of Science of the U.S. Department of Energy under Contract No. DE-AC02-05CH11231 through ERCAP;
facilities of the USQCD Collaboration, which are funded by the Office of Science of the U.S. Department of Energy,
and supported in part by Michigan State University through computational resources provided by the Institute for Cyber-Enabled Research (iCER).
The work of HL is partially supported by the US National Science Foundation under grant PHY 1653405 ``CAREER: Constraining Parton Distribution Functions for New-Physics Searches''
and by the  Research  Corporation  for  Science  Advancement through the Cottrell Scholar Award ``Unveiling the  Three-Dimensional Structure of Nucleons''.
\end{acknowledgments}

\bibliographystyle{apsrev4-1}

\begin{thebibliography}{144}%
\makeatletter
\providecommand \@ifxundefined [1]{%
 \@ifx{#1\undefined}
}%
\providecommand \@ifnum [1]{%
 \ifnum #1\expandafter \@firstoftwo
 \else \expandafter \@secondoftwo
 \fi
}%
\providecommand \@ifx [1]{%
 \ifx #1\expandafter \@firstoftwo
 \else \expandafter \@secondoftwo
 \fi
}%
\providecommand \natexlab [1]{#1}%
\providecommand \enquote  [1]{``#1''}%
\providecommand \bibnamefont  [1]{#1}%
\providecommand \bibfnamefont [1]{#1}%
\providecommand \citenamefont [1]{#1}%
\providecommand \href@noop [0]{\@secondoftwo}%
\providecommand \href [0]{\begingroup \@sanitize@url \@href}%
\providecommand \@href[1]{\@@startlink{#1}\@@href}%
\providecommand \@@href[1]{\endgroup#1\@@endlink}%
\providecommand \@sanitize@url [0]{\catcode `\\12\catcode `\$12\catcode
  `\&12\catcode `\#12\catcode `\^12\catcode `\_12\catcode `\%12\relax}%
\providecommand \@@startlink[1]{}%
\providecommand \@@endlink[0]{}%
\providecommand \url  [0]{\begingroup\@sanitize@url \@url }%
\providecommand \@url [1]{\endgroup\@href {#1}{\urlprefix }}%
\providecommand \urlprefix  [0]{URL }%
\providecommand \Eprint [0]{\href }%
\providecommand \doibase [0]{http://dx.doi.org/}%
\providecommand \selectlanguage [0]{\@gobble}%
\providecommand \bibinfo  [0]{\@secondoftwo}%
\providecommand \bibfield  [0]{\@secondoftwo}%
\providecommand \translation [1]{[#1]}%
\providecommand \BibitemOpen [0]{}%
\providecommand \bibitemStop [0]{}%
\providecommand \bibitemNoStop [0]{.\EOS\space}%
\providecommand \EOS [0]{\spacefactor3000\relax}%
\providecommand \BibitemShut  [1]{\csname bibitem#1\endcsname}%
\let\auto@bib@innerbib\@empty
\bibitem [{\citenamefont {Müller}\ \emph {et~al.}(1994)\citenamefont
  {Müller}, \citenamefont {Robaschik}, \citenamefont {Geyer}, \citenamefont
  {Dittes},\ and\ \citenamefont {Ho\v{r}ej\v{s}i}}]{Mueller:1998fv}%
  \BibitemOpen
  \bibfield  {author} {\bibinfo {author} {\bibfnamefont {D.}~\bibnamefont
  {Müller}}, \bibinfo {author} {\bibfnamefont {D.}~\bibnamefont {Robaschik}},
  \bibinfo {author} {\bibfnamefont {B.}~\bibnamefont {Geyer}}, \bibinfo
  {author} {\bibfnamefont {F.-M.}\ \bibnamefont {Dittes}}, \ and\ \bibinfo
  {author} {\bibfnamefont {J.}~\bibnamefont {Ho\v{r}ej\v{s}i}},\ }\href
  {\doibase 10.1002/prop.2190420202} {\bibfield  {journal} {\bibinfo  {journal}
  {Fortsch. Phys.}\ }\textbf {\bibinfo {volume} {42}},\ \bibinfo {pages} {101}
  (\bibinfo {year} {1994})},\ \Eprint {http://arxiv.org/abs/hep-ph/9812448}
  {arXiv:hep-ph/9812448} \BibitemShut {NoStop}%
\bibitem [{\citenamefont {Ji}(1997{\natexlab{a}})}]{Ji:1996ek}%
  \BibitemOpen
  \bibfield  {author} {\bibinfo {author} {\bibfnamefont {X.-D.}\ \bibnamefont
  {Ji}},\ }\href {\doibase 10.1103/PhysRevLett.78.610} {\bibfield  {journal}
  {\bibinfo  {journal} {Phys. Rev. Lett.}\ }\textbf {\bibinfo {volume} {78}},\
  \bibinfo {pages} {610} (\bibinfo {year} {1997}{\natexlab{a}})},\ \Eprint
  {http://arxiv.org/abs/hep-ph/9603249} {arXiv:hep-ph/9603249 [hep-ph]}
  \BibitemShut {NoStop}%
\bibitem [{\citenamefont {Radyushkin}(1996)}]{Radyushkin:1996nd}%
  \BibitemOpen
  \bibfield  {author} {\bibinfo {author} {\bibfnamefont {A.}~\bibnamefont
  {Radyushkin}},\ }\href {\doibase 10.1016/0370-2693(96)00528-X} {\bibfield
  {journal} {\bibinfo  {journal} {Phys. Lett. B}\ }\textbf {\bibinfo {volume}
  {380}},\ \bibinfo {pages} {417} (\bibinfo {year} {1996})},\ \Eprint
  {http://arxiv.org/abs/hep-ph/9604317} {arXiv:hep-ph/9604317} \BibitemShut
  {NoStop}%
\bibitem [{\citenamefont {Burkardt}(2000)}]{Burkardt:2000za}%
  \BibitemOpen
  \bibfield  {author} {\bibinfo {author} {\bibfnamefont {M.}~\bibnamefont
  {Burkardt}},\ }\href {\doibase 10.1103/PhysRevD.62.071503,
  10.1103/PhysRevD.66.119903} {\bibfield  {journal} {\bibinfo  {journal} {Phys.
  Rev.}\ }\textbf {\bibinfo {volume} {D62}},\ \bibinfo {pages} {071503}
  (\bibinfo {year} {2000})},\ \bibinfo {note} {[Erratum: Phys.
  Rev.D66,119903(2002)]},\ \Eprint {http://arxiv.org/abs/hep-ph/0005108}
  {arXiv:hep-ph/0005108 [hep-ph]} \BibitemShut {NoStop}%
\bibitem [{\citenamefont {Ji}(1997{\natexlab{b}})}]{Ji:1996nm}%
  \BibitemOpen
  \bibfield  {author} {\bibinfo {author} {\bibfnamefont {X.-D.}\ \bibnamefont
  {Ji}},\ }\href {\doibase 10.1103/PhysRevD.55.7114} {\bibfield  {journal}
  {\bibinfo  {journal} {Phys. Rev.}\ }\textbf {\bibinfo {volume} {D55}},\
  \bibinfo {pages} {7114} (\bibinfo {year} {1997}{\natexlab{b}})},\ \Eprint
  {http://arxiv.org/abs/hep-ph/9609381} {arXiv:hep-ph/9609381 [hep-ph]}
  \BibitemShut {NoStop}%
\bibitem [{\citenamefont {Kriesten}\ \emph {et~al.}(2020)\citenamefont
  {Kriesten}, \citenamefont {Liuti}, \citenamefont {Calero-Diaz}, \citenamefont
  {Keller}, \citenamefont {Meyer}, \citenamefont {Goldstein},\ and\
  \citenamefont {Osvaldo Gonzalez-Hernandez}}]{Kriesten:2019jep}%
  \BibitemOpen
  \bibfield  {author} {\bibinfo {author} {\bibfnamefont {B.}~\bibnamefont
  {Kriesten}}, \bibinfo {author} {\bibfnamefont {S.}~\bibnamefont {Liuti}},
  \bibinfo {author} {\bibfnamefont {L.}~\bibnamefont {Calero-Diaz}}, \bibinfo
  {author} {\bibfnamefont {D.}~\bibnamefont {Keller}}, \bibinfo {author}
  {\bibfnamefont {A.}~\bibnamefont {Meyer}}, \bibinfo {author} {\bibfnamefont
  {G.~R.}\ \bibnamefont {Goldstein}}, \ and\ \bibinfo {author} {\bibfnamefont
  {J.}~\bibnamefont {Osvaldo Gonzalez-Hernandez}},\ }\href {\doibase
  10.1103/PhysRevD.101.054021} {\bibfield  {journal} {\bibinfo  {journal}
  {Phys. Rev. D}\ }\textbf {\bibinfo {volume} {101}},\ \bibinfo {pages}
  {054021} (\bibinfo {year} {2020})},\ \Eprint
  {http://arxiv.org/abs/1903.05742} {arXiv:1903.05742 [hep-ph]} \BibitemShut
  {NoStop}%
\bibitem [{\citenamefont {National Academies~of Sciences}\ and\ \citenamefont
  {Medicine}(2018)}]{NAP25171}%
  \BibitemOpen
  \bibfield  {author} {\bibinfo {author} {\bibfnamefont {E.}~\bibnamefont
  {National Academies~of Sciences}}\ and\ \bibinfo {author} {\bibnamefont
  {Medicine}},\ }\href {\doibase 10.17226/25171} {\emph {\bibinfo {title} {An
  Assessment of U.S.-Based Electron-Ion Collider Science}}}\ (\bibinfo
  {publisher} {The National Academies Press},\ \bibinfo {address} {Washington,
  DC},\ \bibinfo {year} {2018})\BibitemShut {NoStop}%
\bibitem [{\citenamefont {Chen}(2018)}]{Chen:2018wyz}%
  \BibitemOpen
  \bibfield  {author} {\bibinfo {author} {\bibfnamefont {X.}~\bibnamefont
  {Chen}},\ }\href {\doibase 10.22323/1.316.0170} {\bibfield  {journal}
  {\bibinfo  {journal} {PoS}\ }\textbf {\bibinfo {volume} {DIS2018}},\ \bibinfo
  {pages} {170} (\bibinfo {year} {2018})},\ \Eprint
  {http://arxiv.org/abs/1809.00448} {arXiv:1809.00448 [nucl-ex]} \BibitemShut
  {NoStop}%
\bibitem [{\citenamefont {Chen}\ \emph {et~al.}(2020)\citenamefont {Chen},
  \citenamefont {Guo}, \citenamefont {Roberts},\ and\ \citenamefont
  {Wang}}]{Chen:2020ijn}%
  \BibitemOpen
  \bibfield  {author} {\bibinfo {author} {\bibfnamefont {X.}~\bibnamefont
  {Chen}}, \bibinfo {author} {\bibfnamefont {F.-K.}\ \bibnamefont {Guo}},
  \bibinfo {author} {\bibfnamefont {C.~D.}\ \bibnamefont {Roberts}}, \ and\
  \bibinfo {author} {\bibfnamefont {R.}~\bibnamefont {Wang}}\ }(\bibinfo {year}
  {2020})\ \Eprint {http://arxiv.org/abs/2008.00102} {arXiv:2008.00102
  [hep-ph]} \BibitemShut {NoStop}%
\bibitem [{\citenamefont {Abelleira~Fernandez}\ \emph
  {et~al.}(2012)\citenamefont {Abelleira~Fernandez} \emph
  {et~al.}}]{AbelleiraFernandez:2012cc}%
  \BibitemOpen
  \bibfield  {author} {\bibinfo {author} {\bibfnamefont {J.}~\bibnamefont
  {Abelleira~Fernandez}} \emph {et~al.} (\bibinfo {collaboration} {LHeC Study
  Group}),\ }\href {\doibase 10.1088/0954-3899/39/7/075001} {\bibfield
  {journal} {\bibinfo  {journal} {J. Phys. G}\ }\textbf {\bibinfo {volume}
  {39}},\ \bibinfo {pages} {075001} (\bibinfo {year} {2012})},\ \Eprint
  {http://arxiv.org/abs/1206.2913} {arXiv:1206.2913 [physics.acc-ph]}
  \BibitemShut {NoStop}%
\bibitem [{\citenamefont {Agostini}\ \emph {et~al.}(2020)\citenamefont
  {Agostini} \emph {et~al.}}]{Agostini:2020fmq}%
  \BibitemOpen
  \bibfield  {author} {\bibinfo {author} {\bibfnamefont {P.}~\bibnamefont
  {Agostini}} \emph {et~al.} (\bibinfo {collaboration} {LHeC, FCC-he Study
  Group}),\ }\href@noop {} {\  (\bibinfo {year} {2020})},\ \Eprint
  {http://arxiv.org/abs/2007.14491} {arXiv:2007.14491 [hep-ex]} \BibitemShut
  {NoStop}%
\bibitem [{\citenamefont {Ji}(2013)}]{Ji:2013dva}%
  \BibitemOpen
  \bibfield  {author} {\bibinfo {author} {\bibfnamefont {X.}~\bibnamefont
  {Ji}},\ }\href {\doibase 10.1103/PhysRevLett.110.262002} {\bibfield
  {journal} {\bibinfo  {journal} {Phys. Rev. Lett.}\ }\textbf {\bibinfo
  {volume} {110}},\ \bibinfo {pages} {262002} (\bibinfo {year} {2013})},\
  \Eprint {http://arxiv.org/abs/1305.1539} {arXiv:1305.1539 [hep-ph]}
  \BibitemShut {NoStop}%
\bibitem [{\citenamefont {Ji}(2014)}]{Ji:2014gla}%
  \BibitemOpen
  \bibfield  {author} {\bibinfo {author} {\bibfnamefont {X.}~\bibnamefont
  {Ji}},\ }\href {\doibase 10.1007/s11433-014-5492-3} {\bibfield  {journal}
  {\bibinfo  {journal} {Sci. China Phys. Mech. Astron.}\ }\textbf {\bibinfo
  {volume} {57}},\ \bibinfo {pages} {1407} (\bibinfo {year} {2014})},\ \Eprint
  {http://arxiv.org/abs/1404.6680} {arXiv:1404.6680 [hep-ph]} \BibitemShut
  {NoStop}%
\bibitem [{\citenamefont {Ji}\ \emph {et~al.}(2017)\citenamefont {Ji},
  \citenamefont {Zhang},\ and\ \citenamefont {Zhao}}]{Ji:2017rah}%
  \BibitemOpen
  \bibfield  {author} {\bibinfo {author} {\bibfnamefont {X.}~\bibnamefont
  {Ji}}, \bibinfo {author} {\bibfnamefont {J.-H.}\ \bibnamefont {Zhang}}, \
  and\ \bibinfo {author} {\bibfnamefont {Y.}~\bibnamefont {Zhao}},\ }\href
  {\doibase 10.1016/j.nuclphysb.2017.09.001} {\bibfield  {journal} {\bibinfo
  {journal} {Nucl. Phys.}\ }\textbf {\bibinfo {volume} {B924}},\ \bibinfo
  {pages} {366} (\bibinfo {year} {2017})},\ \Eprint
  {http://arxiv.org/abs/1706.07416} {arXiv:1706.07416 [hep-ph]} \BibitemShut
  {NoStop}%
\bibitem [{\citenamefont {Lin}(2014{\natexlab{a}})}]{Lin:2013yra}%
  \BibitemOpen
  \bibfield  {author} {\bibinfo {author} {\bibfnamefont {H.-W.}\ \bibnamefont
  {Lin}},\ }\href {\doibase 10.22323/1.187.0293} {\bibfield  {journal}
  {\bibinfo  {journal} {PoS}\ }\textbf {\bibinfo {volume} {LATTICE2013}},\
  \bibinfo {pages} {293} (\bibinfo {year} {2014}{\natexlab{a}})}\BibitemShut
  {NoStop}%
\bibitem [{\citenamefont {Lin}\ \emph {et~al.}(2015)\citenamefont {Lin},
  \citenamefont {Chen}, \citenamefont {Cohen},\ and\ \citenamefont
  {Ji}}]{Lin:2014zya}%
  \BibitemOpen
  \bibfield  {author} {\bibinfo {author} {\bibfnamefont {H.-W.}\ \bibnamefont
  {Lin}}, \bibinfo {author} {\bibfnamefont {J.-W.}\ \bibnamefont {Chen}},
  \bibinfo {author} {\bibfnamefont {S.~D.}\ \bibnamefont {Cohen}}, \ and\
  \bibinfo {author} {\bibfnamefont {X.}~\bibnamefont {Ji}},\ }\href {\doibase
  10.1103/PhysRevD.91.054510} {\bibfield  {journal} {\bibinfo  {journal} {Phys.
  Rev.}\ }\textbf {\bibinfo {volume} {D91}},\ \bibinfo {pages} {054510}
  (\bibinfo {year} {2015})},\ \Eprint {http://arxiv.org/abs/1402.1462}
  {arXiv:1402.1462 [hep-ph]} \BibitemShut {NoStop}%
\bibitem [{\citenamefont {Chen}\ \emph {et~al.}(2016)\citenamefont {Chen},
  \citenamefont {Cohen}, \citenamefont {Ji}, \citenamefont {Lin},\ and\
  \citenamefont {Zhang}}]{Chen:2016utp}%
  \BibitemOpen
  \bibfield  {author} {\bibinfo {author} {\bibfnamefont {J.-W.}\ \bibnamefont
  {Chen}}, \bibinfo {author} {\bibfnamefont {S.~D.}\ \bibnamefont {Cohen}},
  \bibinfo {author} {\bibfnamefont {X.}~\bibnamefont {Ji}}, \bibinfo {author}
  {\bibfnamefont {H.-W.}\ \bibnamefont {Lin}}, \ and\ \bibinfo {author}
  {\bibfnamefont {J.-H.}\ \bibnamefont {Zhang}},\ }\href {\doibase
  10.1016/j.nuclphysb.2016.07.033} {\bibfield  {journal} {\bibinfo  {journal}
  {Nucl. Phys.}\ }\textbf {\bibinfo {volume} {B911}},\ \bibinfo {pages} {246}
  (\bibinfo {year} {2016})},\ \Eprint {http://arxiv.org/abs/1603.06664}
  {arXiv:1603.06664 [hep-ph]} \BibitemShut {NoStop}%
\bibitem [{\citenamefont {Lin}\ \emph {et~al.}(2018{\natexlab{a}})\citenamefont
  {Lin}, \citenamefont {Chen}, \citenamefont {Ishikawa},\ and\ \citenamefont
  {Zhang}}]{Lin:2017ani}%
  \BibitemOpen
  \bibfield  {author} {\bibinfo {author} {\bibfnamefont {H.-W.}\ \bibnamefont
  {Lin}}, \bibinfo {author} {\bibfnamefont {J.-W.}\ \bibnamefont {Chen}},
  \bibinfo {author} {\bibfnamefont {T.}~\bibnamefont {Ishikawa}}, \ and\
  \bibinfo {author} {\bibfnamefont {J.-H.}\ \bibnamefont {Zhang}} (\bibinfo
  {collaboration} {LP3}),\ }\href {\doibase 10.1103/PhysRevD.98.054504}
  {\bibfield  {journal} {\bibinfo  {journal} {Phys. Rev.}\ }\textbf {\bibinfo
  {volume} {D98}},\ \bibinfo {pages} {054504} (\bibinfo {year}
  {2018}{\natexlab{a}})},\ \Eprint {http://arxiv.org/abs/1708.05301}
  {arXiv:1708.05301 [hep-lat]} \BibitemShut {NoStop}%
\bibitem [{\citenamefont {Alexandrou}\ \emph {et~al.}(2015)\citenamefont
  {Alexandrou}, \citenamefont {Cichy}, \citenamefont {Drach}, \citenamefont
  {Garcia-Ramos}, \citenamefont {Hadjiyiannakou}, \citenamefont {Jansen},
  \citenamefont {Steffens},\ and\ \citenamefont {Wiese}}]{Alexandrou:2015rja}%
  \BibitemOpen
  \bibfield  {author} {\bibinfo {author} {\bibfnamefont {C.}~\bibnamefont
  {Alexandrou}}, \bibinfo {author} {\bibfnamefont {K.}~\bibnamefont {Cichy}},
  \bibinfo {author} {\bibfnamefont {V.}~\bibnamefont {Drach}}, \bibinfo
  {author} {\bibfnamefont {E.}~\bibnamefont {Garcia-Ramos}}, \bibinfo {author}
  {\bibfnamefont {K.}~\bibnamefont {Hadjiyiannakou}}, \bibinfo {author}
  {\bibfnamefont {K.}~\bibnamefont {Jansen}}, \bibinfo {author} {\bibfnamefont
  {F.}~\bibnamefont {Steffens}}, \ and\ \bibinfo {author} {\bibfnamefont
  {C.}~\bibnamefont {Wiese}},\ }\href {\doibase 10.1103/PhysRevD.92.014502}
  {\bibfield  {journal} {\bibinfo  {journal} {Phys. Rev.}\ }\textbf {\bibinfo
  {volume} {D92}},\ \bibinfo {pages} {014502} (\bibinfo {year} {2015})},\
  \Eprint {http://arxiv.org/abs/1504.07455} {arXiv:1504.07455 [hep-lat]}
  \BibitemShut {NoStop}%
\bibitem [{\citenamefont {Alexandrou}\ \emph
  {et~al.}(2017{\natexlab{a}})\citenamefont {Alexandrou}, \citenamefont
  {Cichy}, \citenamefont {Constantinou}, \citenamefont {Hadjiyiannakou},
  \citenamefont {Jansen}, \citenamefont {Steffens},\ and\ \citenamefont
  {Wiese}}]{Alexandrou:2016jqi}%
  \BibitemOpen
  \bibfield  {author} {\bibinfo {author} {\bibfnamefont {C.}~\bibnamefont
  {Alexandrou}}, \bibinfo {author} {\bibfnamefont {K.}~\bibnamefont {Cichy}},
  \bibinfo {author} {\bibfnamefont {M.}~\bibnamefont {Constantinou}}, \bibinfo
  {author} {\bibfnamefont {K.}~\bibnamefont {Hadjiyiannakou}}, \bibinfo
  {author} {\bibfnamefont {K.}~\bibnamefont {Jansen}}, \bibinfo {author}
  {\bibfnamefont {F.}~\bibnamefont {Steffens}}, \ and\ \bibinfo {author}
  {\bibfnamefont {C.}~\bibnamefont {Wiese}},\ }\href {\doibase
  10.1103/PhysRevD.96.014513} {\bibfield  {journal} {\bibinfo  {journal} {Phys.
  Rev.}\ }\textbf {\bibinfo {volume} {D96}},\ \bibinfo {pages} {014513}
  (\bibinfo {year} {2017}{\natexlab{a}})},\ \Eprint
  {http://arxiv.org/abs/1610.03689} {arXiv:1610.03689 [hep-lat]} \BibitemShut
  {NoStop}%
\bibitem [{\citenamefont {Alexandrou}\ \emph
  {et~al.}(2017{\natexlab{b}})\citenamefont {Alexandrou}, \citenamefont
  {Cichy}, \citenamefont {Constantinou}, \citenamefont {Hadjiyiannakou},
  \citenamefont {Jansen}, \citenamefont {Panagopoulos},\ and\ \citenamefont
  {Steffens}}]{Alexandrou:2017huk}%
  \BibitemOpen
  \bibfield  {author} {\bibinfo {author} {\bibfnamefont {C.}~\bibnamefont
  {Alexandrou}}, \bibinfo {author} {\bibfnamefont {K.}~\bibnamefont {Cichy}},
  \bibinfo {author} {\bibfnamefont {M.}~\bibnamefont {Constantinou}}, \bibinfo
  {author} {\bibfnamefont {K.}~\bibnamefont {Hadjiyiannakou}}, \bibinfo
  {author} {\bibfnamefont {K.}~\bibnamefont {Jansen}}, \bibinfo {author}
  {\bibfnamefont {H.}~\bibnamefont {Panagopoulos}}, \ and\ \bibinfo {author}
  {\bibfnamefont {F.}~\bibnamefont {Steffens}},\ }\href {\doibase
  10.1016/j.nuclphysb.2017.08.012} {\bibfield  {journal} {\bibinfo  {journal}
  {Nucl. Phys.}\ }\textbf {\bibinfo {volume} {B923}},\ \bibinfo {pages} {394}
  (\bibinfo {year} {2017}{\natexlab{b}})},\ \Eprint
  {http://arxiv.org/abs/1706.00265} {arXiv:1706.00265 [hep-lat]} \BibitemShut
  {NoStop}%
\bibitem [{\citenamefont {Chen}\ \emph
  {et~al.}(2018{\natexlab{a}})\citenamefont {Chen}, \citenamefont {Ishikawa},
  \citenamefont {Jin}, \citenamefont {Lin}, \citenamefont {Yang}, \citenamefont
  {Zhang},\ and\ \citenamefont {Zhao}}]{Chen:2017mzz}%
  \BibitemOpen
  \bibfield  {author} {\bibinfo {author} {\bibfnamefont {J.-W.}\ \bibnamefont
  {Chen}}, \bibinfo {author} {\bibfnamefont {T.}~\bibnamefont {Ishikawa}},
  \bibinfo {author} {\bibfnamefont {L.}~\bibnamefont {Jin}}, \bibinfo {author}
  {\bibfnamefont {H.-W.}\ \bibnamefont {Lin}}, \bibinfo {author} {\bibfnamefont
  {Y.-B.}\ \bibnamefont {Yang}}, \bibinfo {author} {\bibfnamefont {J.-H.}\
  \bibnamefont {Zhang}}, \ and\ \bibinfo {author} {\bibfnamefont
  {Y.}~\bibnamefont {Zhao}},\ }\href {\doibase 10.1103/PhysRevD.97.014505}
  {\bibfield  {journal} {\bibinfo  {journal} {Phys. Rev.}\ }\textbf {\bibinfo
  {volume} {D97}},\ \bibinfo {pages} {014505} (\bibinfo {year}
  {2018}{\natexlab{a}})},\ \Eprint {http://arxiv.org/abs/1706.01295}
  {arXiv:1706.01295 [hep-lat]} \BibitemShut {NoStop}%
\bibitem [{\citenamefont {Alexandrou}\ \emph
  {et~al.}(2018{\natexlab{a}})\citenamefont {Alexandrou}, \citenamefont
  {Cichy}, \citenamefont {Constantinou}, \citenamefont {Jansen}, \citenamefont
  {Scapellato},\ and\ \citenamefont {Steffens}}]{Alexandrou:2018pbm}%
  \BibitemOpen
  \bibfield  {author} {\bibinfo {author} {\bibfnamefont {C.}~\bibnamefont
  {Alexandrou}}, \bibinfo {author} {\bibfnamefont {K.}~\bibnamefont {Cichy}},
  \bibinfo {author} {\bibfnamefont {M.}~\bibnamefont {Constantinou}}, \bibinfo
  {author} {\bibfnamefont {K.}~\bibnamefont {Jansen}}, \bibinfo {author}
  {\bibfnamefont {A.}~\bibnamefont {Scapellato}}, \ and\ \bibinfo {author}
  {\bibfnamefont {F.}~\bibnamefont {Steffens}},\ }\href {\doibase
  10.1103/PhysRevLett.121.112001} {\bibfield  {journal} {\bibinfo  {journal}
  {Phys. Rev. Lett.}\ }\textbf {\bibinfo {volume} {121}},\ \bibinfo {pages}
  {112001} (\bibinfo {year} {2018}{\natexlab{a}})},\ \Eprint
  {http://arxiv.org/abs/1803.02685} {arXiv:1803.02685 [hep-lat]} \BibitemShut
  {NoStop}%
\bibitem [{\citenamefont {Chen}\ \emph
  {et~al.}(2018{\natexlab{b}})\citenamefont {Chen}, \citenamefont {Jin},
  \citenamefont {Lin}, \citenamefont {Liu}, \citenamefont {Yang}, \citenamefont
  {Zhang},\ and\ \citenamefont {Zhao}}]{Chen:2018xof}%
  \BibitemOpen
  \bibfield  {author} {\bibinfo {author} {\bibfnamefont {J.-W.}\ \bibnamefont
  {Chen}}, \bibinfo {author} {\bibfnamefont {L.}~\bibnamefont {Jin}}, \bibinfo
  {author} {\bibfnamefont {H.-W.}\ \bibnamefont {Lin}}, \bibinfo {author}
  {\bibfnamefont {Y.-S.}\ \bibnamefont {Liu}}, \bibinfo {author} {\bibfnamefont
  {Y.-B.}\ \bibnamefont {Yang}}, \bibinfo {author} {\bibfnamefont {J.-H.}\
  \bibnamefont {Zhang}}, \ and\ \bibinfo {author} {\bibfnamefont
  {Y.}~\bibnamefont {Zhao}},\ }\href@noop {} {\  (\bibinfo {year}
  {2018}{\natexlab{b}})},\ \Eprint {http://arxiv.org/abs/1803.04393}
  {arXiv:1803.04393 [hep-lat]} \BibitemShut {NoStop}%
\bibitem [{\citenamefont {Zhang}\ \emph
  {et~al.}(2019{\natexlab{a}})\citenamefont {Zhang}, \citenamefont {Chen},
  \citenamefont {Jin}, \citenamefont {Lin}, \citenamefont {Schäfer},\ and\
  \citenamefont {Zhao}}]{Chen:2018fwa}%
  \BibitemOpen
  \bibfield  {author} {\bibinfo {author} {\bibfnamefont {J.-H.}\ \bibnamefont
  {Zhang}}, \bibinfo {author} {\bibfnamefont {J.-W.}\ \bibnamefont {Chen}},
  \bibinfo {author} {\bibfnamefont {L.}~\bibnamefont {Jin}}, \bibinfo {author}
  {\bibfnamefont {H.-W.}\ \bibnamefont {Lin}}, \bibinfo {author} {\bibfnamefont
  {A.}~\bibnamefont {Schäfer}}, \ and\ \bibinfo {author} {\bibfnamefont
  {Y.}~\bibnamefont {Zhao}},\ }\href {\doibase 10.1103/PhysRevD.100.034505}
  {\bibfield  {journal} {\bibinfo  {journal} {Phys. Rev. D}\ }\textbf {\bibinfo
  {volume} {100}},\ \bibinfo {pages} {034505} (\bibinfo {year}
  {2019}{\natexlab{a}})},\ \Eprint {http://arxiv.org/abs/1804.01483}
  {arXiv:1804.01483 [hep-lat]} \BibitemShut {NoStop}%
\bibitem [{\citenamefont {Alexandrou}\ \emph
  {et~al.}(2018{\natexlab{b}})\citenamefont {Alexandrou}, \citenamefont
  {Cichy}, \citenamefont {Constantinou}, \citenamefont {Jansen}, \citenamefont
  {Scapellato},\ and\ \citenamefont {Steffens}}]{Alexandrou:2018eet}%
  \BibitemOpen
  \bibfield  {author} {\bibinfo {author} {\bibfnamefont {C.}~\bibnamefont
  {Alexandrou}}, \bibinfo {author} {\bibfnamefont {K.}~\bibnamefont {Cichy}},
  \bibinfo {author} {\bibfnamefont {M.}~\bibnamefont {Constantinou}}, \bibinfo
  {author} {\bibfnamefont {K.}~\bibnamefont {Jansen}}, \bibinfo {author}
  {\bibfnamefont {A.}~\bibnamefont {Scapellato}}, \ and\ \bibinfo {author}
  {\bibfnamefont {F.}~\bibnamefont {Steffens}},\ }\href {\doibase
  10.1103/PhysRevD.98.091503} {\bibfield  {journal} {\bibinfo  {journal} {Phys.
  Rev.}\ }\textbf {\bibinfo {volume} {D98}},\ \bibinfo {pages} {091503}
  (\bibinfo {year} {2018}{\natexlab{b}})},\ \Eprint
  {http://arxiv.org/abs/1807.00232} {arXiv:1807.00232 [hep-lat]} \BibitemShut
  {NoStop}%
\bibitem [{\citenamefont {Lin}\ \emph {et~al.}(2018{\natexlab{b}})\citenamefont
  {Lin}, \citenamefont {Chen}, \citenamefont {Ji}, \citenamefont {Jin},
  \citenamefont {Li}, \citenamefont {Liu}, \citenamefont {Yang}, \citenamefont
  {Zhang},\ and\ \citenamefont {Zhao}}]{Lin:2018qky}%
  \BibitemOpen
  \bibfield  {author} {\bibinfo {author} {\bibfnamefont {H.-W.}\ \bibnamefont
  {Lin}}, \bibinfo {author} {\bibfnamefont {J.-W.}\ \bibnamefont {Chen}},
  \bibinfo {author} {\bibfnamefont {X.}~\bibnamefont {Ji}}, \bibinfo {author}
  {\bibfnamefont {L.}~\bibnamefont {Jin}}, \bibinfo {author} {\bibfnamefont
  {R.}~\bibnamefont {Li}}, \bibinfo {author} {\bibfnamefont {Y.-S.}\
  \bibnamefont {Liu}}, \bibinfo {author} {\bibfnamefont {Y.-B.}\ \bibnamefont
  {Yang}}, \bibinfo {author} {\bibfnamefont {J.-H.}\ \bibnamefont {Zhang}}, \
  and\ \bibinfo {author} {\bibfnamefont {Y.}~\bibnamefont {Zhao}},\ }\href
  {\doibase 10.1103/PhysRevLett.121.242003} {\bibfield  {journal} {\bibinfo
  {journal} {Phys. Rev. Lett.}\ }\textbf {\bibinfo {volume} {121}},\ \bibinfo
  {pages} {242003} (\bibinfo {year} {2018}{\natexlab{b}})},\ \Eprint
  {http://arxiv.org/abs/1807.07431} {arXiv:1807.07431 [hep-lat]} \BibitemShut
  {NoStop}%
\bibitem [{\citenamefont {Fan}\ \emph {et~al.}(2018)\citenamefont {Fan},
  \citenamefont {Yang}, \citenamefont {Anthony}, \citenamefont {Lin},\ and\
  \citenamefont {Liu}}]{Fan:2018dxu}%
  \BibitemOpen
  \bibfield  {author} {\bibinfo {author} {\bibfnamefont {Z.-Y.}\ \bibnamefont
  {Fan}}, \bibinfo {author} {\bibfnamefont {Y.-B.}\ \bibnamefont {Yang}},
  \bibinfo {author} {\bibfnamefont {A.}~\bibnamefont {Anthony}}, \bibinfo
  {author} {\bibfnamefont {H.-W.}\ \bibnamefont {Lin}}, \ and\ \bibinfo
  {author} {\bibfnamefont {K.-F.}\ \bibnamefont {Liu}},\ }\href {\doibase
  10.1103/PhysRevLett.121.242001} {\bibfield  {journal} {\bibinfo  {journal}
  {Phys. Rev. Lett.}\ }\textbf {\bibinfo {volume} {121}},\ \bibinfo {pages}
  {242001} (\bibinfo {year} {2018})},\ \Eprint
  {http://arxiv.org/abs/1808.02077} {arXiv:1808.02077 [hep-lat]} \BibitemShut
  {NoStop}%
\bibitem [{\citenamefont {Liu}\ \emph {et~al.}(2018{\natexlab{a}})\citenamefont
  {Liu}, \citenamefont {Chen}, \citenamefont {Jin}, \citenamefont {Li},
  \citenamefont {Lin}, \citenamefont {Yang}, \citenamefont {Zhang},\ and\
  \citenamefont {Zhao}}]{Liu:2018hxv}%
  \BibitemOpen
  \bibfield  {author} {\bibinfo {author} {\bibfnamefont {Y.-S.}\ \bibnamefont
  {Liu}}, \bibinfo {author} {\bibfnamefont {J.-W.}\ \bibnamefont {Chen}},
  \bibinfo {author} {\bibfnamefont {L.}~\bibnamefont {Jin}}, \bibinfo {author}
  {\bibfnamefont {R.}~\bibnamefont {Li}}, \bibinfo {author} {\bibfnamefont
  {H.-W.}\ \bibnamefont {Lin}}, \bibinfo {author} {\bibfnamefont {Y.-B.}\
  \bibnamefont {Yang}}, \bibinfo {author} {\bibfnamefont {J.-H.}\ \bibnamefont
  {Zhang}}, \ and\ \bibinfo {author} {\bibfnamefont {Y.}~\bibnamefont {Zhao}},\
  }\href@noop {} {\  (\bibinfo {year} {2018}{\natexlab{a}})},\ \Eprint
  {http://arxiv.org/abs/1810.05043} {arXiv:1810.05043 [hep-lat]} \BibitemShut
  {NoStop}%
\bibitem [{\citenamefont {Wang}\ \emph {et~al.}(2019)\citenamefont {Wang},
  \citenamefont {Zhang}, \citenamefont {Zhao},\ and\ \citenamefont
  {Zhu}}]{Wang:2019tgg}%
  \BibitemOpen
  \bibfield  {author} {\bibinfo {author} {\bibfnamefont {W.}~\bibnamefont
  {Wang}}, \bibinfo {author} {\bibfnamefont {J.-H.}\ \bibnamefont {Zhang}},
  \bibinfo {author} {\bibfnamefont {S.}~\bibnamefont {Zhao}}, \ and\ \bibinfo
  {author} {\bibfnamefont {R.}~\bibnamefont {Zhu}},\ }\href@noop {} {\
  (\bibinfo {year} {2019})},\ \Eprint {http://arxiv.org/abs/1904.00978}
  {arXiv:1904.00978 [hep-ph]} \BibitemShut {NoStop}%
\bibitem [{\citenamefont {Lin}\ and\ \citenamefont
  {Zhang}(2019)}]{Lin:2019ocg}%
  \BibitemOpen
  \bibfield  {author} {\bibinfo {author} {\bibfnamefont {H.-W.}\ \bibnamefont
  {Lin}}\ and\ \bibinfo {author} {\bibfnamefont {R.}~\bibnamefont {Zhang}},\
  }\href {\doibase 10.1103/PhysRevD.100.074502} {\bibfield  {journal} {\bibinfo
   {journal} {Phys. Rev.}\ }\textbf {\bibinfo {volume} {D100}},\ \bibinfo
  {pages} {074502} (\bibinfo {year} {2019})}\BibitemShut {NoStop}%
\bibitem [{\citenamefont {Chen}\ \emph {et~al.}(2019)\citenamefont {Chen},
  \citenamefont {Lin},\ and\ \citenamefont {Zhang}}]{Chen:2019lcm}%
  \BibitemOpen
  \bibfield  {author} {\bibinfo {author} {\bibfnamefont {J.-W.}\ \bibnamefont
  {Chen}}, \bibinfo {author} {\bibfnamefont {H.-W.}\ \bibnamefont {Lin}}, \
  and\ \bibinfo {author} {\bibfnamefont {J.-H.}\ \bibnamefont {Zhang}},\ }\href
  {\doibase 10.1016/j.nuclphysb.2020.114940} {\  (\bibinfo {year} {2019}),\
  10.1016/j.nuclphysb.2020.114940},\ \Eprint {http://arxiv.org/abs/1904.12376}
  {arXiv:1904.12376 [hep-lat]} \BibitemShut {NoStop}%
\bibitem [{\citenamefont {Xiong}\ \emph {et~al.}(2014)\citenamefont {Xiong},
  \citenamefont {Ji}, \citenamefont {Zhang},\ and\ \citenamefont
  {Zhao}}]{Xiong:2013bka}%
  \BibitemOpen
  \bibfield  {author} {\bibinfo {author} {\bibfnamefont {X.}~\bibnamefont
  {Xiong}}, \bibinfo {author} {\bibfnamefont {X.}~\bibnamefont {Ji}}, \bibinfo
  {author} {\bibfnamefont {J.-H.}\ \bibnamefont {Zhang}}, \ and\ \bibinfo
  {author} {\bibfnamefont {Y.}~\bibnamefont {Zhao}},\ }\href {\doibase
  10.1103/PhysRevD.90.014051} {\bibfield  {journal} {\bibinfo  {journal} {Phys.
  Rev.}\ }\textbf {\bibinfo {volume} {D90}},\ \bibinfo {pages} {014051}
  (\bibinfo {year} {2014})},\ \Eprint {http://arxiv.org/abs/1310.7471}
  {arXiv:1310.7471 [hep-ph]} \BibitemShut {NoStop}%
\bibitem [{\citenamefont {Ji}\ and\ \citenamefont {Zhang}(2015)}]{Ji:2015jwa}%
  \BibitemOpen
  \bibfield  {author} {\bibinfo {author} {\bibfnamefont {X.}~\bibnamefont
  {Ji}}\ and\ \bibinfo {author} {\bibfnamefont {J.-H.}\ \bibnamefont {Zhang}},\
  }\href {\doibase 10.1103/PhysRevD.92.034006} {\bibfield  {journal} {\bibinfo
  {journal} {Phys. Rev.}\ }\textbf {\bibinfo {volume} {D92}},\ \bibinfo {pages}
  {034006} (\bibinfo {year} {2015})},\ \Eprint
  {http://arxiv.org/abs/1505.07699} {arXiv:1505.07699 [hep-ph]} \BibitemShut
  {NoStop}%
\bibitem [{\citenamefont {Ji}\ \emph {et~al.}(2015{\natexlab{a}})\citenamefont
  {Ji}, \citenamefont {Schäfer}, \citenamefont {Xiong},\ and\ \citenamefont
  {Zhang}}]{Ji:2015qla}%
  \BibitemOpen
  \bibfield  {author} {\bibinfo {author} {\bibfnamefont {X.}~\bibnamefont
  {Ji}}, \bibinfo {author} {\bibfnamefont {A.}~\bibnamefont {Schäfer}},
  \bibinfo {author} {\bibfnamefont {X.}~\bibnamefont {Xiong}}, \ and\ \bibinfo
  {author} {\bibfnamefont {J.-H.}\ \bibnamefont {Zhang}},\ }\href {\doibase
  10.1103/PhysRevD.92.014039} {\bibfield  {journal} {\bibinfo  {journal} {Phys.
  Rev. D}\ }\textbf {\bibinfo {volume} {92}},\ \bibinfo {pages} {014039}
  (\bibinfo {year} {2015}{\natexlab{a}})},\ \Eprint
  {http://arxiv.org/abs/1506.00248} {arXiv:1506.00248 [hep-ph]} \BibitemShut
  {NoStop}%
\bibitem [{\citenamefont {Xiong}\ and\ \citenamefont
  {Zhang}(2015)}]{Xiong:2015nua}%
  \BibitemOpen
  \bibfield  {author} {\bibinfo {author} {\bibfnamefont {X.}~\bibnamefont
  {Xiong}}\ and\ \bibinfo {author} {\bibfnamefont {J.-H.}\ \bibnamefont
  {Zhang}},\ }\href {\doibase 10.1103/PhysRevD.92.054037} {\bibfield  {journal}
  {\bibinfo  {journal} {Phys. Rev.}\ }\textbf {\bibinfo {volume} {D92}},\
  \bibinfo {pages} {054037} (\bibinfo {year} {2015})},\ \Eprint
  {http://arxiv.org/abs/1509.08016} {arXiv:1509.08016 [hep-ph]} \BibitemShut
  {NoStop}%
\bibitem [{\citenamefont {Ji}\ \emph {et~al.}(2015{\natexlab{b}})\citenamefont
  {Ji}, \citenamefont {Sun}, \citenamefont {Xiong},\ and\ \citenamefont
  {Yuan}}]{Ji:2014hxa}%
  \BibitemOpen
  \bibfield  {author} {\bibinfo {author} {\bibfnamefont {X.}~\bibnamefont
  {Ji}}, \bibinfo {author} {\bibfnamefont {P.}~\bibnamefont {Sun}}, \bibinfo
  {author} {\bibfnamefont {X.}~\bibnamefont {Xiong}}, \ and\ \bibinfo {author}
  {\bibfnamefont {F.}~\bibnamefont {Yuan}},\ }\href {\doibase
  10.1103/PhysRevD.91.074009} {\bibfield  {journal} {\bibinfo  {journal} {Phys.
  Rev.}\ }\textbf {\bibinfo {volume} {D91}},\ \bibinfo {pages} {074009}
  (\bibinfo {year} {2015}{\natexlab{b}})},\ \Eprint
  {http://arxiv.org/abs/1405.7640} {arXiv:1405.7640 [hep-ph]} \BibitemShut
  {NoStop}%
\bibitem [{\citenamefont {Lin}(2014{\natexlab{b}})}]{Lin:2014yra}%
  \BibitemOpen
  \bibfield  {author} {\bibinfo {author} {\bibfnamefont {H.-W.}\ \bibnamefont
  {Lin}},\ }\href {\doibase 10.22323/1.187.0293} {\bibfield  {journal}
  {\bibinfo  {journal} {PoS}\ }\textbf {\bibinfo {volume} {LATTICE2013}},\
  \bibinfo {pages} {293} (\bibinfo {year} {2014}{\natexlab{b}})}\BibitemShut
  {NoStop}%
\bibitem [{\citenamefont {Monahan}(2018)}]{Monahan:2017hpu}%
  \BibitemOpen
  \bibfield  {author} {\bibinfo {author} {\bibfnamefont {C.}~\bibnamefont
  {Monahan}},\ }\href {\doibase 10.1103/PhysRevD.97.054507} {\bibfield
  {journal} {\bibinfo  {journal} {Phys. Rev.}\ }\textbf {\bibinfo {volume}
  {D97}},\ \bibinfo {pages} {054507} (\bibinfo {year} {2018})},\ \Eprint
  {http://arxiv.org/abs/1710.04607} {arXiv:1710.04607 [hep-lat]} \BibitemShut
  {NoStop}%
\bibitem [{\citenamefont {Ji}\ \emph {et~al.}(2018{\natexlab{a}})\citenamefont
  {Ji}, \citenamefont {Jin}, \citenamefont {Yuan}, \citenamefont {Zhang},\ and\
  \citenamefont {Zhao}}]{Ji:2018hvs}%
  \BibitemOpen
  \bibfield  {author} {\bibinfo {author} {\bibfnamefont {X.}~\bibnamefont
  {Ji}}, \bibinfo {author} {\bibfnamefont {L.-C.}\ \bibnamefont {Jin}},
  \bibinfo {author} {\bibfnamefont {F.}~\bibnamefont {Yuan}}, \bibinfo {author}
  {\bibfnamefont {J.-H.}\ \bibnamefont {Zhang}}, \ and\ \bibinfo {author}
  {\bibfnamefont {Y.}~\bibnamefont {Zhao}},\ }\href@noop {} {\  (\bibinfo
  {year} {2018}{\natexlab{a}})},\ \Eprint {http://arxiv.org/abs/1801.05930}
  {arXiv:1801.05930 [hep-ph]} \BibitemShut {NoStop}%
\bibitem [{\citenamefont {Stewart}\ and\ \citenamefont
  {Zhao}(2018)}]{Stewart:2017tvs}%
  \BibitemOpen
  \bibfield  {author} {\bibinfo {author} {\bibfnamefont {I.~W.}\ \bibnamefont
  {Stewart}}\ and\ \bibinfo {author} {\bibfnamefont {Y.}~\bibnamefont {Zhao}},\
  }\href {\doibase 10.1103/PhysRevD.97.054512} {\bibfield  {journal} {\bibinfo
  {journal} {Phys. Rev.}\ }\textbf {\bibinfo {volume} {D97}},\ \bibinfo {pages}
  {054512} (\bibinfo {year} {2018})},\ \Eprint
  {http://arxiv.org/abs/1709.04933} {arXiv:1709.04933 [hep-ph]} \BibitemShut
  {NoStop}%
\bibitem [{\citenamefont {Constantinou}\ and\ \citenamefont
  {Panagopoulos}(2017)}]{Constantinou:2017sej}%
  \BibitemOpen
  \bibfield  {author} {\bibinfo {author} {\bibfnamefont {M.}~\bibnamefont
  {Constantinou}}\ and\ \bibinfo {author} {\bibfnamefont {H.}~\bibnamefont
  {Panagopoulos}},\ }\href {\doibase 10.1103/PhysRevD.96.054506} {\bibfield
  {journal} {\bibinfo  {journal} {Phys. Rev.}\ }\textbf {\bibinfo {volume}
  {D96}},\ \bibinfo {pages} {054506} (\bibinfo {year} {2017})},\ \Eprint
  {http://arxiv.org/abs/1705.11193} {arXiv:1705.11193 [hep-lat]} \BibitemShut
  {NoStop}%
\bibitem [{\citenamefont {Green}\ \emph {et~al.}(2018)\citenamefont {Green},
  \citenamefont {Jansen},\ and\ \citenamefont {Steffens}}]{Green:2017xeu}%
  \BibitemOpen
  \bibfield  {author} {\bibinfo {author} {\bibfnamefont {J.}~\bibnamefont
  {Green}}, \bibinfo {author} {\bibfnamefont {K.}~\bibnamefont {Jansen}}, \
  and\ \bibinfo {author} {\bibfnamefont {F.}~\bibnamefont {Steffens}},\ }\href
  {\doibase 10.1103/PhysRevLett.121.022004} {\bibfield  {journal} {\bibinfo
  {journal} {Phys. Rev. Lett.}\ }\textbf {\bibinfo {volume} {121}},\ \bibinfo
  {pages} {022004} (\bibinfo {year} {2018})},\ \Eprint
  {http://arxiv.org/abs/1707.07152} {arXiv:1707.07152 [hep-lat]} \BibitemShut
  {NoStop}%
\bibitem [{\citenamefont {Izubuchi}\ \emph {et~al.}(2018)\citenamefont
  {Izubuchi}, \citenamefont {Ji}, \citenamefont {Jin}, \citenamefont
  {Stewart},\ and\ \citenamefont {Zhao}}]{Izubuchi:2018srq}%
  \BibitemOpen
  \bibfield  {author} {\bibinfo {author} {\bibfnamefont {T.}~\bibnamefont
  {Izubuchi}}, \bibinfo {author} {\bibfnamefont {X.}~\bibnamefont {Ji}},
  \bibinfo {author} {\bibfnamefont {L.}~\bibnamefont {Jin}}, \bibinfo {author}
  {\bibfnamefont {I.~W.}\ \bibnamefont {Stewart}}, \ and\ \bibinfo {author}
  {\bibfnamefont {Y.}~\bibnamefont {Zhao}},\ }\href {\doibase
  10.1103/PhysRevD.98.056004} {\bibfield  {journal} {\bibinfo  {journal} {Phys.
  Rev.}\ }\textbf {\bibinfo {volume} {D98}},\ \bibinfo {pages} {056004}
  (\bibinfo {year} {2018})},\ \Eprint {http://arxiv.org/abs/1801.03917}
  {arXiv:1801.03917 [hep-ph]} \BibitemShut {NoStop}%
\bibitem [{\citenamefont {Xiong}\ \emph {et~al.}(2017)\citenamefont {Xiong},
  \citenamefont {Luu},\ and\ \citenamefont {Meißner}}]{Xiong:2017jtn}%
  \BibitemOpen
  \bibfield  {author} {\bibinfo {author} {\bibfnamefont {X.}~\bibnamefont
  {Xiong}}, \bibinfo {author} {\bibfnamefont {T.}~\bibnamefont {Luu}}, \ and\
  \bibinfo {author} {\bibfnamefont {U.-G.}\ \bibnamefont {Meißner}},\
  }\href@noop {} {\  (\bibinfo {year} {2017})},\ \Eprint
  {http://arxiv.org/abs/1705.00246} {arXiv:1705.00246 [hep-ph]} \BibitemShut
  {NoStop}%
\bibitem [{\citenamefont {Wang}\ \emph {et~al.}(2018)\citenamefont {Wang},
  \citenamefont {Zhao},\ and\ \citenamefont {Zhu}}]{Wang:2017qyg}%
  \BibitemOpen
  \bibfield  {author} {\bibinfo {author} {\bibfnamefont {W.}~\bibnamefont
  {Wang}}, \bibinfo {author} {\bibfnamefont {S.}~\bibnamefont {Zhao}}, \ and\
  \bibinfo {author} {\bibfnamefont {R.}~\bibnamefont {Zhu}},\ }\href {\doibase
  10.1140/epjc/s10052-018-5617-3} {\bibfield  {journal} {\bibinfo  {journal}
  {Eur. Phys. J.}\ }\textbf {\bibinfo {volume} {C78}},\ \bibinfo {pages} {147}
  (\bibinfo {year} {2018})},\ \Eprint {http://arxiv.org/abs/1708.02458}
  {arXiv:1708.02458 [hep-ph]} \BibitemShut {NoStop}%
\bibitem [{\citenamefont {Wang}\ and\ \citenamefont
  {Zhao}(2018)}]{Wang:2017eel}%
  \BibitemOpen
  \bibfield  {author} {\bibinfo {author} {\bibfnamefont {W.}~\bibnamefont
  {Wang}}\ and\ \bibinfo {author} {\bibfnamefont {S.}~\bibnamefont {Zhao}},\
  }\href {\doibase 10.1007/JHEP05(2018)142} {\bibfield  {journal} {\bibinfo
  {journal} {JHEP}\ }\textbf {\bibinfo {volume} {05}},\ \bibinfo {pages} {142}
  (\bibinfo {year} {2018})},\ \Eprint {http://arxiv.org/abs/1712.09247}
  {arXiv:1712.09247 [hep-ph]} \BibitemShut {NoStop}%
\bibitem [{\citenamefont {Xu}\ \emph {et~al.}(2018{\natexlab{a}})\citenamefont
  {Xu}, \citenamefont {Zhang},\ and\ \citenamefont {Zhao}}]{Xu:2018mpf}%
  \BibitemOpen
  \bibfield  {author} {\bibinfo {author} {\bibfnamefont {J.}~\bibnamefont
  {Xu}}, \bibinfo {author} {\bibfnamefont {Q.-A.}\ \bibnamefont {Zhang}}, \
  and\ \bibinfo {author} {\bibfnamefont {S.}~\bibnamefont {Zhao}},\ }\href
  {\doibase 10.1103/PhysRevD.97.114026} {\bibfield  {journal} {\bibinfo
  {journal} {Phys. Rev.}\ }\textbf {\bibinfo {volume} {D97}},\ \bibinfo {pages}
  {114026} (\bibinfo {year} {2018}{\natexlab{a}})},\ \Eprint
  {http://arxiv.org/abs/1804.01042} {arXiv:1804.01042 [hep-ph]} \BibitemShut
  {NoStop}%
\bibitem [{\citenamefont {Zhang}\ \emph {et~al.}(2017)\citenamefont {Zhang},
  \citenamefont {Chen}, \citenamefont {Ji}, \citenamefont {Jin},\ and\
  \citenamefont {Lin}}]{Zhang:2017bzy}%
  \BibitemOpen
  \bibfield  {author} {\bibinfo {author} {\bibfnamefont {J.-H.}\ \bibnamefont
  {Zhang}}, \bibinfo {author} {\bibfnamefont {J.-W.}\ \bibnamefont {Chen}},
  \bibinfo {author} {\bibfnamefont {X.}~\bibnamefont {Ji}}, \bibinfo {author}
  {\bibfnamefont {L.}~\bibnamefont {Jin}}, \ and\ \bibinfo {author}
  {\bibfnamefont {H.-W.}\ \bibnamefont {Lin}},\ }\href {\doibase
  10.1103/PhysRevD.95.094514} {\bibfield  {journal} {\bibinfo  {journal} {Phys.
  Rev.}\ }\textbf {\bibinfo {volume} {D95}},\ \bibinfo {pages} {094514}
  (\bibinfo {year} {2017})},\ \Eprint {http://arxiv.org/abs/1702.00008}
  {arXiv:1702.00008 [hep-lat]} \BibitemShut {NoStop}%
\bibitem [{\citenamefont {Ishikawa}\ \emph {et~al.}(2016)\citenamefont
  {Ishikawa}, \citenamefont {Ma}, \citenamefont {Qiu},\ and\ \citenamefont
  {Yoshida}}]{Ishikawa:2016znu}%
  \BibitemOpen
  \bibfield  {author} {\bibinfo {author} {\bibfnamefont {T.}~\bibnamefont
  {Ishikawa}}, \bibinfo {author} {\bibfnamefont {Y.-Q.}\ \bibnamefont {Ma}},
  \bibinfo {author} {\bibfnamefont {J.-W.}\ \bibnamefont {Qiu}}, \ and\
  \bibinfo {author} {\bibfnamefont {S.}~\bibnamefont {Yoshida}},\ }\href@noop
  {} {\  (\bibinfo {year} {2016})},\ \Eprint {http://arxiv.org/abs/1609.02018}
  {arXiv:1609.02018 [hep-lat]} \BibitemShut {NoStop}%
\bibitem [{\citenamefont {Chen}\ \emph {et~al.}(2017)\citenamefont {Chen},
  \citenamefont {Ji},\ and\ \citenamefont {Zhang}}]{Chen:2016fxx}%
  \BibitemOpen
  \bibfield  {author} {\bibinfo {author} {\bibfnamefont {J.-W.}\ \bibnamefont
  {Chen}}, \bibinfo {author} {\bibfnamefont {X.}~\bibnamefont {Ji}}, \ and\
  \bibinfo {author} {\bibfnamefont {J.-H.}\ \bibnamefont {Zhang}},\ }\href
  {\doibase 10.1016/j.nuclphysb.2016.12.004} {\bibfield  {journal} {\bibinfo
  {journal} {Nucl. Phys.}\ }\textbf {\bibinfo {volume} {B915}},\ \bibinfo
  {pages} {1} (\bibinfo {year} {2017})},\ \Eprint
  {http://arxiv.org/abs/1609.08102} {arXiv:1609.08102 [hep-ph]} \BibitemShut
  {NoStop}%
\bibitem [{\citenamefont {Ji}\ \emph {et~al.}(2018{\natexlab{b}})\citenamefont
  {Ji}, \citenamefont {Zhang},\ and\ \citenamefont {Zhao}}]{Ji:2017oey}%
  \BibitemOpen
  \bibfield  {author} {\bibinfo {author} {\bibfnamefont {X.}~\bibnamefont
  {Ji}}, \bibinfo {author} {\bibfnamefont {J.-H.}\ \bibnamefont {Zhang}}, \
  and\ \bibinfo {author} {\bibfnamefont {Y.}~\bibnamefont {Zhao}},\ }\href
  {\doibase 10.1103/PhysRevLett.120.112001} {\bibfield  {journal} {\bibinfo
  {journal} {Phys. Rev. Lett.}\ }\textbf {\bibinfo {volume} {120}},\ \bibinfo
  {pages} {112001} (\bibinfo {year} {2018}{\natexlab{b}})},\ \Eprint
  {http://arxiv.org/abs/1706.08962} {arXiv:1706.08962 [hep-ph]} \BibitemShut
  {NoStop}%
\bibitem [{\citenamefont {Ishikawa}\ \emph {et~al.}(2017)\citenamefont
  {Ishikawa}, \citenamefont {Ma}, \citenamefont {Qiu},\ and\ \citenamefont
  {Yoshida}}]{Ishikawa:2017faj}%
  \BibitemOpen
  \bibfield  {author} {\bibinfo {author} {\bibfnamefont {T.}~\bibnamefont
  {Ishikawa}}, \bibinfo {author} {\bibfnamefont {Y.-Q.}\ \bibnamefont {Ma}},
  \bibinfo {author} {\bibfnamefont {J.-W.}\ \bibnamefont {Qiu}}, \ and\
  \bibinfo {author} {\bibfnamefont {S.}~\bibnamefont {Yoshida}},\ }\href
  {\doibase 10.1103/PhysRevD.96.094019} {\bibfield  {journal} {\bibinfo
  {journal} {Phys. Rev.}\ }\textbf {\bibinfo {volume} {D96}},\ \bibinfo {pages}
  {094019} (\bibinfo {year} {2017})},\ \Eprint
  {http://arxiv.org/abs/1707.03107} {arXiv:1707.03107 [hep-ph]} \BibitemShut
  {NoStop}%
\bibitem [{\citenamefont {Ishikawa}\ \emph {et~al.}(2019)\citenamefont
  {Ishikawa}, \citenamefont {Jin}, \citenamefont {Lin}, \citenamefont
  {Schäfer}, \citenamefont {Yang}, \citenamefont {Zhang},\ and\ \citenamefont
  {Zhao}}]{Ishikawa:2019flg}%
  \BibitemOpen
  \bibfield  {author} {\bibinfo {author} {\bibfnamefont {T.}~\bibnamefont
  {Ishikawa}}, \bibinfo {author} {\bibfnamefont {L.}~\bibnamefont {Jin}},
  \bibinfo {author} {\bibfnamefont {H.-W.}\ \bibnamefont {Lin}}, \bibinfo
  {author} {\bibfnamefont {A.}~\bibnamefont {Schäfer}}, \bibinfo {author}
  {\bibfnamefont {Y.-B.}\ \bibnamefont {Yang}}, \bibinfo {author}
  {\bibfnamefont {J.-H.}\ \bibnamefont {Zhang}}, \ and\ \bibinfo {author}
  {\bibfnamefont {Y.}~\bibnamefont {Zhao}},\ }\href {\doibase
  10.1007/s11433-018-9375-1} {\bibfield  {journal} {\bibinfo  {journal} {Sci.
  China Phys. Mech. Astron.}\ }\textbf {\bibinfo {volume} {62}},\ \bibinfo
  {pages} {991021} (\bibinfo {year} {2019})},\ \Eprint
  {http://arxiv.org/abs/1711.07858} {arXiv:1711.07858 [hep-ph]} \BibitemShut
  {NoStop}%
\bibitem [{\citenamefont {Li}(2016)}]{Li:2016amo}%
  \BibitemOpen
  \bibfield  {author} {\bibinfo {author} {\bibfnamefont {H.-n.}\ \bibnamefont
  {Li}},\ }\href {\doibase 10.1103/PhysRevD.94.074036} {\bibfield  {journal}
  {\bibinfo  {journal} {Phys. Rev.}\ }\textbf {\bibinfo {volume} {D94}},\
  \bibinfo {pages} {074036} (\bibinfo {year} {2016})},\ \Eprint
  {http://arxiv.org/abs/1602.07575} {arXiv:1602.07575 [hep-ph]} \BibitemShut
  {NoStop}%
\bibitem [{\citenamefont {Monahan}\ and\ \citenamefont
  {Orginos}(2017)}]{Monahan:2016bvm}%
  \BibitemOpen
  \bibfield  {author} {\bibinfo {author} {\bibfnamefont {C.}~\bibnamefont
  {Monahan}}\ and\ \bibinfo {author} {\bibfnamefont {K.}~\bibnamefont
  {Orginos}},\ }\href {\doibase 10.1007/JHEP03(2017)116} {\bibfield  {journal}
  {\bibinfo  {journal} {JHEP}\ }\textbf {\bibinfo {volume} {03}},\ \bibinfo
  {pages} {116} (\bibinfo {year} {2017})},\ \Eprint
  {http://arxiv.org/abs/1612.01584} {arXiv:1612.01584 [hep-lat]} \BibitemShut
  {NoStop}%
\bibitem [{\citenamefont
  {Radyushkin}(2017{\natexlab{a}})}]{Radyushkin:2016hsy}%
  \BibitemOpen
  \bibfield  {author} {\bibinfo {author} {\bibfnamefont {A.}~\bibnamefont
  {Radyushkin}},\ }\href {\doibase 10.1016/j.physletb.2017.02.019} {\bibfield
  {journal} {\bibinfo  {journal} {Phys. Lett.}\ }\textbf {\bibinfo {volume}
  {B767}},\ \bibinfo {pages} {314} (\bibinfo {year} {2017}{\natexlab{a}})},\
  \Eprint {http://arxiv.org/abs/1612.05170} {arXiv:1612.05170 [hep-ph]}
  \BibitemShut {NoStop}%
\bibitem [{\citenamefont {Rossi}\ and\ \citenamefont
  {Testa}(2017)}]{Rossi:2017muf}%
  \BibitemOpen
  \bibfield  {author} {\bibinfo {author} {\bibfnamefont {G.~C.}\ \bibnamefont
  {Rossi}}\ and\ \bibinfo {author} {\bibfnamefont {M.}~\bibnamefont {Testa}},\
  }\href {\doibase 10.1103/PhysRevD.96.014507} {\bibfield  {journal} {\bibinfo
  {journal} {Phys. Rev.}\ }\textbf {\bibinfo {volume} {D96}},\ \bibinfo {pages}
  {014507} (\bibinfo {year} {2017})},\ \Eprint
  {http://arxiv.org/abs/1706.04428} {arXiv:1706.04428 [hep-lat]} \BibitemShut
  {NoStop}%
\bibitem [{\citenamefont {Carlson}\ and\ \citenamefont
  {Freid}(2017)}]{Carlson:2017gpk}%
  \BibitemOpen
  \bibfield  {author} {\bibinfo {author} {\bibfnamefont {C.~E.}\ \bibnamefont
  {Carlson}}\ and\ \bibinfo {author} {\bibfnamefont {M.}~\bibnamefont
  {Freid}},\ }\href {\doibase 10.1103/PhysRevD.95.094504} {\bibfield  {journal}
  {\bibinfo  {journal} {Phys. Rev.}\ }\textbf {\bibinfo {volume} {D95}},\
  \bibinfo {pages} {094504} (\bibinfo {year} {2017})},\ \Eprint
  {http://arxiv.org/abs/1702.05775} {arXiv:1702.05775 [hep-ph]} \BibitemShut
  {NoStop}%
\bibitem [{\citenamefont {Briceño}\ \emph {et~al.}(2018)\citenamefont
  {Briceño}, \citenamefont {Guerrero}, \citenamefont {Hansen},\ and\
  \citenamefont {Monahan}}]{Briceno:2018lfj}%
  \BibitemOpen
  \bibfield  {author} {\bibinfo {author} {\bibfnamefont {R.~A.}\ \bibnamefont
  {Briceño}}, \bibinfo {author} {\bibfnamefont {J.~V.}\ \bibnamefont
  {Guerrero}}, \bibinfo {author} {\bibfnamefont {M.~T.}\ \bibnamefont
  {Hansen}}, \ and\ \bibinfo {author} {\bibfnamefont {C.~J.}\ \bibnamefont
  {Monahan}},\ }\href {\doibase 10.1103/PhysRevD.98.014511} {\bibfield
  {journal} {\bibinfo  {journal} {Phys. Rev. D}\ }\textbf {\bibinfo {volume}
  {98}},\ \bibinfo {pages} {014511} (\bibinfo {year} {2018})},\ \Eprint
  {http://arxiv.org/abs/1805.01034} {arXiv:1805.01034 [hep-lat]} \BibitemShut
  {NoStop}%
\bibitem [{\citenamefont {Hobbs}(2018)}]{Hobbs:2017xtq}%
  \BibitemOpen
  \bibfield  {author} {\bibinfo {author} {\bibfnamefont {T.~J.}\ \bibnamefont
  {Hobbs}},\ }\href {\doibase 10.1103/PhysRevD.97.054028} {\bibfield  {journal}
  {\bibinfo  {journal} {Phys. Rev.}\ }\textbf {\bibinfo {volume} {D97}},\
  \bibinfo {pages} {054028} (\bibinfo {year} {2018})},\ \Eprint
  {http://arxiv.org/abs/1708.05463} {arXiv:1708.05463 [hep-ph]} \BibitemShut
  {NoStop}%
\bibitem [{\citenamefont {Jia}\ \emph {et~al.}(2017)\citenamefont {Jia},
  \citenamefont {Liang}, \citenamefont {Li},\ and\ \citenamefont
  {Xiong}}]{Jia:2017uul}%
  \BibitemOpen
  \bibfield  {author} {\bibinfo {author} {\bibfnamefont {Y.}~\bibnamefont
  {Jia}}, \bibinfo {author} {\bibfnamefont {S.}~\bibnamefont {Liang}}, \bibinfo
  {author} {\bibfnamefont {L.}~\bibnamefont {Li}}, \ and\ \bibinfo {author}
  {\bibfnamefont {X.}~\bibnamefont {Xiong}},\ }\href {\doibase
  10.1007/JHEP11(2017)151} {\bibfield  {journal} {\bibinfo  {journal} {JHEP}\
  }\textbf {\bibinfo {volume} {11}},\ \bibinfo {pages} {151} (\bibinfo {year}
  {2017})},\ \Eprint {http://arxiv.org/abs/1708.09379} {arXiv:1708.09379
  [hep-ph]} \BibitemShut {NoStop}%
\bibitem [{\citenamefont {Xu}\ \emph {et~al.}(2018{\natexlab{b}})\citenamefont
  {Xu}, \citenamefont {Chang}, \citenamefont {Roberts},\ and\ \citenamefont
  {Zong}}]{Xu:2018eii}%
  \BibitemOpen
  \bibfield  {author} {\bibinfo {author} {\bibfnamefont {S.-S.}\ \bibnamefont
  {Xu}}, \bibinfo {author} {\bibfnamefont {L.}~\bibnamefont {Chang}}, \bibinfo
  {author} {\bibfnamefont {C.~D.}\ \bibnamefont {Roberts}}, \ and\ \bibinfo
  {author} {\bibfnamefont {H.-S.}\ \bibnamefont {Zong}},\ }\href {\doibase
  10.1103/PhysRevD.97.094014} {\bibfield  {journal} {\bibinfo  {journal} {Phys.
  Rev.}\ }\textbf {\bibinfo {volume} {D97}},\ \bibinfo {pages} {094014}
  (\bibinfo {year} {2018}{\natexlab{b}})},\ \Eprint
  {http://arxiv.org/abs/1802.09552} {arXiv:1802.09552 [nucl-th]} \BibitemShut
  {NoStop}%
\bibitem [{\citenamefont {Jia}\ \emph {et~al.}(2018)\citenamefont {Jia},
  \citenamefont {Liang}, \citenamefont {Xiong},\ and\ \citenamefont
  {Yu}}]{Jia:2018qee}%
  \BibitemOpen
  \bibfield  {author} {\bibinfo {author} {\bibfnamefont {Y.}~\bibnamefont
  {Jia}}, \bibinfo {author} {\bibfnamefont {S.}~\bibnamefont {Liang}}, \bibinfo
  {author} {\bibfnamefont {X.}~\bibnamefont {Xiong}}, \ and\ \bibinfo {author}
  {\bibfnamefont {R.}~\bibnamefont {Yu}},\ }\href {\doibase
  10.1103/PhysRevD.98.054011} {\bibfield  {journal} {\bibinfo  {journal} {Phys.
  Rev.}\ }\textbf {\bibinfo {volume} {D98}},\ \bibinfo {pages} {054011}
  (\bibinfo {year} {2018})},\ \Eprint {http://arxiv.org/abs/1804.04644}
  {arXiv:1804.04644 [hep-th]} \BibitemShut {NoStop}%
\bibitem [{\citenamefont {Spanoudes}\ and\ \citenamefont
  {Panagopoulos}(2018)}]{Spanoudes:2018zya}%
  \BibitemOpen
  \bibfield  {author} {\bibinfo {author} {\bibfnamefont {G.}~\bibnamefont
  {Spanoudes}}\ and\ \bibinfo {author} {\bibfnamefont {H.}~\bibnamefont
  {Panagopoulos}},\ }\href {\doibase 10.1103/PhysRevD.98.014509} {\bibfield
  {journal} {\bibinfo  {journal} {Phys. Rev.}\ }\textbf {\bibinfo {volume}
  {D98}},\ \bibinfo {pages} {014509} (\bibinfo {year} {2018})},\ \Eprint
  {http://arxiv.org/abs/1805.01164} {arXiv:1805.01164 [hep-lat]} \BibitemShut
  {NoStop}%
\bibitem [{\citenamefont {Rossi}\ and\ \citenamefont
  {Testa}(2018)}]{Rossi:2018zkn}%
  \BibitemOpen
  \bibfield  {author} {\bibinfo {author} {\bibfnamefont {G.}~\bibnamefont
  {Rossi}}\ and\ \bibinfo {author} {\bibfnamefont {M.}~\bibnamefont {Testa}},\
  }\href {\doibase 10.1103/PhysRevD.98.054028} {\bibfield  {journal} {\bibinfo
  {journal} {Phys. Rev.}\ }\textbf {\bibinfo {volume} {D98}},\ \bibinfo {pages}
  {054028} (\bibinfo {year} {2018})},\ \Eprint
  {http://arxiv.org/abs/1806.00808} {arXiv:1806.00808 [hep-lat]} \BibitemShut
  {NoStop}%
\bibitem [{\citenamefont {Liu}\ \emph {et~al.}(2018{\natexlab{b}})\citenamefont
  {Liu}, \citenamefont {Chen}, \citenamefont {Jin}, \citenamefont {Lin},
  \citenamefont {Yang}, \citenamefont {Zhang},\ and\ \citenamefont
  {Zhao}}]{Liu:2018uuj}%
  \BibitemOpen
  \bibfield  {author} {\bibinfo {author} {\bibfnamefont {Y.-S.}\ \bibnamefont
  {Liu}}, \bibinfo {author} {\bibfnamefont {J.-W.}\ \bibnamefont {Chen}},
  \bibinfo {author} {\bibfnamefont {L.}~\bibnamefont {Jin}}, \bibinfo {author}
  {\bibfnamefont {H.-W.}\ \bibnamefont {Lin}}, \bibinfo {author} {\bibfnamefont
  {Y.-B.}\ \bibnamefont {Yang}}, \bibinfo {author} {\bibfnamefont {J.-H.}\
  \bibnamefont {Zhang}}, \ and\ \bibinfo {author} {\bibfnamefont
  {Y.}~\bibnamefont {Zhao}},\ }\href@noop {} {\  (\bibinfo {year}
  {2018}{\natexlab{b}})},\ \Eprint {http://arxiv.org/abs/1807.06566}
  {arXiv:1807.06566 [hep-lat]} \BibitemShut {NoStop}%
\bibitem [{\citenamefont {Ji}\ \emph {et~al.}(2019{\natexlab{a}})\citenamefont
  {Ji}, \citenamefont {Liu},\ and\ \citenamefont {Zahed}}]{Ji:2018waw}%
  \BibitemOpen
  \bibfield  {author} {\bibinfo {author} {\bibfnamefont {X.}~\bibnamefont
  {Ji}}, \bibinfo {author} {\bibfnamefont {Y.}~\bibnamefont {Liu}}, \ and\
  \bibinfo {author} {\bibfnamefont {I.}~\bibnamefont {Zahed}},\ }\href
  {\doibase 10.1103/PhysRevD.99.054008} {\bibfield  {journal} {\bibinfo
  {journal} {Phys. Rev.}\ }\textbf {\bibinfo {volume} {D99}},\ \bibinfo {pages}
  {054008} (\bibinfo {year} {2019}{\natexlab{a}})},\ \Eprint
  {http://arxiv.org/abs/1807.07528} {arXiv:1807.07528 [hep-ph]} \BibitemShut
  {NoStop}%
\bibitem [{\citenamefont {Bhattacharya}\ \emph {et~al.}(2019)\citenamefont
  {Bhattacharya}, \citenamefont {Cocuzza},\ and\ \citenamefont
  {Metz}}]{Bhattacharya:2018zxi}%
  \BibitemOpen
  \bibfield  {author} {\bibinfo {author} {\bibfnamefont {S.}~\bibnamefont
  {Bhattacharya}}, \bibinfo {author} {\bibfnamefont {C.}~\bibnamefont
  {Cocuzza}}, \ and\ \bibinfo {author} {\bibfnamefont {A.}~\bibnamefont
  {Metz}},\ }\href {\doibase 10.1016/j.physletb.2018.09.061} {\bibfield
  {journal} {\bibinfo  {journal} {Phys. Lett.}\ }\textbf {\bibinfo {volume}
  {B788}},\ \bibinfo {pages} {453} (\bibinfo {year} {2019})},\ \Eprint
  {http://arxiv.org/abs/1808.01437} {arXiv:1808.01437 [hep-ph]} \BibitemShut
  {NoStop}%
\bibitem [{\citenamefont {Radyushkin}(2019)}]{Radyushkin:2018nbf}%
  \BibitemOpen
  \bibfield  {author} {\bibinfo {author} {\bibfnamefont {A.~V.}\ \bibnamefont
  {Radyushkin}},\ }\href {\doibase 10.1016/j.physletb.2018.11.047} {\bibfield
  {journal} {\bibinfo  {journal} {Phys. Lett.}\ }\textbf {\bibinfo {volume}
  {B788}},\ \bibinfo {pages} {380} (\bibinfo {year} {2019})},\ \Eprint
  {http://arxiv.org/abs/1807.07509} {arXiv:1807.07509 [hep-ph]} \BibitemShut
  {NoStop}%
\bibitem [{\citenamefont {Zhang}\ \emph
  {et~al.}(2019{\natexlab{b}})\citenamefont {Zhang}, \citenamefont {Ji},
  \citenamefont {Schäfer}, \citenamefont {Wang},\ and\ \citenamefont
  {Zhao}}]{Zhang:2018diq}%
  \BibitemOpen
  \bibfield  {author} {\bibinfo {author} {\bibfnamefont {J.-H.}\ \bibnamefont
  {Zhang}}, \bibinfo {author} {\bibfnamefont {X.}~\bibnamefont {Ji}}, \bibinfo
  {author} {\bibfnamefont {A.}~\bibnamefont {Schäfer}}, \bibinfo {author}
  {\bibfnamefont {W.}~\bibnamefont {Wang}}, \ and\ \bibinfo {author}
  {\bibfnamefont {S.}~\bibnamefont {Zhao}},\ }\href {\doibase
  10.1103/PhysRevLett.122.142001} {\bibfield  {journal} {\bibinfo  {journal}
  {Phys. Rev. Lett.}\ }\textbf {\bibinfo {volume} {122}},\ \bibinfo {pages}
  {142001} (\bibinfo {year} {2019}{\natexlab{b}})},\ \Eprint
  {http://arxiv.org/abs/1808.10824} {arXiv:1808.10824 [hep-ph]} \BibitemShut
  {NoStop}%
\bibitem [{\citenamefont {Li}\ \emph {et~al.}(2019)\citenamefont {Li},
  \citenamefont {Ma},\ and\ \citenamefont {Qiu}}]{Li:2018tpe}%
  \BibitemOpen
  \bibfield  {author} {\bibinfo {author} {\bibfnamefont {Z.-Y.}\ \bibnamefont
  {Li}}, \bibinfo {author} {\bibfnamefont {Y.-Q.}\ \bibnamefont {Ma}}, \ and\
  \bibinfo {author} {\bibfnamefont {J.-W.}\ \bibnamefont {Qiu}},\ }\href
  {\doibase 10.1103/PhysRevLett.122.062002} {\bibfield  {journal} {\bibinfo
  {journal} {Phys. Rev. Lett.}\ }\textbf {\bibinfo {volume} {122}},\ \bibinfo
  {pages} {062002} (\bibinfo {year} {2019})},\ \Eprint
  {http://arxiv.org/abs/1809.01836} {arXiv:1809.01836 [hep-ph]} \BibitemShut
  {NoStop}%
\bibitem [{\citenamefont {Braun}\ \emph {et~al.}(2019)\citenamefont {Braun},
  \citenamefont {Vladimirov},\ and\ \citenamefont {Zhang}}]{Braun:2018brg}%
  \BibitemOpen
  \bibfield  {author} {\bibinfo {author} {\bibfnamefont {V.~M.}\ \bibnamefont
  {Braun}}, \bibinfo {author} {\bibfnamefont {A.}~\bibnamefont {Vladimirov}}, \
  and\ \bibinfo {author} {\bibfnamefont {J.-H.}\ \bibnamefont {Zhang}},\ }\href
  {\doibase 10.1103/PhysRevD.99.014013} {\bibfield  {journal} {\bibinfo
  {journal} {Phys. Rev.}\ }\textbf {\bibinfo {volume} {D99}},\ \bibinfo {pages}
  {014013} (\bibinfo {year} {2019})},\ \Eprint
  {http://arxiv.org/abs/1810.00048} {arXiv:1810.00048 [hep-ph]} \BibitemShut
  {NoStop}%
\bibitem [{\citenamefont {Detmold}\ \emph {et~al.}(2019)\citenamefont
  {Detmold}, \citenamefont {Edwards}, \citenamefont {Dudek}, \citenamefont
  {Engelhardt}, \citenamefont {Lin}, \citenamefont {Meinel}, \citenamefont
  {Orginos},\ and\ \citenamefont {Shanahan}}]{Detmold:2019ghl}%
  \BibitemOpen
  \bibfield  {author} {\bibinfo {author} {\bibfnamefont {W.}~\bibnamefont
  {Detmold}}, \bibinfo {author} {\bibfnamefont {R.~G.}\ \bibnamefont
  {Edwards}}, \bibinfo {author} {\bibfnamefont {J.~J.}\ \bibnamefont {Dudek}},
  \bibinfo {author} {\bibfnamefont {M.}~\bibnamefont {Engelhardt}}, \bibinfo
  {author} {\bibfnamefont {H.-W.}\ \bibnamefont {Lin}}, \bibinfo {author}
  {\bibfnamefont {S.}~\bibnamefont {Meinel}}, \bibinfo {author} {\bibfnamefont
  {K.}~\bibnamefont {Orginos}}, \ and\ \bibinfo {author} {\bibfnamefont
  {P.}~\bibnamefont {Shanahan}} (\bibinfo {collaboration} {USQCD}),\ }\href
  {\doibase 10.1140/epja/i2019-12902-4} {\bibfield  {journal} {\bibinfo
  {journal} {Eur. Phys. J. A}\ }\textbf {\bibinfo {volume} {55}},\ \bibinfo
  {pages} {193} (\bibinfo {year} {2019})},\ \Eprint
  {http://arxiv.org/abs/1904.09512} {arXiv:1904.09512 [hep-lat]} \BibitemShut
  {NoStop}%
\bibitem [{\citenamefont {Ebert}\ \emph
  {et~al.}(2020{\natexlab{a}})\citenamefont {Ebert}, \citenamefont {Stewart},\
  and\ \citenamefont {Zhao}}]{Ebert:2019tvc}%
  \BibitemOpen
  \bibfield  {author} {\bibinfo {author} {\bibfnamefont {M.~A.}\ \bibnamefont
  {Ebert}}, \bibinfo {author} {\bibfnamefont {I.~W.}\ \bibnamefont {Stewart}},
  \ and\ \bibinfo {author} {\bibfnamefont {Y.}~\bibnamefont {Zhao}},\ }\href
  {\doibase 10.1007/JHEP03(2020)099} {\bibfield  {journal} {\bibinfo  {journal}
  {JHEP}\ }\textbf {\bibinfo {volume} {03}},\ \bibinfo {pages} {099} (\bibinfo
  {year} {2020}{\natexlab{a}})},\ \Eprint {http://arxiv.org/abs/1910.08569}
  {arXiv:1910.08569 [hep-ph]} \BibitemShut {NoStop}%
\bibitem [{\citenamefont {Ji}\ \emph {et~al.}(2019{\natexlab{b}})\citenamefont
  {Ji}, \citenamefont {Liu},\ and\ \citenamefont {Liu}}]{Ji:2019ewn}%
  \BibitemOpen
  \bibfield  {author} {\bibinfo {author} {\bibfnamefont {X.}~\bibnamefont
  {Ji}}, \bibinfo {author} {\bibfnamefont {Y.}~\bibnamefont {Liu}}, \ and\
  \bibinfo {author} {\bibfnamefont {Y.-S.}\ \bibnamefont {Liu}},\ }\href@noop
  {} {\  (\bibinfo {year} {2019}{\natexlab{b}})},\ \Eprint
  {http://arxiv.org/abs/1911.03840} {arXiv:1911.03840 [hep-ph]} \BibitemShut
  {NoStop}%
\bibitem [{\citenamefont {Bali}\ \emph
  {et~al.}(2018{\natexlab{a}})\citenamefont {Bali} \emph
  {et~al.}}]{Bali:2017gfr}%
  \BibitemOpen
  \bibfield  {author} {\bibinfo {author} {\bibfnamefont {G.~S.}\ \bibnamefont
  {Bali}} \emph {et~al.},\ }\bibfield  {booktitle} {\emph {\bibinfo {booktitle}
  {{Proceedings, 35th International Symposium on Lattice Field Theory (Lattice
  2017): Granada, Spain, June 18-24, 2017}}},\ }\href {\doibase
  10.1140/epjc/s10052-018-5700-9} {\bibfield  {journal} {\bibinfo  {journal}
  {Eur. Phys. J.}\ }\textbf {\bibinfo {volume} {C78}},\ \bibinfo {pages} {217}
  (\bibinfo {year} {2018}{\natexlab{a}})},\ \Eprint
  {http://arxiv.org/abs/1709.04325} {arXiv:1709.04325 [hep-lat]} \BibitemShut
  {NoStop}%
\bibitem [{\citenamefont {Bali}\ \emph
  {et~al.}(2018{\natexlab{b}})\citenamefont {Bali}, \citenamefont {Braun},
  \citenamefont {Gläßle}, \citenamefont {Göckeler}, \citenamefont {Gruber},
  \citenamefont {Hutzler}, \citenamefont {Korcyl}, \citenamefont {Schäfer},
  \citenamefont {Wein},\ and\ \citenamefont {Zhang}}]{Bali:2018spj}%
  \BibitemOpen
  \bibfield  {author} {\bibinfo {author} {\bibfnamefont {G.~S.}\ \bibnamefont
  {Bali}}, \bibinfo {author} {\bibfnamefont {V.~M.}\ \bibnamefont {Braun}},
  \bibinfo {author} {\bibfnamefont {B.}~\bibnamefont {Gläßle}}, \bibinfo
  {author} {\bibfnamefont {M.}~\bibnamefont {Göckeler}}, \bibinfo {author}
  {\bibfnamefont {M.}~\bibnamefont {Gruber}}, \bibinfo {author} {\bibfnamefont
  {F.}~\bibnamefont {Hutzler}}, \bibinfo {author} {\bibfnamefont
  {P.}~\bibnamefont {Korcyl}}, \bibinfo {author} {\bibfnamefont
  {A.}~\bibnamefont {Schäfer}}, \bibinfo {author} {\bibfnamefont
  {P.}~\bibnamefont {Wein}}, \ and\ \bibinfo {author} {\bibfnamefont {J.-H.}\
  \bibnamefont {Zhang}},\ }\href {\doibase 10.1103/PhysRevD.98.094507}
  {\bibfield  {journal} {\bibinfo  {journal} {Phys. Rev. D}\ }\textbf {\bibinfo
  {volume} {98}},\ \bibinfo {pages} {094507} (\bibinfo {year}
  {2018}{\natexlab{b}})},\ \Eprint {http://arxiv.org/abs/1807.06671}
  {arXiv:1807.06671 [hep-lat]} \BibitemShut {NoStop}%
\bibitem [{\citenamefont {Sufian}\ \emph {et~al.}(2019)\citenamefont {Sufian},
  \citenamefont {Karpie}, \citenamefont {Egerer}, \citenamefont {Orginos},
  \citenamefont {Qiu},\ and\ \citenamefont {Richards}}]{Sufian:2019bol}%
  \BibitemOpen
  \bibfield  {author} {\bibinfo {author} {\bibfnamefont {R.~S.}\ \bibnamefont
  {Sufian}}, \bibinfo {author} {\bibfnamefont {J.}~\bibnamefont {Karpie}},
  \bibinfo {author} {\bibfnamefont {C.}~\bibnamefont {Egerer}}, \bibinfo
  {author} {\bibfnamefont {K.}~\bibnamefont {Orginos}}, \bibinfo {author}
  {\bibfnamefont {J.-W.}\ \bibnamefont {Qiu}}, \ and\ \bibinfo {author}
  {\bibfnamefont {D.~G.}\ \bibnamefont {Richards}},\ }\href {\doibase
  10.1103/PhysRevD.99.074507} {\bibfield  {journal} {\bibinfo  {journal} {Phys.
  Rev. D}\ }\textbf {\bibinfo {volume} {99}},\ \bibinfo {pages} {074507}
  (\bibinfo {year} {2019})},\ \Eprint {http://arxiv.org/abs/1901.03921}
  {arXiv:1901.03921 [hep-lat]} \BibitemShut {NoStop}%
\bibitem [{\citenamefont {Bali}\ \emph
  {et~al.}(2019{\natexlab{a}})\citenamefont {Bali} \emph
  {et~al.}}]{Bali:2019ecy}%
  \BibitemOpen
  \bibfield  {author} {\bibinfo {author} {\bibfnamefont {G.~S.}\ \bibnamefont
  {Bali}} \emph {et~al.} (\bibinfo {collaboration} {RQCD}),\ }\href {\doibase
  10.1140/epja/i2019-12803-6} {\bibfield  {journal} {\bibinfo  {journal} {Eur.
  Phys. J. A}\ }\textbf {\bibinfo {volume} {55}},\ \bibinfo {pages} {116}
  (\bibinfo {year} {2019}{\natexlab{a}})},\ \Eprint
  {http://arxiv.org/abs/1903.12590} {arXiv:1903.12590 [hep-lat]} \BibitemShut
  {NoStop}%
\bibitem [{\citenamefont {Orginos}\ \emph {et~al.}(2017)\citenamefont
  {Orginos}, \citenamefont {Radyushkin}, \citenamefont {Karpie},\ and\
  \citenamefont {Zafeiropoulos}}]{Orginos:2017kos}%
  \BibitemOpen
  \bibfield  {author} {\bibinfo {author} {\bibfnamefont {K.}~\bibnamefont
  {Orginos}}, \bibinfo {author} {\bibfnamefont {A.}~\bibnamefont {Radyushkin}},
  \bibinfo {author} {\bibfnamefont {J.}~\bibnamefont {Karpie}}, \ and\ \bibinfo
  {author} {\bibfnamefont {S.}~\bibnamefont {Zafeiropoulos}},\ }\href {\doibase
  10.1103/PhysRevD.96.094503} {\bibfield  {journal} {\bibinfo  {journal} {Phys.
  Rev.}\ }\textbf {\bibinfo {volume} {D96}},\ \bibinfo {pages} {094503}
  (\bibinfo {year} {2017})},\ \Eprint {http://arxiv.org/abs/1706.05373}
  {arXiv:1706.05373 [hep-ph]} \BibitemShut {NoStop}%
\bibitem [{\citenamefont {Karpie}\ \emph
  {et~al.}(2018{\natexlab{a}})\citenamefont {Karpie}, \citenamefont {Orginos},
  \citenamefont {Radyushkin},\ and\ \citenamefont
  {Zafeiropoulos}}]{Karpie:2017bzm}%
  \BibitemOpen
  \bibfield  {author} {\bibinfo {author} {\bibfnamefont {J.}~\bibnamefont
  {Karpie}}, \bibinfo {author} {\bibfnamefont {K.}~\bibnamefont {Orginos}},
  \bibinfo {author} {\bibfnamefont {A.}~\bibnamefont {Radyushkin}}, \ and\
  \bibinfo {author} {\bibfnamefont {S.}~\bibnamefont {Zafeiropoulos}},\ }\href
  {\doibase 10.1051/epjconf/201817506032} {\bibfield  {journal} {\bibinfo
  {journal} {EPJ Web Conf.}\ }\textbf {\bibinfo {volume} {175}},\ \bibinfo
  {pages} {06032} (\bibinfo {year} {2018}{\natexlab{a}})},\ \Eprint
  {http://arxiv.org/abs/1710.08288} {arXiv:1710.08288 [hep-lat]} \BibitemShut
  {NoStop}%
\bibitem [{\citenamefont {Karpie}\ \emph
  {et~al.}(2018{\natexlab{b}})\citenamefont {Karpie}, \citenamefont {Orginos},\
  and\ \citenamefont {Zafeiropoulos}}]{Karpie:2018zaz}%
  \BibitemOpen
  \bibfield  {author} {\bibinfo {author} {\bibfnamefont {J.}~\bibnamefont
  {Karpie}}, \bibinfo {author} {\bibfnamefont {K.}~\bibnamefont {Orginos}}, \
  and\ \bibinfo {author} {\bibfnamefont {S.}~\bibnamefont {Zafeiropoulos}},\
  }\href {\doibase 10.1007/JHEP11(2018)178} {\bibfield  {journal} {\bibinfo
  {journal} {JHEP}\ }\textbf {\bibinfo {volume} {11}},\ \bibinfo {pages} {178}
  (\bibinfo {year} {2018}{\natexlab{b}})},\ \Eprint
  {http://arxiv.org/abs/1807.10933} {arXiv:1807.10933 [hep-lat]} \BibitemShut
  {NoStop}%
\bibitem [{\citenamefont {Karpie}\ \emph {et~al.}(2019)\citenamefont {Karpie},
  \citenamefont {Orginos}, \citenamefont {Rothkopf},\ and\ \citenamefont
  {Zafeiropoulos}}]{Karpie:2019eiq}%
  \BibitemOpen
  \bibfield  {author} {\bibinfo {author} {\bibfnamefont {J.}~\bibnamefont
  {Karpie}}, \bibinfo {author} {\bibfnamefont {K.}~\bibnamefont {Orginos}},
  \bibinfo {author} {\bibfnamefont {A.}~\bibnamefont {Rothkopf}}, \ and\
  \bibinfo {author} {\bibfnamefont {S.}~\bibnamefont {Zafeiropoulos}},\ }\href
  {\doibase 10.1007/JHEP04(2019)057} {\bibfield  {journal} {\bibinfo  {journal}
  {JHEP}\ }\textbf {\bibinfo {volume} {04}},\ \bibinfo {pages} {057} (\bibinfo
  {year} {2019})},\ \Eprint {http://arxiv.org/abs/1901.05408} {arXiv:1901.05408
  [hep-lat]} \BibitemShut {NoStop}%
\bibitem [{\citenamefont {Joó}\ \emph
  {et~al.}(2019{\natexlab{a}})\citenamefont {Joó}, \citenamefont {Karpie},
  \citenamefont {Orginos}, \citenamefont {Radyushkin}, \citenamefont
  {Richards},\ and\ \citenamefont {Zafeiropoulos}}]{Joo:2019jct}%
  \BibitemOpen
  \bibfield  {author} {\bibinfo {author} {\bibfnamefont {B.}~\bibnamefont
  {Joó}}, \bibinfo {author} {\bibfnamefont {J.}~\bibnamefont {Karpie}},
  \bibinfo {author} {\bibfnamefont {K.}~\bibnamefont {Orginos}}, \bibinfo
  {author} {\bibfnamefont {A.}~\bibnamefont {Radyushkin}}, \bibinfo {author}
  {\bibfnamefont {D.}~\bibnamefont {Richards}}, \ and\ \bibinfo {author}
  {\bibfnamefont {S.}~\bibnamefont {Zafeiropoulos}},\ }\href {\doibase
  10.1007/JHEP12(2019)081} {\bibfield  {journal} {\bibinfo  {journal} {JHEP}\
  }\textbf {\bibinfo {volume} {12}},\ \bibinfo {pages} {081} (\bibinfo {year}
  {2019}{\natexlab{a}})},\ \Eprint {http://arxiv.org/abs/1908.09771}
  {arXiv:1908.09771 [hep-lat]} \BibitemShut {NoStop}%
\bibitem [{\citenamefont {Joó}\ \emph
  {et~al.}(2019{\natexlab{b}})\citenamefont {Joó}, \citenamefont {Karpie},
  \citenamefont {Orginos}, \citenamefont {Radyushkin}, \citenamefont
  {Richards}, \citenamefont {Sufian},\ and\ \citenamefont
  {Zafeiropoulos}}]{Joo:2019bzr}%
  \BibitemOpen
  \bibfield  {author} {\bibinfo {author} {\bibfnamefont {B.}~\bibnamefont
  {Joó}}, \bibinfo {author} {\bibfnamefont {J.}~\bibnamefont {Karpie}},
  \bibinfo {author} {\bibfnamefont {K.}~\bibnamefont {Orginos}}, \bibinfo
  {author} {\bibfnamefont {A.~V.}\ \bibnamefont {Radyushkin}}, \bibinfo
  {author} {\bibfnamefont {D.~G.}\ \bibnamefont {Richards}}, \bibinfo {author}
  {\bibfnamefont {R.~S.}\ \bibnamefont {Sufian}}, \ and\ \bibinfo {author}
  {\bibfnamefont {S.}~\bibnamefont {Zafeiropoulos}},\ }\href {\doibase
  10.1103/PhysRevD.100.114512} {\bibfield  {journal} {\bibinfo  {journal}
  {Phys. Rev. D}\ }\textbf {\bibinfo {volume} {100}},\ \bibinfo {pages}
  {114512} (\bibinfo {year} {2019}{\natexlab{b}})},\ \Eprint
  {http://arxiv.org/abs/1909.08517} {arXiv:1909.08517 [hep-lat]} \BibitemShut
  {NoStop}%
\bibitem [{\citenamefont {Radyushkin}(2018)}]{Radyushkin:2018cvn}%
  \BibitemOpen
  \bibfield  {author} {\bibinfo {author} {\bibfnamefont {A.}~\bibnamefont
  {Radyushkin}},\ }\href {\doibase 10.1103/PhysRevD.98.014019} {\bibfield
  {journal} {\bibinfo  {journal} {Phys. Rev.}\ }\textbf {\bibinfo {volume}
  {D98}},\ \bibinfo {pages} {014019} (\bibinfo {year} {2018})},\ \Eprint
  {http://arxiv.org/abs/1801.02427} {arXiv:1801.02427 [hep-ph]} \BibitemShut
  {NoStop}%
\bibitem [{\citenamefont {Zhang}\ \emph {et~al.}(2018)\citenamefont {Zhang},
  \citenamefont {Chen},\ and\ \citenamefont {Monahan}}]{Zhang:2018ggy}%
  \BibitemOpen
  \bibfield  {author} {\bibinfo {author} {\bibfnamefont {J.-H.}\ \bibnamefont
  {Zhang}}, \bibinfo {author} {\bibfnamefont {J.-W.}\ \bibnamefont {Chen}}, \
  and\ \bibinfo {author} {\bibfnamefont {C.}~\bibnamefont {Monahan}},\ }\href
  {\doibase 10.1103/PhysRevD.97.074508} {\bibfield  {journal} {\bibinfo
  {journal} {Phys. Rev.}\ }\textbf {\bibinfo {volume} {D97}},\ \bibinfo {pages}
  {074508} (\bibinfo {year} {2018})},\ \Eprint
  {http://arxiv.org/abs/1801.03023} {arXiv:1801.03023 [hep-ph]} \BibitemShut
  {NoStop}%
\bibitem [{\citenamefont {Lin}\ \emph {et~al.}(2020)\citenamefont {Lin},
  \citenamefont {Chen}, \citenamefont {Fan}, \citenamefont {Zhang},\ and\
  \citenamefont {Zhang}}]{Lin:2020ssv}%
  \BibitemOpen
  \bibfield  {author} {\bibinfo {author} {\bibfnamefont {H.-W.}\ \bibnamefont
  {Lin}}, \bibinfo {author} {\bibfnamefont {J.-W.}\ \bibnamefont {Chen}},
  \bibinfo {author} {\bibfnamefont {Z.}~\bibnamefont {Fan}}, \bibinfo {author}
  {\bibfnamefont {J.-H.}\ \bibnamefont {Zhang}}, \ and\ \bibinfo {author}
  {\bibfnamefont {R.}~\bibnamefont {Zhang}},\ }\href@noop {} {\  (\bibinfo
  {year} {2020})},\ \Eprint {http://arxiv.org/abs/2003.14128} {arXiv:2003.14128
  [hep-lat]} \BibitemShut {NoStop}%
\bibitem [{\citenamefont {Zhang}\ \emph
  {et~al.}(2020{\natexlab{a}})\citenamefont {Zhang}, \citenamefont {Lin},\ and\
  \citenamefont {Yoon}}]{Zhang:2020dkn}%
  \BibitemOpen
  \bibfield  {author} {\bibinfo {author} {\bibfnamefont {R.}~\bibnamefont
  {Zhang}}, \bibinfo {author} {\bibfnamefont {H.-W.}\ \bibnamefont {Lin}}, \
  and\ \bibinfo {author} {\bibfnamefont {B.}~\bibnamefont {Yoon}},\ }\href@noop
  {} {\  (\bibinfo {year} {2020}{\natexlab{a}})},\ \Eprint
  {http://arxiv.org/abs/2005.01124} {arXiv:2005.01124 [hep-lat]} \BibitemShut
  {NoStop}%
\bibitem [{\citenamefont {Fan}\ \emph {et~al.}(2020{\natexlab{a}})\citenamefont
  {Fan}, \citenamefont {Gao}, \citenamefont {Li}, \citenamefont {Lin},
  \citenamefont {Karthik}, \citenamefont {Mukherjee}, \citenamefont
  {Petreczky}, \citenamefont {Syritsyn}, \citenamefont {Yang},\ and\
  \citenamefont {Zhang}}]{Fan:2020nzz}%
  \BibitemOpen
  \bibfield  {author} {\bibinfo {author} {\bibfnamefont {Z.}~\bibnamefont
  {Fan}}, \bibinfo {author} {\bibfnamefont {X.}~\bibnamefont {Gao}}, \bibinfo
  {author} {\bibfnamefont {R.}~\bibnamefont {Li}}, \bibinfo {author}
  {\bibfnamefont {H.-W.}\ \bibnamefont {Lin}}, \bibinfo {author} {\bibfnamefont
  {N.}~\bibnamefont {Karthik}}, \bibinfo {author} {\bibfnamefont
  {S.}~\bibnamefont {Mukherjee}}, \bibinfo {author} {\bibfnamefont
  {P.}~\bibnamefont {Petreczky}}, \bibinfo {author} {\bibfnamefont
  {S.}~\bibnamefont {Syritsyn}}, \bibinfo {author} {\bibfnamefont {Y.-B.}\
  \bibnamefont {Yang}}, \ and\ \bibinfo {author} {\bibfnamefont
  {R.}~\bibnamefont {Zhang}},\ }\href@noop {} {\  (\bibinfo {year}
  {2020}{\natexlab{a}})},\ \Eprint {http://arxiv.org/abs/2005.12015}
  {arXiv:2005.12015 [hep-lat]} \BibitemShut {NoStop}%
\bibitem [{\citenamefont {Sufian}\ \emph {et~al.}(2020)\citenamefont {Sufian},
  \citenamefont {Egerer}, \citenamefont {Karpie}, \citenamefont {Edwards},
  \citenamefont {Joó}, \citenamefont {Ma}, \citenamefont {Orginos},
  \citenamefont {Qiu},\ and\ \citenamefont {Richards}}]{Sufian:2020vzb}%
  \BibitemOpen
  \bibfield  {author} {\bibinfo {author} {\bibfnamefont {R.~S.}\ \bibnamefont
  {Sufian}}, \bibinfo {author} {\bibfnamefont {C.}~\bibnamefont {Egerer}},
  \bibinfo {author} {\bibfnamefont {J.}~\bibnamefont {Karpie}}, \bibinfo
  {author} {\bibfnamefont {R.~G.}\ \bibnamefont {Edwards}}, \bibinfo {author}
  {\bibfnamefont {B.}~\bibnamefont {Joó}}, \bibinfo {author} {\bibfnamefont
  {Y.-Q.}\ \bibnamefont {Ma}}, \bibinfo {author} {\bibfnamefont
  {K.}~\bibnamefont {Orginos}}, \bibinfo {author} {\bibfnamefont {J.-W.}\
  \bibnamefont {Qiu}}, \ and\ \bibinfo {author} {\bibfnamefont {D.~G.}\
  \bibnamefont {Richards}},\ }\href@noop {} {\  (\bibinfo {year} {2020})},\
  \Eprint {http://arxiv.org/abs/2001.04960} {arXiv:2001.04960 [hep-lat]}
  \BibitemShut {NoStop}%
\bibitem [{\citenamefont {Shugert}\ \emph {et~al.}(2020)\citenamefont
  {Shugert}, \citenamefont {Gao}, \citenamefont {Izubichi}, \citenamefont
  {Jin}, \citenamefont {Kallidonis}, \citenamefont {Karthik}, \citenamefont
  {Mukherjee}, \citenamefont {Petreczky}, \citenamefont {Syritsyn},\ and\
  \citenamefont {Zhao}}]{Shugert:2020tgq}%
  \BibitemOpen
  \bibfield  {author} {\bibinfo {author} {\bibfnamefont {C.}~\bibnamefont
  {Shugert}}, \bibinfo {author} {\bibfnamefont {X.}~\bibnamefont {Gao}},
  \bibinfo {author} {\bibfnamefont {T.}~\bibnamefont {Izubichi}}, \bibinfo
  {author} {\bibfnamefont {L.}~\bibnamefont {Jin}}, \bibinfo {author}
  {\bibfnamefont {C.}~\bibnamefont {Kallidonis}}, \bibinfo {author}
  {\bibfnamefont {N.}~\bibnamefont {Karthik}}, \bibinfo {author} {\bibfnamefont
  {S.}~\bibnamefont {Mukherjee}}, \bibinfo {author} {\bibfnamefont
  {P.}~\bibnamefont {Petreczky}}, \bibinfo {author} {\bibfnamefont
  {S.}~\bibnamefont {Syritsyn}}, \ and\ \bibinfo {author} {\bibfnamefont
  {Y.}~\bibnamefont {Zhao}},\ }in\ \href@noop {} {\emph {\bibinfo {booktitle}
  {{37th International Symposium on Lattice Field Theory}}}}\ (\bibinfo {year}
  {2020})\ \Eprint {http://arxiv.org/abs/2001.11650} {arXiv:2001.11650
  [hep-lat]} \BibitemShut {NoStop}%
\bibitem [{\citenamefont {Green}\ \emph {et~al.}(2020)\citenamefont {Green},
  \citenamefont {Jansen},\ and\ \citenamefont {Steffens}}]{Green:2020xco}%
  \BibitemOpen
  \bibfield  {author} {\bibinfo {author} {\bibfnamefont {J.~R.}\ \bibnamefont
  {Green}}, \bibinfo {author} {\bibfnamefont {K.}~\bibnamefont {Jansen}}, \
  and\ \bibinfo {author} {\bibfnamefont {F.}~\bibnamefont {Steffens}},\ }\href
  {\doibase 10.1103/PhysRevD.101.074509} {\bibfield  {journal} {\bibinfo
  {journal} {Phys. Rev. D}\ }\textbf {\bibinfo {volume} {101}},\ \bibinfo
  {pages} {074509} (\bibinfo {year} {2020})},\ \Eprint
  {http://arxiv.org/abs/2002.09408} {arXiv:2002.09408 [hep-lat]} \BibitemShut
  {NoStop}%
\bibitem [{\citenamefont {Chai}\ \emph {et~al.}(2020)\citenamefont {Chai} \emph
  {et~al.}}]{Chai:2020nxw}%
  \BibitemOpen
  \bibfield  {author} {\bibinfo {author} {\bibfnamefont {Y.}~\bibnamefont
  {Chai}} \emph {et~al.},\ }\href@noop {} {\  (\bibinfo {year} {2020})},\
  \Eprint {http://arxiv.org/abs/2002.12044} {arXiv:2002.12044 [hep-lat]}
  \BibitemShut {NoStop}%
\bibitem [{\citenamefont {Shanahan}\ \emph {et~al.}(2020)\citenamefont
  {Shanahan}, \citenamefont {Wagman},\ and\ \citenamefont
  {Zhao}}]{Shanahan:2020zxr}%
  \BibitemOpen
  \bibfield  {author} {\bibinfo {author} {\bibfnamefont {P.}~\bibnamefont
  {Shanahan}}, \bibinfo {author} {\bibfnamefont {M.}~\bibnamefont {Wagman}}, \
  and\ \bibinfo {author} {\bibfnamefont {Y.}~\bibnamefont {Zhao}},\ }\href@noop
  {} {\  (\bibinfo {year} {2020})},\ \Eprint {http://arxiv.org/abs/2003.06063}
  {arXiv:2003.06063 [hep-lat]} \BibitemShut {NoStop}%
\bibitem [{\citenamefont {Braun}\ \emph {et~al.}(2020)\citenamefont {Braun},
  \citenamefont {Chetyrkin},\ and\ \citenamefont {Kniehl}}]{Braun:2020ymy}%
  \BibitemOpen
  \bibfield  {author} {\bibinfo {author} {\bibfnamefont {V.}~\bibnamefont
  {Braun}}, \bibinfo {author} {\bibfnamefont {K.}~\bibnamefont {Chetyrkin}}, \
  and\ \bibinfo {author} {\bibfnamefont {B.}~\bibnamefont {Kniehl}},\
  }\href@noop {} {\  (\bibinfo {year} {2020})},\ \Eprint
  {http://arxiv.org/abs/2004.01043} {arXiv:2004.01043 [hep-ph]} \BibitemShut
  {NoStop}%
\bibitem [{\citenamefont {Bhattacharya}\ \emph {et~al.}(2020)\citenamefont
  {Bhattacharya}, \citenamefont {Cichy}, \citenamefont {Constantinou},
  \citenamefont {Metz}, \citenamefont {Scapellato},\ and\ \citenamefont
  {Steffens}}]{Bhattacharya:2020cen}%
  \BibitemOpen
  \bibfield  {author} {\bibinfo {author} {\bibfnamefont {S.}~\bibnamefont
  {Bhattacharya}}, \bibinfo {author} {\bibfnamefont {K.}~\bibnamefont {Cichy}},
  \bibinfo {author} {\bibfnamefont {M.}~\bibnamefont {Constantinou}}, \bibinfo
  {author} {\bibfnamefont {A.}~\bibnamefont {Metz}}, \bibinfo {author}
  {\bibfnamefont {A.}~\bibnamefont {Scapellato}}, \ and\ \bibinfo {author}
  {\bibfnamefont {F.}~\bibnamefont {Steffens}},\ }\href@noop {} {\  (\bibinfo
  {year} {2020})},\ \Eprint {http://arxiv.org/abs/2004.04130} {arXiv:2004.04130
  [hep-lat]} \BibitemShut {NoStop}%
\bibitem [{\citenamefont {Ji}\ \emph {et~al.}(2020{\natexlab{a}})\citenamefont
  {Ji}, \citenamefont {Liu}, \citenamefont {Liu}, \citenamefont {Zhang},\ and\
  \citenamefont {Zhao}}]{Ji:2020ect}%
  \BibitemOpen
  \bibfield  {author} {\bibinfo {author} {\bibfnamefont {X.}~\bibnamefont
  {Ji}}, \bibinfo {author} {\bibfnamefont {Y.-S.}\ \bibnamefont {Liu}},
  \bibinfo {author} {\bibfnamefont {Y.}~\bibnamefont {Liu}}, \bibinfo {author}
  {\bibfnamefont {J.-H.}\ \bibnamefont {Zhang}}, \ and\ \bibinfo {author}
  {\bibfnamefont {Y.}~\bibnamefont {Zhao}},\ }\href@noop {} {\  (\bibinfo
  {year} {2020}{\natexlab{a}})},\ \Eprint {http://arxiv.org/abs/2004.03543}
  {arXiv:2004.03543 [hep-ph]} \BibitemShut {NoStop}%
\bibitem [{\citenamefont {Ebert}\ \emph
  {et~al.}(2020{\natexlab{b}})\citenamefont {Ebert}, \citenamefont {Schindler},
  \citenamefont {Stewart},\ and\ \citenamefont {Zhao}}]{Ebert:2020gxr}%
  \BibitemOpen
  \bibfield  {author} {\bibinfo {author} {\bibfnamefont {M.~A.}\ \bibnamefont
  {Ebert}}, \bibinfo {author} {\bibfnamefont {S.~T.}\ \bibnamefont
  {Schindler}}, \bibinfo {author} {\bibfnamefont {I.~W.}\ \bibnamefont
  {Stewart}}, \ and\ \bibinfo {author} {\bibfnamefont {Y.}~\bibnamefont
  {Zhao}},\ }\href@noop {} {\  (\bibinfo {year} {2020}{\natexlab{b}})},\
  \Eprint {http://arxiv.org/abs/2004.14831} {arXiv:2004.14831 [hep-ph]}
  \BibitemShut {NoStop}%
\bibitem [{\citenamefont {Lin}(2020{\natexlab{a}})}]{Lin:2020ijm}%
  \BibitemOpen
  \bibfield  {author} {\bibinfo {author} {\bibfnamefont {H.-W.}\ \bibnamefont
  {Lin}},\ }\href {\doibase 10.1142/S0217751X20300069} {\bibfield  {journal}
  {\bibinfo  {journal} {Int. J. Mod. Phys. A}\ }\textbf {\bibinfo {volume}
  {35}},\ \bibinfo {pages} {2030006} (\bibinfo {year}
  {2020}{\natexlab{a}})}\BibitemShut {NoStop}%
\bibitem [{\citenamefont {Joó}\ \emph {et~al.}(2020)\citenamefont {Joó},
  \citenamefont {Karpie}, \citenamefont {Orginos}, \citenamefont {Radyushkin},
  \citenamefont {Richards},\ and\ \citenamefont {Zafeiropoulos}}]{Joo:2020spy}%
  \BibitemOpen
  \bibfield  {author} {\bibinfo {author} {\bibfnamefont {B.}~\bibnamefont
  {Joó}}, \bibinfo {author} {\bibfnamefont {J.}~\bibnamefont {Karpie}},
  \bibinfo {author} {\bibfnamefont {K.}~\bibnamefont {Orginos}}, \bibinfo
  {author} {\bibfnamefont {A.~V.}\ \bibnamefont {Radyushkin}}, \bibinfo
  {author} {\bibfnamefont {D.~G.}\ \bibnamefont {Richards}}, \ and\ \bibinfo
  {author} {\bibfnamefont {S.}~\bibnamefont {Zafeiropoulos}},\ }\href@noop {}
  {\  (\bibinfo {year} {2020})},\ \Eprint {http://arxiv.org/abs/2004.01687}
  {arXiv:2004.01687 [hep-lat]} \BibitemShut {NoStop}%
\bibitem [{\citenamefont {Bhat}\ \emph {et~al.}(2020)\citenamefont {Bhat},
  \citenamefont {Cichy}, \citenamefont {Constantinou},\ and\ \citenamefont
  {Scapellato}}]{Bhat:2020ktg}%
  \BibitemOpen
  \bibfield  {author} {\bibinfo {author} {\bibfnamefont {M.}~\bibnamefont
  {Bhat}}, \bibinfo {author} {\bibfnamefont {K.}~\bibnamefont {Cichy}},
  \bibinfo {author} {\bibfnamefont {M.}~\bibnamefont {Constantinou}}, \ and\
  \bibinfo {author} {\bibfnamefont {A.}~\bibnamefont {Scapellato}},\
  }\href@noop {} {\  (\bibinfo {year} {2020})},\ \Eprint
  {http://arxiv.org/abs/2005.02102} {arXiv:2005.02102 [hep-lat]} \BibitemShut
  {NoStop}%
\bibitem [{\citenamefont {Fan}\ \emph {et~al.}(2020{\natexlab{b}})\citenamefont
  {Fan}, \citenamefont {Zhang},\ and\ \citenamefont {Lin}}]{Fan:2020cpa}%
  \BibitemOpen
  \bibfield  {author} {\bibinfo {author} {\bibfnamefont {Z.}~\bibnamefont
  {Fan}}, \bibinfo {author} {\bibfnamefont {R.}~\bibnamefont {Zhang}}, \ and\
  \bibinfo {author} {\bibfnamefont {H.-W.}\ \bibnamefont {Lin}},\ }\href@noop
  {} {\  (\bibinfo {year} {2020}{\natexlab{b}})},\ \Eprint
  {http://arxiv.org/abs/2007.16113} {arXiv:2007.16113 [hep-lat]} \BibitemShut
  {NoStop}%
\bibitem [{\citenamefont {Zhang}\ \emph
  {et~al.}(2020{\natexlab{b}})\citenamefont {Zhang}, \citenamefont {Honkala},
  \citenamefont {Lin},\ and\ \citenamefont {Chen}}]{Zhang:2020gaj}%
  \BibitemOpen
  \bibfield  {author} {\bibinfo {author} {\bibfnamefont {R.}~\bibnamefont
  {Zhang}}, \bibinfo {author} {\bibfnamefont {C.}~\bibnamefont {Honkala}},
  \bibinfo {author} {\bibfnamefont {H.-W.}\ \bibnamefont {Lin}}, \ and\
  \bibinfo {author} {\bibfnamefont {J.-W.}\ \bibnamefont {Chen}},\ }\href@noop
  {} {\  (\bibinfo {year} {2020}{\natexlab{b}})},\ \Eprint
  {http://arxiv.org/abs/2005.13955} {arXiv:2005.13955 [hep-lat]} \BibitemShut
  {NoStop}%
\bibitem [{\citenamefont {Zhang}\ \emph
  {et~al.}(2020{\natexlab{c}})\citenamefont {Zhang}, \citenamefont {Fan},
  \citenamefont {Li}, \citenamefont {Lin},\ and\ \citenamefont
  {Yoon}}]{Zhang:2019qiq}%
  \BibitemOpen
  \bibfield  {author} {\bibinfo {author} {\bibfnamefont {R.}~\bibnamefont
  {Zhang}}, \bibinfo {author} {\bibfnamefont {Z.}~\bibnamefont {Fan}}, \bibinfo
  {author} {\bibfnamefont {R.}~\bibnamefont {Li}}, \bibinfo {author}
  {\bibfnamefont {H.-W.}\ \bibnamefont {Lin}}, \ and\ \bibinfo {author}
  {\bibfnamefont {B.}~\bibnamefont {Yoon}},\ }\href {\doibase
  10.1103/PhysRevD.101.034516} {\bibfield  {journal} {\bibinfo  {journal}
  {Phys. Rev.}\ }\textbf {\bibinfo {volume} {D101}},\ \bibinfo {pages} {034516}
  (\bibinfo {year} {2020}{\natexlab{c}})},\ \Eprint
  {http://arxiv.org/abs/1909.10990} {arXiv:1909.10990 [hep-lat]} \BibitemShut
  {NoStop}%
\bibitem [{\citenamefont {Alexandrou}\ \emph
  {et~al.}(2019{\natexlab{a}})\citenamefont {Alexandrou}, \citenamefont
  {Cichy}, \citenamefont {Constantinou}, \citenamefont {Hadjiyiannakou},
  \citenamefont {Jansen}, \citenamefont {Scapellato},\ and\ \citenamefont
  {Steffens}}]{Alexandrou:2019dax}%
  \BibitemOpen
  \bibfield  {author} {\bibinfo {author} {\bibfnamefont {C.}~\bibnamefont
  {Alexandrou}}, \bibinfo {author} {\bibfnamefont {K.}~\bibnamefont {Cichy}},
  \bibinfo {author} {\bibfnamefont {M.}~\bibnamefont {Constantinou}}, \bibinfo
  {author} {\bibfnamefont {K.}~\bibnamefont {Hadjiyiannakou}}, \bibinfo
  {author} {\bibfnamefont {K.}~\bibnamefont {Jansen}}, \bibinfo {author}
  {\bibfnamefont {A.}~\bibnamefont {Scapellato}}, \ and\ \bibinfo {author}
  {\bibfnamefont {F.}~\bibnamefont {Steffens}},\ }in\ \href@noop {} {\emph
  {\bibinfo {booktitle} {{37th International Symposium on Lattice Field
  Theory}}}}\ (\bibinfo {year} {2019})\ \Eprint
  {http://arxiv.org/abs/1910.13229} {arXiv:1910.13229 [hep-lat]} \BibitemShut
  {NoStop}%
\bibitem [{\citenamefont {Alexandrou}\ \emph
  {et~al.}(2020{\natexlab{a}})\citenamefont {Alexandrou}, \citenamefont
  {Cichy}, \citenamefont {Constantinou}, \citenamefont {Hadjiyiannakou},
  \citenamefont {Jansen}, \citenamefont {Scapellato},\ and\ \citenamefont
  {Steffens}}]{Alexandrou:2020zbe}%
  \BibitemOpen
  \bibfield  {author} {\bibinfo {author} {\bibfnamefont {C.}~\bibnamefont
  {Alexandrou}}, \bibinfo {author} {\bibfnamefont {K.}~\bibnamefont {Cichy}},
  \bibinfo {author} {\bibfnamefont {M.}~\bibnamefont {Constantinou}}, \bibinfo
  {author} {\bibfnamefont {K.}~\bibnamefont {Hadjiyiannakou}}, \bibinfo
  {author} {\bibfnamefont {K.}~\bibnamefont {Jansen}}, \bibinfo {author}
  {\bibfnamefont {A.}~\bibnamefont {Scapellato}}, \ and\ \bibinfo {author}
  {\bibfnamefont {F.}~\bibnamefont {Steffens}},\ }\href {\doibase
  10.1103/PhysRevLett.125.262001} {\bibfield  {journal} {\bibinfo  {journal}
  {Phys. Rev. Lett.}\ }\textbf {\bibinfo {volume} {125}},\ \bibinfo {pages}
  {262001} (\bibinfo {year} {2020}{\natexlab{a}})},\ \Eprint
  {http://arxiv.org/abs/2008.10573} {arXiv:2008.10573 [hep-lat]} \BibitemShut
  {NoStop}%
\bibitem [{\citenamefont {Follana}\ \emph {et~al.}(2007)\citenamefont
  {Follana}, \citenamefont {Mason}, \citenamefont {Davies}, \citenamefont
  {Hornbostel}, \citenamefont {Lepage}, \citenamefont {Shigemitsu},
  \citenamefont {Trottier},\ and\ \citenamefont {Wong}}]{Follana:2006rc}%
  \BibitemOpen
  \bibfield  {author} {\bibinfo {author} {\bibfnamefont {E.}~\bibnamefont
  {Follana}}, \bibinfo {author} {\bibfnamefont {Q.}~\bibnamefont {Mason}},
  \bibinfo {author} {\bibfnamefont {C.}~\bibnamefont {Davies}}, \bibinfo
  {author} {\bibfnamefont {K.}~\bibnamefont {Hornbostel}}, \bibinfo {author}
  {\bibfnamefont {G.~P.}\ \bibnamefont {Lepage}}, \bibinfo {author}
  {\bibfnamefont {J.}~\bibnamefont {Shigemitsu}}, \bibinfo {author}
  {\bibfnamefont {H.}~\bibnamefont {Trottier}}, \ and\ \bibinfo {author}
  {\bibfnamefont {K.}~\bibnamefont {Wong}} (\bibinfo {collaboration} {HPQCD,
  UKQCD}),\ }\href {\doibase 10.1103/PhysRevD.75.054502} {\bibfield  {journal}
  {\bibinfo  {journal} {Phys. Rev.}\ }\textbf {\bibinfo {volume} {D75}},\
  \bibinfo {pages} {054502} (\bibinfo {year} {2007})},\ \Eprint
  {http://arxiv.org/abs/hep-lat/0610092} {arXiv:hep-lat/0610092 [hep-lat]}
  \BibitemShut {NoStop}%
\bibitem [{\citenamefont {Bazavov}\ \emph {et~al.}(2013)\citenamefont {Bazavov}
  \emph {et~al.}}]{Bazavov:2012xda}%
  \BibitemOpen
  \bibfield  {author} {\bibinfo {author} {\bibfnamefont {A.}~\bibnamefont
  {Bazavov}} \emph {et~al.} (\bibinfo {collaboration} {MILC}),\ }\href
  {\doibase 10.1103/PhysRevD.87.054505} {\bibfield  {journal} {\bibinfo
  {journal} {Phys. Rev.}\ }\textbf {\bibinfo {volume} {D87}},\ \bibinfo {pages}
  {054505} (\bibinfo {year} {2013})},\ \Eprint {http://arxiv.org/abs/1212.4768}
  {arXiv:1212.4768 [hep-lat]} \BibitemShut {NoStop}%
\bibitem [{\citenamefont {Hasenfratz}\ and\ \citenamefont
  {Knechtli}(2001)}]{Hasenfratz:2001hp}%
  \BibitemOpen
  \bibfield  {author} {\bibinfo {author} {\bibfnamefont {A.}~\bibnamefont
  {Hasenfratz}}\ and\ \bibinfo {author} {\bibfnamefont {F.}~\bibnamefont
  {Knechtli}},\ }\href {\doibase 10.1103/PhysRevD.64.034504} {\bibfield
  {journal} {\bibinfo  {journal} {Phys. Rev.}\ }\textbf {\bibinfo {volume}
  {D64}},\ \bibinfo {pages} {034504} (\bibinfo {year} {2001})},\ \Eprint
  {http://arxiv.org/abs/hep-lat/0103029} {arXiv:hep-lat/0103029 [hep-lat]}
  \BibitemShut {NoStop}%
\bibitem [{\citenamefont {Mondal}\ \emph {et~al.}(2020)\citenamefont {Mondal},
  \citenamefont {Gupta}, \citenamefont {Park}, \citenamefont {Yoon},
  \citenamefont {Bhattacharya},\ and\ \citenamefont {Lin}}]{Mondal:2020cmt}%
  \BibitemOpen
  \bibfield  {author} {\bibinfo {author} {\bibfnamefont {S.}~\bibnamefont
  {Mondal}}, \bibinfo {author} {\bibfnamefont {R.}~\bibnamefont {Gupta}},
  \bibinfo {author} {\bibfnamefont {S.}~\bibnamefont {Park}}, \bibinfo {author}
  {\bibfnamefont {B.}~\bibnamefont {Yoon}}, \bibinfo {author} {\bibfnamefont
  {T.}~\bibnamefont {Bhattacharya}}, \ and\ \bibinfo {author} {\bibfnamefont
  {H.-W.}\ \bibnamefont {Lin}},\ }\href@noop {} {\  (\bibinfo {year} {2020})},\
  \Eprint {http://arxiv.org/abs/2005.13779} {arXiv:2005.13779 [hep-lat]}
  \BibitemShut {NoStop}%
\bibitem [{\citenamefont {Jang}\ \emph {et~al.}(2020)\citenamefont {Jang},
  \citenamefont {Gupta}, \citenamefont {Lin}, \citenamefont {Yoon},\ and\
  \citenamefont {Bhattacharya}}]{Jang:2019jkn}%
  \BibitemOpen
  \bibfield  {author} {\bibinfo {author} {\bibfnamefont {Y.-C.}\ \bibnamefont
  {Jang}}, \bibinfo {author} {\bibfnamefont {R.}~\bibnamefont {Gupta}},
  \bibinfo {author} {\bibfnamefont {H.-W.}\ \bibnamefont {Lin}}, \bibinfo
  {author} {\bibfnamefont {B.}~\bibnamefont {Yoon}}, \ and\ \bibinfo {author}
  {\bibfnamefont {T.}~\bibnamefont {Bhattacharya}},\ }\href {\doibase
  10.1103/PhysRevD.101.014507} {\bibfield  {journal} {\bibinfo  {journal}
  {Phys. Rev. D}\ }\textbf {\bibinfo {volume} {101}},\ \bibinfo {pages}
  {014507} (\bibinfo {year} {2020})},\ \Eprint
  {http://arxiv.org/abs/1906.07217} {arXiv:1906.07217 [hep-lat]} \BibitemShut
  {NoStop}%
\bibitem [{\citenamefont {Gupta}\ \emph
  {et~al.}(2018{\natexlab{a}})\citenamefont {Gupta}, \citenamefont {Yoon},
  \citenamefont {Bhattacharya}, \citenamefont {Cirigliano}, \citenamefont
  {Jang},\ and\ \citenamefont {Lin}}]{Gupta:2018lvp}%
  \BibitemOpen
  \bibfield  {author} {\bibinfo {author} {\bibfnamefont {R.}~\bibnamefont
  {Gupta}}, \bibinfo {author} {\bibfnamefont {B.}~\bibnamefont {Yoon}},
  \bibinfo {author} {\bibfnamefont {T.}~\bibnamefont {Bhattacharya}}, \bibinfo
  {author} {\bibfnamefont {V.}~\bibnamefont {Cirigliano}}, \bibinfo {author}
  {\bibfnamefont {Y.-C.}\ \bibnamefont {Jang}}, \ and\ \bibinfo {author}
  {\bibfnamefont {H.-W.}\ \bibnamefont {Lin}},\ }\href {\doibase
  10.1103/PhysRevD.98.091501} {\bibfield  {journal} {\bibinfo  {journal} {Phys.
  Rev. D}\ }\textbf {\bibinfo {volume} {98}},\ \bibinfo {pages} {091501}
  (\bibinfo {year} {2018}{\natexlab{a}})},\ \Eprint
  {http://arxiv.org/abs/1808.07597} {arXiv:1808.07597 [hep-lat]} \BibitemShut
  {NoStop}%
\bibitem [{\citenamefont {Lin}\ \emph {et~al.}(2018{\natexlab{c}})\citenamefont
  {Lin}, \citenamefont {Gupta}, \citenamefont {Yoon}, \citenamefont {Jang},\
  and\ \citenamefont {Bhattacharya}}]{Lin:2018obj}%
  \BibitemOpen
  \bibfield  {author} {\bibinfo {author} {\bibfnamefont {H.-W.}\ \bibnamefont
  {Lin}}, \bibinfo {author} {\bibfnamefont {R.}~\bibnamefont {Gupta}}, \bibinfo
  {author} {\bibfnamefont {B.}~\bibnamefont {Yoon}}, \bibinfo {author}
  {\bibfnamefont {Y.-C.}\ \bibnamefont {Jang}}, \ and\ \bibinfo {author}
  {\bibfnamefont {T.}~\bibnamefont {Bhattacharya}},\ }\href {\doibase
  10.1103/PhysRevD.98.094512} {\bibfield  {journal} {\bibinfo  {journal} {Phys.
  Rev. D}\ }\textbf {\bibinfo {volume} {98}},\ \bibinfo {pages} {094512}
  (\bibinfo {year} {2018}{\natexlab{c}})},\ \Eprint
  {http://arxiv.org/abs/1806.10604} {arXiv:1806.10604 [hep-lat]} \BibitemShut
  {NoStop}%
\bibitem [{\citenamefont {Gupta}\ \emph
  {et~al.}(2018{\natexlab{b}})\citenamefont {Gupta}, \citenamefont {Jang},
  \citenamefont {Yoon}, \citenamefont {Lin}, \citenamefont {Cirigliano},\ and\
  \citenamefont {Bhattacharya}}]{Gupta:2018qil}%
  \BibitemOpen
  \bibfield  {author} {\bibinfo {author} {\bibfnamefont {R.}~\bibnamefont
  {Gupta}}, \bibinfo {author} {\bibfnamefont {Y.-C.}\ \bibnamefont {Jang}},
  \bibinfo {author} {\bibfnamefont {B.}~\bibnamefont {Yoon}}, \bibinfo {author}
  {\bibfnamefont {H.-W.}\ \bibnamefont {Lin}}, \bibinfo {author} {\bibfnamefont
  {V.}~\bibnamefont {Cirigliano}}, \ and\ \bibinfo {author} {\bibfnamefont
  {T.}~\bibnamefont {Bhattacharya}},\ }\href {\doibase
  10.1103/PhysRevD.98.034503} {\bibfield  {journal} {\bibinfo  {journal} {Phys.
  Rev.}\ }\textbf {\bibinfo {volume} {D98}},\ \bibinfo {pages} {034503}
  (\bibinfo {year} {2018}{\natexlab{b}})},\ \Eprint
  {http://arxiv.org/abs/1806.09006} {arXiv:1806.09006 [hep-lat]} \BibitemShut
  {NoStop}%
\bibitem [{\citenamefont {Gupta}\ \emph {et~al.}(2017)\citenamefont {Gupta},
  \citenamefont {Jang}, \citenamefont {Lin}, \citenamefont {Yoon},\ and\
  \citenamefont {Bhattacharya}}]{Rajan:2017lxk}%
  \BibitemOpen
  \bibfield  {author} {\bibinfo {author} {\bibfnamefont {R.}~\bibnamefont
  {Gupta}}, \bibinfo {author} {\bibfnamefont {Y.-C.}\ \bibnamefont {Jang}},
  \bibinfo {author} {\bibfnamefont {H.-W.}\ \bibnamefont {Lin}}, \bibinfo
  {author} {\bibfnamefont {B.}~\bibnamefont {Yoon}}, \ and\ \bibinfo {author}
  {\bibfnamefont {T.}~\bibnamefont {Bhattacharya}},\ }\href {\doibase
  10.1103/PhysRevD.96.114503} {\bibfield  {journal} {\bibinfo  {journal} {Phys.
  Rev.}\ }\textbf {\bibinfo {volume} {D96}},\ \bibinfo {pages} {114503}
  (\bibinfo {year} {2017})},\ \Eprint {http://arxiv.org/abs/1705.06834}
  {arXiv:1705.06834 [hep-lat]} \BibitemShut {NoStop}%
\bibitem [{\citenamefont {Yoon}\ \emph {et~al.}(2017)\citenamefont {Yoon} \emph
  {et~al.}}]{Yoon:2016jzj}%
  \BibitemOpen
  \bibfield  {author} {\bibinfo {author} {\bibfnamefont {B.}~\bibnamefont
  {Yoon}} \emph {et~al.},\ }\href {\doibase 10.1103/PhysRevD.95.074508}
  {\bibfield  {journal} {\bibinfo  {journal} {Phys. Rev. D}\ }\textbf {\bibinfo
  {volume} {95}},\ \bibinfo {pages} {074508} (\bibinfo {year} {2017})},\
  \Eprint {http://arxiv.org/abs/1611.07452} {arXiv:1611.07452 [hep-lat]}
  \BibitemShut {NoStop}%
\bibitem [{\citenamefont {Bhattacharya}\ \emph {et~al.}(2016)\citenamefont
  {Bhattacharya}, \citenamefont {Cirigliano}, \citenamefont {Cohen},
  \citenamefont {Gupta}, \citenamefont {Lin},\ and\ \citenamefont
  {Yoon}}]{Bhattacharya:2016zcn}%
  \BibitemOpen
  \bibfield  {author} {\bibinfo {author} {\bibfnamefont {T.}~\bibnamefont
  {Bhattacharya}}, \bibinfo {author} {\bibfnamefont {V.}~\bibnamefont
  {Cirigliano}}, \bibinfo {author} {\bibfnamefont {S.}~\bibnamefont {Cohen}},
  \bibinfo {author} {\bibfnamefont {R.}~\bibnamefont {Gupta}}, \bibinfo
  {author} {\bibfnamefont {H.-W.}\ \bibnamefont {Lin}}, \ and\ \bibinfo
  {author} {\bibfnamefont {B.}~\bibnamefont {Yoon}},\ }\href {\doibase
  10.1103/PhysRevD.94.054508} {\bibfield  {journal} {\bibinfo  {journal} {Phys.
  Rev. D}\ }\textbf {\bibinfo {volume} {94}},\ \bibinfo {pages} {054508}
  (\bibinfo {year} {2016})},\ \Eprint {http://arxiv.org/abs/1606.07049}
  {arXiv:1606.07049 [hep-lat]} \BibitemShut {NoStop}%
\bibitem [{\citenamefont {Bhattacharya}\ \emph {et~al.}(2015)\citenamefont
  {Bhattacharya}, \citenamefont {Cirigliano}, \citenamefont {Gupta},
  \citenamefont {Lin},\ and\ \citenamefont {Yoon}}]{Bhattacharya:2015esa}%
  \BibitemOpen
  \bibfield  {author} {\bibinfo {author} {\bibfnamefont {T.}~\bibnamefont
  {Bhattacharya}}, \bibinfo {author} {\bibfnamefont {V.}~\bibnamefont
  {Cirigliano}}, \bibinfo {author} {\bibfnamefont {R.}~\bibnamefont {Gupta}},
  \bibinfo {author} {\bibfnamefont {H.-W.}\ \bibnamefont {Lin}}, \ and\
  \bibinfo {author} {\bibfnamefont {B.}~\bibnamefont {Yoon}},\ }\href {\doibase
  10.1103/PhysRevLett.115.212002} {\bibfield  {journal} {\bibinfo  {journal}
  {Phys. Rev. Lett.}\ }\textbf {\bibinfo {volume} {115}},\ \bibinfo {pages}
  {212002} (\bibinfo {year} {2015})},\ \Eprint
  {http://arxiv.org/abs/1506.04196} {arXiv:1506.04196 [hep-lat]} \BibitemShut
  {NoStop}%
\bibitem [{\citenamefont {Bhattacharya}\ \emph {et~al.}(2014)\citenamefont
  {Bhattacharya}, \citenamefont {Cohen}, \citenamefont {Gupta}, \citenamefont
  {Joseph}, \citenamefont {Lin},\ and\ \citenamefont
  {Yoon}}]{Bhattacharya:2013ehc}%
  \BibitemOpen
  \bibfield  {author} {\bibinfo {author} {\bibfnamefont {T.}~\bibnamefont
  {Bhattacharya}}, \bibinfo {author} {\bibfnamefont {S.~D.}\ \bibnamefont
  {Cohen}}, \bibinfo {author} {\bibfnamefont {R.}~\bibnamefont {Gupta}},
  \bibinfo {author} {\bibfnamefont {A.}~\bibnamefont {Joseph}}, \bibinfo
  {author} {\bibfnamefont {H.-W.}\ \bibnamefont {Lin}}, \ and\ \bibinfo
  {author} {\bibfnamefont {B.}~\bibnamefont {Yoon}},\ }\href {\doibase
  10.1103/PhysRevD.89.094502} {\bibfield  {journal} {\bibinfo  {journal} {Phys.
  Rev.}\ }\textbf {\bibinfo {volume} {D89}},\ \bibinfo {pages} {094502}
  (\bibinfo {year} {2014})},\ \Eprint {http://arxiv.org/abs/1306.5435}
  {arXiv:1306.5435 [hep-lat]} \BibitemShut {NoStop}%
\bibitem [{\citenamefont {Briceno}\ \emph {et~al.}(2012)\citenamefont
  {Briceno}, \citenamefont {Lin},\ and\ \citenamefont
  {Bolton}}]{Briceno:2012wt}%
  \BibitemOpen
  \bibfield  {author} {\bibinfo {author} {\bibfnamefont {R.~A.}\ \bibnamefont
  {Briceno}}, \bibinfo {author} {\bibfnamefont {H.-W.}\ \bibnamefont {Lin}}, \
  and\ \bibinfo {author} {\bibfnamefont {D.~R.}\ \bibnamefont {Bolton}},\
  }\href {\doibase 10.1103/PhysRevD.86.094504} {\bibfield  {journal} {\bibinfo
  {journal} {Phys. Rev. D}\ }\textbf {\bibinfo {volume} {86}},\ \bibinfo
  {pages} {094504} (\bibinfo {year} {2012})},\ \Eprint
  {http://arxiv.org/abs/1207.3536} {arXiv:1207.3536 [hep-lat]} \BibitemShut
  {NoStop}%
\bibitem [{\citenamefont {Bhattacharya}\ \emph {et~al.}(2012)\citenamefont
  {Bhattacharya}, \citenamefont {Cirigliano}, \citenamefont {Cohen},
  \citenamefont {Filipuzzi}, \citenamefont {Gonzalez-Alonso}, \citenamefont
  {Graesser}, \citenamefont {Gupta},\ and\ \citenamefont
  {Lin}}]{Bhattacharya:2011qm}%
  \BibitemOpen
  \bibfield  {author} {\bibinfo {author} {\bibfnamefont {T.}~\bibnamefont
  {Bhattacharya}}, \bibinfo {author} {\bibfnamefont {V.}~\bibnamefont
  {Cirigliano}}, \bibinfo {author} {\bibfnamefont {S.~D.}\ \bibnamefont
  {Cohen}}, \bibinfo {author} {\bibfnamefont {A.}~\bibnamefont {Filipuzzi}},
  \bibinfo {author} {\bibfnamefont {M.}~\bibnamefont {Gonzalez-Alonso}},
  \bibinfo {author} {\bibfnamefont {M.~L.}\ \bibnamefont {Graesser}}, \bibinfo
  {author} {\bibfnamefont {R.}~\bibnamefont {Gupta}}, \ and\ \bibinfo {author}
  {\bibfnamefont {H.-W.}\ \bibnamefont {Lin}},\ }\href {\doibase
  10.1103/PhysRevD.85.054512} {\bibfield  {journal} {\bibinfo  {journal} {Phys.
  Rev. D}\ }\textbf {\bibinfo {volume} {85}},\ \bibinfo {pages} {054512}
  (\bibinfo {year} {2012})},\ \Eprint {http://arxiv.org/abs/1110.6448}
  {arXiv:1110.6448 [hep-ph]} \BibitemShut {NoStop}%
\bibitem [{\citenamefont {Lin}(2020{\natexlab{b}})}]{Lin:2020reh}%
  \BibitemOpen
  \bibfield  {author} {\bibinfo {author} {\bibfnamefont {H.-W.}\ \bibnamefont
  {Lin}},\ }\href {\doibase 10.1142/S0217751X20300069} {\bibfield  {journal}
  {\bibinfo  {journal} {Int. J. Mod. Phys. A}\ }\textbf {\bibinfo {volume}
  {35}},\ \bibinfo {pages} {2030006} (\bibinfo {year}
  {2020}{\natexlab{b}})}\BibitemShut {NoStop}%
\bibitem [{\citenamefont {Bali}\ \emph {et~al.}(2016)\citenamefont {Bali},
  \citenamefont {Lang}, \citenamefont {Musch},\ and\ \citenamefont
  {Schäfer}}]{Bali:2016lva}%
  \BibitemOpen
  \bibfield  {author} {\bibinfo {author} {\bibfnamefont {G.~S.}\ \bibnamefont
  {Bali}}, \bibinfo {author} {\bibfnamefont {B.}~\bibnamefont {Lang}}, \bibinfo
  {author} {\bibfnamefont {B.~U.}\ \bibnamefont {Musch}}, \ and\ \bibinfo
  {author} {\bibfnamefont {A.}~\bibnamefont {Schäfer}},\ }\href {\doibase
  10.1103/PhysRevD.93.094515} {\bibfield  {journal} {\bibinfo  {journal} {Phys.
  Rev.}\ }\textbf {\bibinfo {volume} {D93}},\ \bibinfo {pages} {094515}
  (\bibinfo {year} {2016})},\ \Eprint {http://arxiv.org/abs/1602.05525}
  {arXiv:1602.05525 [hep-lat]} \BibitemShut {NoStop}%
\bibitem [{\citenamefont {Yoon}\ \emph {et~al.}(2016)\citenamefont {Yoon} \emph
  {et~al.}}]{Yoon:2016dij}%
  \BibitemOpen
  \bibfield  {author} {\bibinfo {author} {\bibfnamefont {B.}~\bibnamefont
  {Yoon}} \emph {et~al.},\ }\href {\doibase 10.1103/PhysRevD.93.114506}
  {\bibfield  {journal} {\bibinfo  {journal} {Phys. Rev. D}\ }\textbf {\bibinfo
  {volume} {93}},\ \bibinfo {pages} {114506} (\bibinfo {year} {2016})},\
  \Eprint {http://arxiv.org/abs/1602.07737} {arXiv:1602.07737 [hep-lat]}
  \BibitemShut {NoStop}%
\bibitem [{\citenamefont {Ji}\ \emph {et~al.}(2020{\natexlab{b}})\citenamefont
  {Ji}, \citenamefont {Liu}, \citenamefont {Sch\"afer}, \citenamefont {Wang},
  \citenamefont {Yang}, \citenamefont {Zhang},\ and\ \citenamefont
  {Zhao}}]{Ji:2020brr}%
  \BibitemOpen
  \bibfield  {author} {\bibinfo {author} {\bibfnamefont {X.}~\bibnamefont
  {Ji}}, \bibinfo {author} {\bibfnamefont {Y.}~\bibnamefont {Liu}}, \bibinfo
  {author} {\bibfnamefont {A.}~\bibnamefont {Sch\"afer}}, \bibinfo {author}
  {\bibfnamefont {W.}~\bibnamefont {Wang}}, \bibinfo {author} {\bibfnamefont
  {Y.-B.}\ \bibnamefont {Yang}}, \bibinfo {author} {\bibfnamefont {J.-H.}\
  \bibnamefont {Zhang}}, \ and\ \bibinfo {author} {\bibfnamefont
  {Y.}~\bibnamefont {Zhao}},\ }\href@noop {} {\  (\bibinfo {year}
  {2020}{\natexlab{b}})},\ \Eprint {http://arxiv.org/abs/2008.03886}
  {arXiv:2008.03886 [hep-ph]} \BibitemShut {NoStop}%
\bibitem [{\citenamefont {Ji}(1998)}]{Ji:1998pc}%
  \BibitemOpen
  \bibfield  {author} {\bibinfo {author} {\bibfnamefont {X.-D.}\ \bibnamefont
  {Ji}},\ }\href {\doibase 10.1088/0954-3899/24/7/002} {\bibfield  {journal}
  {\bibinfo  {journal} {J. Phys. G}\ }\textbf {\bibinfo {volume} {24}},\
  \bibinfo {pages} {1181} (\bibinfo {year} {1998})},\ \Eprint
  {http://arxiv.org/abs/hep-ph/9807358} {arXiv:hep-ph/9807358} \BibitemShut
  {NoStop}%
\bibitem [{\citenamefont {Hagler}(2010)}]{Hagler:2009ni}%
  \BibitemOpen
  \bibfield  {author} {\bibinfo {author} {\bibfnamefont {P.}~\bibnamefont
  {Hagler}},\ }\href {\doibase 10.1016/j.physrep.2009.12.008} {\bibfield
  {journal} {\bibinfo  {journal} {Phys. Rept.}\ }\textbf {\bibinfo {volume}
  {490}},\ \bibinfo {pages} {49} (\bibinfo {year} {2010})},\ \Eprint
  {http://arxiv.org/abs/0912.5483} {arXiv:0912.5483 [hep-lat]} \BibitemShut
  {NoStop}%
\bibitem [{\citenamefont {Alexandrou}\ \emph
  {et~al.}(2020{\natexlab{b}})\citenamefont {Alexandrou} \emph
  {et~al.}}]{Alexandrou:2019ali}%
  \BibitemOpen
  \bibfield  {author} {\bibinfo {author} {\bibfnamefont {C.}~\bibnamefont
  {Alexandrou}} \emph {et~al.},\ }\href {\doibase 10.1103/PhysRevD.101.034519}
  {\bibfield  {journal} {\bibinfo  {journal} {Phys. Rev.}\ }\textbf {\bibinfo
  {volume} {D101}},\ \bibinfo {pages} {034519} (\bibinfo {year}
  {2020}{\natexlab{b}})},\ \Eprint {http://arxiv.org/abs/1908.10706}
  {arXiv:1908.10706 [hep-lat]} \BibitemShut {NoStop}%
\bibitem [{\citenamefont {Bali}\ \emph
  {et~al.}(2019{\natexlab{b}})\citenamefont {Bali}, \citenamefont {Collins},
  \citenamefont {G\"ockeler}, \citenamefont {R\"odl}, \citenamefont
  {Sch\"afer},\ and\ \citenamefont {Sternbeck}}]{Bali:2018zgl}%
  \BibitemOpen
  \bibfield  {author} {\bibinfo {author} {\bibfnamefont {G.~S.}\ \bibnamefont
  {Bali}}, \bibinfo {author} {\bibfnamefont {S.}~\bibnamefont {Collins}},
  \bibinfo {author} {\bibfnamefont {M.}~\bibnamefont {G\"ockeler}}, \bibinfo
  {author} {\bibfnamefont {R.}~\bibnamefont {R\"odl}}, \bibinfo {author}
  {\bibfnamefont {A.}~\bibnamefont {Sch\"afer}}, \ and\ \bibinfo {author}
  {\bibfnamefont {A.}~\bibnamefont {Sternbeck}},\ }\href {\doibase
  10.1103/PhysRevD.100.014507} {\bibfield  {journal} {\bibinfo  {journal}
  {Phys. Rev. D}\ }\textbf {\bibinfo {volume} {100}},\ \bibinfo {pages}
  {014507} (\bibinfo {year} {2019}{\natexlab{b}})},\ \Eprint
  {http://arxiv.org/abs/1812.08256} {arXiv:1812.08256 [hep-lat]} \BibitemShut
  {NoStop}%
\bibitem [{\citenamefont {Alexandrou}\ \emph
  {et~al.}(2017{\natexlab{c}})\citenamefont {Alexandrou}, \citenamefont
  {Constantinou}, \citenamefont {Hadjiyiannakou}, \citenamefont {Jansen},
  \citenamefont {Kallidonis}, \citenamefont {Koutsou},\ and\ \citenamefont
  {Vaquero Aviles-Casco}}]{Alexandrou:2017ypw}%
  \BibitemOpen
  \bibfield  {author} {\bibinfo {author} {\bibfnamefont {C.}~\bibnamefont
  {Alexandrou}}, \bibinfo {author} {\bibfnamefont {M.}~\bibnamefont
  {Constantinou}}, \bibinfo {author} {\bibfnamefont {K.}~\bibnamefont
  {Hadjiyiannakou}}, \bibinfo {author} {\bibfnamefont {K.}~\bibnamefont
  {Jansen}}, \bibinfo {author} {\bibfnamefont {C.}~\bibnamefont {Kallidonis}},
  \bibinfo {author} {\bibfnamefont {G.}~\bibnamefont {Koutsou}}, \ and\
  \bibinfo {author} {\bibfnamefont {A.}~\bibnamefont {Vaquero Aviles-Casco}},\
  }\href {\doibase 10.1103/PhysRevD.96.034503} {\bibfield  {journal} {\bibinfo
  {journal} {Phys. Rev.}\ }\textbf {\bibinfo {volume} {D96}},\ \bibinfo {pages}
  {034503} (\bibinfo {year} {2017}{\natexlab{c}})},\ \Eprint
  {http://arxiv.org/abs/1706.00469} {arXiv:1706.00469 [hep-lat]} \BibitemShut
  {NoStop}%
\bibitem [{\citenamefont {Green}\ \emph {et~al.}(2014)\citenamefont {Green},
  \citenamefont {Negele}, \citenamefont {Pochinsky}, \citenamefont {Syritsyn},
  \citenamefont {Engelhardt},\ and\ \citenamefont {Krieg}}]{Green:2014xba}%
  \BibitemOpen
  \bibfield  {author} {\bibinfo {author} {\bibfnamefont {J.~R.}\ \bibnamefont
  {Green}}, \bibinfo {author} {\bibfnamefont {J.~W.}\ \bibnamefont {Negele}},
  \bibinfo {author} {\bibfnamefont {A.~V.}\ \bibnamefont {Pochinsky}}, \bibinfo
  {author} {\bibfnamefont {S.~N.}\ \bibnamefont {Syritsyn}}, \bibinfo {author}
  {\bibfnamefont {M.}~\bibnamefont {Engelhardt}}, \ and\ \bibinfo {author}
  {\bibfnamefont {S.}~\bibnamefont {Krieg}},\ }\href {\doibase
  10.1103/PhysRevD.90.074507} {\bibfield  {journal} {\bibinfo  {journal} {Phys.
  Rev.}\ }\textbf {\bibinfo {volume} {D90}},\ \bibinfo {pages} {074507}
  (\bibinfo {year} {2014})},\ \Eprint {http://arxiv.org/abs/1404.4029}
  {arXiv:1404.4029 [hep-lat]} \BibitemShut {NoStop}%
\bibitem [{\citenamefont {Hasan}\ \emph {et~al.}(2018)\citenamefont {Hasan},
  \citenamefont {Green}, \citenamefont {Meinel}, \citenamefont {Engelhardt},
  \citenamefont {Krieg}, \citenamefont {Negele}, \citenamefont {Pochinsky},\
  and\ \citenamefont {Syritsyn}}]{Hasan:2017wwt}%
  \BibitemOpen
  \bibfield  {author} {\bibinfo {author} {\bibfnamefont {N.}~\bibnamefont
  {Hasan}}, \bibinfo {author} {\bibfnamefont {J.}~\bibnamefont {Green}},
  \bibinfo {author} {\bibfnamefont {S.}~\bibnamefont {Meinel}}, \bibinfo
  {author} {\bibfnamefont {M.}~\bibnamefont {Engelhardt}}, \bibinfo {author}
  {\bibfnamefont {S.}~\bibnamefont {Krieg}}, \bibinfo {author} {\bibfnamefont
  {J.}~\bibnamefont {Negele}}, \bibinfo {author} {\bibfnamefont
  {A.}~\bibnamefont {Pochinsky}}, \ and\ \bibinfo {author} {\bibfnamefont
  {S.}~\bibnamefont {Syritsyn}},\ }\href {\doibase 10.1103/PhysRevD.97.034504}
  {\bibfield  {journal} {\bibinfo  {journal} {Phys. Rev.}\ }\textbf {\bibinfo
  {volume} {D97}},\ \bibinfo {pages} {034504} (\bibinfo {year} {2018})},\
  \Eprint {http://arxiv.org/abs/1711.11385} {arXiv:1711.11385 [hep-lat]}
  \BibitemShut {NoStop}%
\bibitem [{\citenamefont {Shintani}\ \emph {et~al.}(2019)\citenamefont
  {Shintani}, \citenamefont {Ishikawa}, \citenamefont {Kuramashi},
  \citenamefont {Sasaki},\ and\ \citenamefont {Yamazaki}}]{Shintani:2018ozy}%
  \BibitemOpen
  \bibfield  {author} {\bibinfo {author} {\bibfnamefont {E.}~\bibnamefont
  {Shintani}}, \bibinfo {author} {\bibfnamefont {K.-I.}\ \bibnamefont
  {Ishikawa}}, \bibinfo {author} {\bibfnamefont {Y.}~\bibnamefont {Kuramashi}},
  \bibinfo {author} {\bibfnamefont {S.}~\bibnamefont {Sasaki}}, \ and\ \bibinfo
  {author} {\bibfnamefont {T.}~\bibnamefont {Yamazaki}},\ }\href {\doibase
  10.1103/PhysRevD.99.014510} {\bibfield  {journal} {\bibinfo  {journal} {Phys.
  Rev.}\ }\textbf {\bibinfo {volume} {D99}},\ \bibinfo {pages} {014510}
  (\bibinfo {year} {2019})},\ \Eprint {http://arxiv.org/abs/1811.07292}
  {arXiv:1811.07292 [hep-lat]} \BibitemShut {NoStop}%
\bibitem [{\citenamefont {Alexandrou}\ \emph
  {et~al.}(2019{\natexlab{b}})\citenamefont {Alexandrou}, \citenamefont
  {Bacchio}, \citenamefont {Constantinou}, \citenamefont {Finkenrath},
  \citenamefont {Hadjiyiannakou}, \citenamefont {Jansen}, \citenamefont
  {Koutsou},\ and\ \citenamefont {Vaquero Aviles-Casco}}]{Alexandrou:2018sjm}%
  \BibitemOpen
  \bibfield  {author} {\bibinfo {author} {\bibfnamefont {C.}~\bibnamefont
  {Alexandrou}}, \bibinfo {author} {\bibfnamefont {S.}~\bibnamefont {Bacchio}},
  \bibinfo {author} {\bibfnamefont {M.}~\bibnamefont {Constantinou}}, \bibinfo
  {author} {\bibfnamefont {J.}~\bibnamefont {Finkenrath}}, \bibinfo {author}
  {\bibfnamefont {K.}~\bibnamefont {Hadjiyiannakou}}, \bibinfo {author}
  {\bibfnamefont {K.}~\bibnamefont {Jansen}}, \bibinfo {author} {\bibfnamefont
  {G.}~\bibnamefont {Koutsou}}, \ and\ \bibinfo {author} {\bibfnamefont
  {A.}~\bibnamefont {Vaquero Aviles-Casco}},\ }\href {\doibase
  10.1103/PhysRevD.100.014509} {\bibfield  {journal} {\bibinfo  {journal}
  {Phys. Rev.}\ }\textbf {\bibinfo {volume} {D100}},\ \bibinfo {pages} {014509}
  (\bibinfo {year} {2019}{\natexlab{b}})},\ \Eprint
  {http://arxiv.org/abs/1812.10311} {arXiv:1812.10311 [hep-lat]} \BibitemShut
  {NoStop}%
\bibitem [{\citenamefont {Jang}\ \emph {et~al.}(2018)\citenamefont {Jang},
  \citenamefont {Bhattacharya}, \citenamefont {Gupta}, \citenamefont {Lin},\
  and\ \citenamefont {Yoon}}]{Jang:2018djx}%
  \BibitemOpen
  \bibfield  {author} {\bibinfo {author} {\bibfnamefont {Y.-C.}\ \bibnamefont
  {Jang}}, \bibinfo {author} {\bibfnamefont {T.}~\bibnamefont {Bhattacharya}},
  \bibinfo {author} {\bibfnamefont {R.}~\bibnamefont {Gupta}}, \bibinfo
  {author} {\bibfnamefont {H.-W.}\ \bibnamefont {Lin}}, \ and\ \bibinfo
  {author} {\bibfnamefont {B.}~\bibnamefont {Yoon}} (\bibinfo {collaboration}
  {PNDME}),\ }\bibfield  {booktitle} {\emph {\bibinfo {booktitle}
  {{Proceedings, 36th International Symposium on Lattice Field Theory (Lattice
  2018): East Lansing, MI, United States, July 22-28, 2018}}},\ }\href
  {\doibase 10.22323/1.334.0123} {\bibfield  {journal} {\bibinfo  {journal}
  {PoS}\ }\textbf {\bibinfo {volume} {LATTICE2018}},\ \bibinfo {pages} {123}
  (\bibinfo {year} {2018})},\ \Eprint {http://arxiv.org/abs/1901.00060}
  {arXiv:1901.00060 [hep-lat]} \BibitemShut {NoStop}%
\bibitem [{\citenamefont {Hou}\ \emph {et~al.}(2019)\citenamefont {Hou} \emph
  {et~al.}}]{Hou:2019efy}%
  \BibitemOpen
  \bibfield  {author} {\bibinfo {author} {\bibfnamefont {T.-J.}\ \bibnamefont
  {Hou}} \emph {et~al.},\ }\href@noop {} {\  (\bibinfo {year} {2019})},\
  \Eprint {http://arxiv.org/abs/1912.10053} {arXiv:1912.10053 [hep-ph]}
  \BibitemShut {NoStop}%
\bibitem [{\citenamefont {Burkardt}(2003)}]{Burkardt:2002hr}%
  \BibitemOpen
  \bibfield  {author} {\bibinfo {author} {\bibfnamefont {M.}~\bibnamefont
  {Burkardt}},\ }\href {\doibase 10.1142/S0217751X03012370} {\bibfield
  {journal} {\bibinfo  {journal} {Int. J. Mod. Phys.}\ }\textbf {\bibinfo
  {volume} {A18}},\ \bibinfo {pages} {173} (\bibinfo {year} {2003})},\ \Eprint
  {http://arxiv.org/abs/hep-ph/0207047} {arXiv:hep-ph/0207047 [hep-ph]}
  \BibitemShut {NoStop}%
\bibitem [{\citenamefont {Alexandrou}\ \emph
  {et~al.}(2020{\natexlab{c}})\citenamefont {Alexandrou}, \citenamefont
  {Cichy}, \citenamefont {Constantinou}, \citenamefont {Green}, \citenamefont
  {Hadjiyiannakou}, \citenamefont {Jansen}, \citenamefont {Manigrasso},
  \citenamefont {Scapellato},\ and\ \citenamefont
  {Steffens}}]{Alexandrou:2020qtt}%
  \BibitemOpen
  \bibfield  {author} {\bibinfo {author} {\bibfnamefont {C.}~\bibnamefont
  {Alexandrou}}, \bibinfo {author} {\bibfnamefont {K.}~\bibnamefont {Cichy}},
  \bibinfo {author} {\bibfnamefont {M.}~\bibnamefont {Constantinou}}, \bibinfo
  {author} {\bibfnamefont {J.~R.}\ \bibnamefont {Green}}, \bibinfo {author}
  {\bibfnamefont {K.}~\bibnamefont {Hadjiyiannakou}}, \bibinfo {author}
  {\bibfnamefont {K.}~\bibnamefont {Jansen}}, \bibinfo {author} {\bibfnamefont
  {F.}~\bibnamefont {Manigrasso}}, \bibinfo {author} {\bibfnamefont
  {A.}~\bibnamefont {Scapellato}}, \ and\ \bibinfo {author} {\bibfnamefont
  {F.}~\bibnamefont {Steffens}},\ }\href@noop {} {\  (\bibinfo {year}
  {2020}{\natexlab{c}})},\ \Eprint {http://arxiv.org/abs/2011.00964}
  {arXiv:2011.00964 [hep-lat]} \BibitemShut {NoStop}%
\bibitem [{\citenamefont {Edwards}\ and\ \citenamefont
  {Joo}(2005)}]{Edwards:2004sx}%
  \BibitemOpen
  \bibfield  {author} {\bibinfo {author} {\bibfnamefont {R.~G.}\ \bibnamefont
  {Edwards}}\ and\ \bibinfo {author} {\bibfnamefont {B.}~\bibnamefont {Joo}}
  (\bibinfo {collaboration} {SciDAC, LHPC, UKQCD}),\ }\href {\doibase
  10.1016/j.nuclphysbps.2004.11.254} {\bibfield  {journal} {\bibinfo  {journal}
  {Nucl. Phys. B Proc. Suppl.}\ }\textbf {\bibinfo {volume} {140}},\ \bibinfo
  {pages} {832} (\bibinfo {year} {2005})},\ \Eprint
  {http://arxiv.org/abs/hep-lat/0409003} {arXiv:hep-lat/0409003} \BibitemShut
  {NoStop}%
\bibitem [{\citenamefont
  {Radyushkin}(2017{\natexlab{b}})}]{Radyushkin:2017cyf}%
  \BibitemOpen
  \bibfield  {author} {\bibinfo {author} {\bibfnamefont {A.~V.}\ \bibnamefont
  {Radyushkin}},\ }\href {\doibase 10.1103/PhysRevD.96.034025} {\bibfield
  {journal} {\bibinfo  {journal} {Phys. Rev.}\ }\textbf {\bibinfo {volume}
  {D96}},\ \bibinfo {pages} {034025} (\bibinfo {year} {2017}{\natexlab{b}})},\
  \Eprint {http://arxiv.org/abs/1705.01488} {arXiv:1705.01488 [hep-ph]}
  \BibitemShut {NoStop}%
\bibitem [{\citenamefont {Luscher}\ and\ \citenamefont
  {Wolff}(1990)}]{Luscher:1990ck}%
  \BibitemOpen
  \bibfield  {author} {\bibinfo {author} {\bibfnamefont {M.}~\bibnamefont
  {Luscher}}\ and\ \bibinfo {author} {\bibfnamefont {U.}~\bibnamefont
  {Wolff}},\ }\href {\doibase 10.1016/0550-3213(90)90540-T} {\bibfield
  {journal} {\bibinfo  {journal} {Nucl. Phys. B}\ }\textbf {\bibinfo {volume}
  {339}},\ \bibinfo {pages} {222} (\bibinfo {year} {1990})}\BibitemShut
  {NoStop}%
\bibitem [{\citenamefont {Liu}\ \emph {et~al.}(2019)\citenamefont {Liu},
  \citenamefont {Wang}, \citenamefont {Xu}, \citenamefont {Zhang},
  \citenamefont {Zhang}, \citenamefont {Zhao},\ and\ \citenamefont
  {Zhao}}]{Liu:2019urm}%
  \BibitemOpen
  \bibfield  {author} {\bibinfo {author} {\bibfnamefont {Y.-S.}\ \bibnamefont
  {Liu}}, \bibinfo {author} {\bibfnamefont {W.}~\bibnamefont {Wang}}, \bibinfo
  {author} {\bibfnamefont {J.}~\bibnamefont {Xu}}, \bibinfo {author}
  {\bibfnamefont {Q.-A.}\ \bibnamefont {Zhang}}, \bibinfo {author}
  {\bibfnamefont {J.-H.}\ \bibnamefont {Zhang}}, \bibinfo {author}
  {\bibfnamefont {S.}~\bibnamefont {Zhao}}, \ and\ \bibinfo {author}
  {\bibfnamefont {Y.}~\bibnamefont {Zhao}},\ }\href@noop {} {\  (\bibinfo
  {year} {2019})},\ \Eprint {http://arxiv.org/abs/1902.00307} {arXiv:1902.00307
  [hep-ph]} \BibitemShut {NoStop}%
\end{thebibliography}

\clearpage
\clearpage
\section*{Supplemental materials}

\subsection*{Details of the lattice calculations}
\begin{figure*}[tb]
\includegraphics[width=0.67\textwidth]{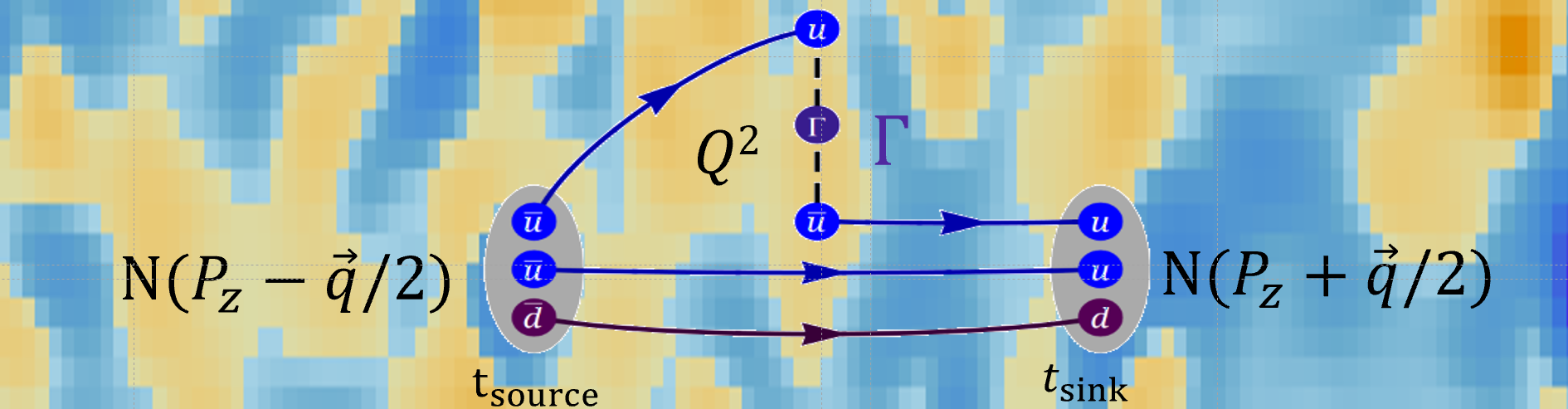}
\caption{Illustration of the lattice matrix-element calculation on top of the QCD vacuum. 
\label{fig:GPD-diagram}}
\end{figure*}

In the LaMET (or ``quasi-PDF'') approach, we calculate time-independent spatially displaced matrix elements that can be connected to the parton distributions.
The operator choice is not unique at finite nucleon momentum;
a convenient choice for leading-twist PDFs is to take the average of the hadron momentum $(0,0, P_z)$ and the quark-antiquark separation to be along the $z$ direction
\begin{equation}\label{eq:qGPDME}
h_\Gamma(z,P_z) =
\langle N(P_f,s^\prime)\left|\bar\psi(\lambda \hat{z}) W(\lambda \hat{z},0)\Gamma \psi(0)\right|N(P_i,s))\rangle,
\end{equation}
where $|N(P_{i,f})\rangle$ denotes a nucleon state with momenta $P_i^\mu=(E_i,-q_x/2,-q_y/2,P_z)$ and $P_f^\mu=(E_f,q_x/2,q_y/2,P_z)$,
$\psi$ ($\bar {\psi}$) are the (anti-)quark fields, 
$n^\mu$ is a unit vector and 
$W(\lambda \hat{z},0)$ is the Wilson link along the $z$ direction. 
See Fig.~\ref{fig:GPD-diagram} for an illustration. 
There are multiple choices of operator in this framework that will recover the same lightcone PDFs when the large-momentum limit is taken. 
For example, $\Gamma$ can be $\gamma_z$ or $\gamma_t$~\cite{Xiong:2013bka,Radyushkin:2016hsy,Radyushkin:2017cyf,Orginos:2017kos};
both will give the unpolarized PDFs in the infinite-momentum frame. 
In this work, we use $\Gamma=\gamma_t$ for the unpolarized GPDs calculations. 
Since the systematics of the LaMET methods, such as the higher-twist effects at $O\left( \frac{\Lambda_\text{QCD}^2}{P_z^2}\right)$, 
decrease as the momentum increases, we use $P_z \approx 2.2$~GeV in this calculation. (A study of the $P_z$ convergence of the LaMET approach on the same ensembles can be found in Refs.~\cite{Chen:2018xof,Lin:2018qky,Liu:2018hxv}.)
This also allows us to reach smaller-$x$ regions of the GPDs functions.

We use Gaussian momentum smearing~\cite{Bali:2016lva} $\psi(x) + \alpha \sum_j U_j(x) e^{ik\hat{e}_j} \psi(x+\hat{e}_j)$, where $k$ is the input momentum-smearing parameter, $U_j(x)$ are the gauge links in the $j$ direction, and $\alpha$ is a tunable parameter as in traditional Gaussian smearing.
We vary the $\alpha$ values in this study so that we can have multiple ways of removing the excited-state contributions from matrix elements later on. 
To better extract the boosted-momentum ground-state nucleon, we apply the variational method~\cite{Luscher:1990ck} to extract the principal correlators corresponding to pure energy eigenstates from a matrix of correlators. 
We use a $3 \times 3$ smeared-smeared two-point correlation matrix, which can be decomposed as
\begin{equation}
  C^{ij}_\text{2pt}(t)=\sum_{n=0} v^i_{n*} v^j_n e^{-t E_n}
\end{equation}
with eigenvalues $\lambda_n(t,t_r)=e^{-(t-t_r)E_n}$ by solving the generalized eigensystem problem $C^{ij}_\text{2pt}(t)V=\lambda(t,t_r)C^{ij}_\text{2pt}(t_r)V$,
where $V$ is the matrix of eigenvectors and $t_r$ is a reference time slice. 
The resulting 3 eigenvalues (principal correlators) $\lambda_n(t,t_r)$ are then further analyzed to extract the energy levels $E_n$. 
Since they have been projected onto pure eigenstates of the Hamiltonian, the principal correlator should be fit well by a single exponential and double checked for the consistency of the obtained energies.
The leading contamination due to higher-lying states is another exponential having higher energy;
we use a two-exponential fit to help remove this contamination.
The overlap factors ($A_n$) between the interpolating operators and the $n^\text{th}$ state are derived from the eigenvectors obtained in the variational method. 

\begin{figure*}[tb]
\includegraphics[width=0.75\textwidth]{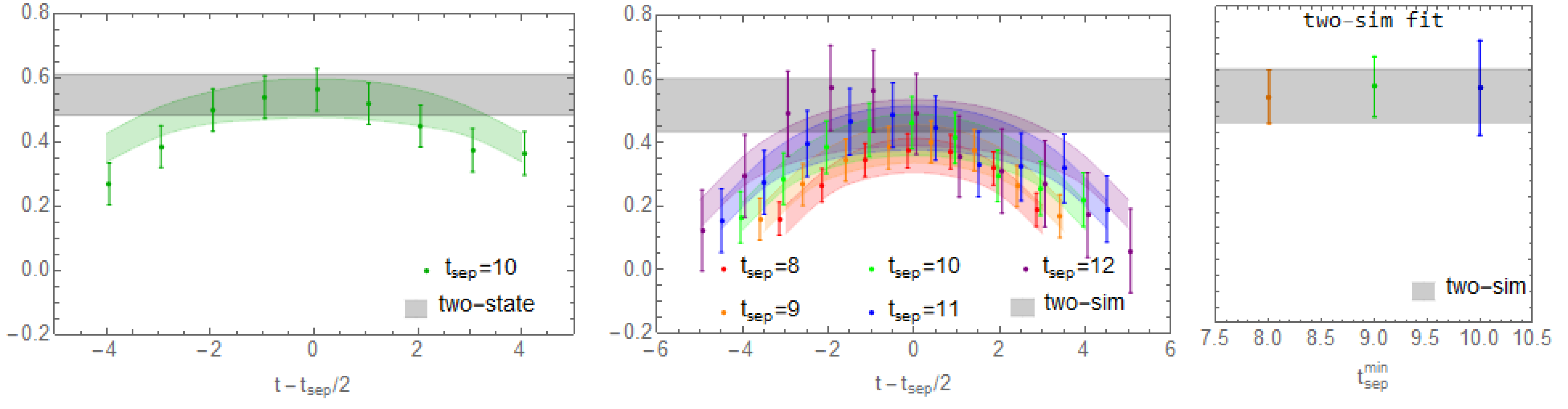}
\caption{
An example of the ground-state matrix element determination using variational method using rotated two- and three-point correlators (left), 
two-state fit to multiple source-sink separation (middle), along with the ratio plots data and reconstructed fit to the ratio points. 
The rightmost plot shows how much the ground-state matrix elements change as we reduce the inputs to the fits;
we see some fluctuations and increase of errors due to the reduction of the available data but overall steady determination of the ground-state matrix element. 
The example used here is from $\vec{P_f}=\{1,0,10\}\frac{2\pi}{L}$ and $\vec{P_i}=\{-1,0,10\}\frac{2\pi}{L}$ with projection operator $\frac{1+\gamma_t}{2}(1+i \gamma_5\gamma_{z})$. 
\label{fig:excited-states}}
\end{figure*}

To calculate the GPD matrix elements at nonzero momentum transfer, first calculate the matrix element $\langle \chi_N (\vec{P}_f) | O^\mu | \chi_N (\vec{P}_i) \rangle$, where $\chi_N$ is the nucleon spin-1/2 interpolating field, $\epsilon^{abc} [q^{a\top}(x)C\gamma_5q^b(x)]q^c(x)$.  
$O_\mu=\overline{\psi}\gamma_\mu W(z) \psi$ is the LaMET Wilson-line displacement operator with $\psi$ being either an up or down quark field, and $\vec{P}_{\{i,f\}}$ are the initial and final nucleon momenta. 
We integrate out the spatial dependence and project the baryonic spin, using projection operators $\mathbb{P}_\rho=\frac{1+\gamma_t}{2}(1+i \gamma_5\gamma_{\rho})$ with $\rho\in\{x,y,z\}$, leaving a time-dependent three-point correlator, $C_\text{3pt}$.  
\begin{widetext}
\begin{eqnarray}\label{eq:general-3pt}
\Gamma^{(3),\mathbb{P}_\rho}_{\mu,AB}(t_i,t,t_f,\vec{p}_i,\vec{P}_f) &=&
Z_O \sum_n \sum_{n^\prime} f_{n,n^\prime}(P_f,P_i,E_n^\prime,E_n,t,t_i,t_f)\nonumber \\
&\times& \sum_{s,s^\prime}
(\mathbb{P}_\rho)_{\alpha\beta} u_{n^\prime}(\vec{P}_f,s^\prime)_\beta
\langle N_{n^\prime}(\vec{P}_f,s^\prime)\left|O_\mu\right|N_n(\vec{P}_i,s)\rangle\overline{u}_n(\vec{P}_i,s)_\alpha,
\end{eqnarray}
\end{widetext}
where $f_{n,n^\prime}(P_f,P_i,E_n^\prime,E_n,t,t_i,t_f)$ contains kinematic factors involving the energies $E_n$ and overlap factors $A_n$ obtained in the two-point variational method, $n$ and $n^\prime$ are the indices of different energy states and $Z_O$ is the operator renormalization constant (which is determined nonperturbatively). 
We also use high-statistics measurements, 501,760 total over 1960 configurations, to drive down the increased statistical noise at high boost momenta, $P_z=\left|\frac{\vec{P_i}+\vec{P_f}}{2}\right| = |\frac{2\pi}{L}\{0,0,10\}a^{-1}|$, and vary spatial momentum transfer $\vec{q}=\vec{P_f}-\vec{P_i}=\frac{2\pi}{L}\{n_x,n_y,0\}a^{-1}$ with all possible integer $n_{x,y}$ and $n_x^2+n_y^2 \in \{0,4,8,16,20\}$ with transfer four-momentum squared $Q^2=-q_\mu q^\mu=\{{0, 0.19, 0.39, 0.77, 0.97} \}$~GeV$^2$. 
Figure~\ref{fig:excited-states} shows an example of the ground-state matrix elements (shown as the gray band) from the variational analysis along with the two-state fitted results with $\vec{P_f}=\{1,0,10\}\frac{2\pi}{L}$ and $\vec{P_i}=\{-1,0,10\}\frac{2\pi}{L}$ with projection operator $\frac{1+\gamma_t}{2}(1+i \gamma_5\gamma_{z})$.
The ground-state matrix elements can be extracted from $v_0^T C_\text{3pt}^{ij}(t) v_0$ using the eigenvectors from the two-point variational-method analysis (shown in the left panel of Fig.~\ref{fig:excited-states}), simultaneous two-state fitted results using source-sink separation of $t_\text{sep} =[8,12]$ lattice units, two-state fitted results as functions of  $t_\text{sep}^\text{min}$.
The consistency of these methods demonstrates that our extracted ground-state matrix elements are stably determined.
For more details on the lattice study of nucleon matrix elements comparing the variational and two-state fit methods, we refer readers to Ref.~\cite{Yoon:2016dij}, which contains a very detailed discussion.

\begin{figure*}[tb]
\includegraphics[width=0.42\textwidth]{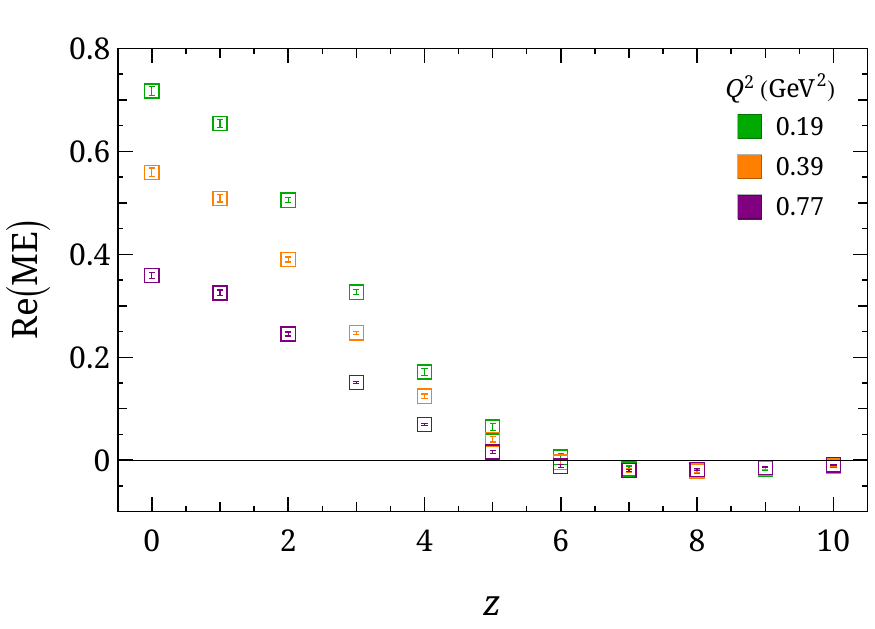}
\includegraphics[width=0.42\textwidth]{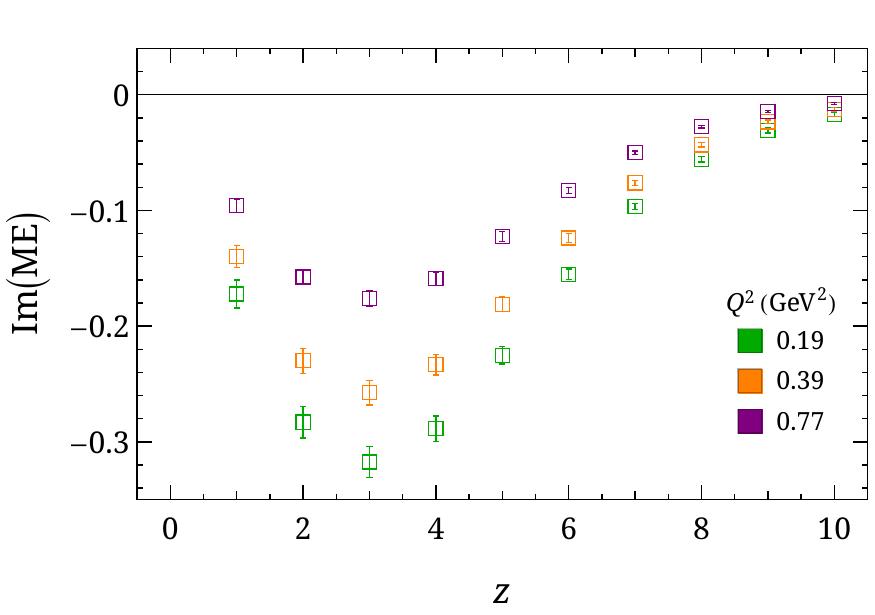}
\includegraphics[width=0.42\textwidth]{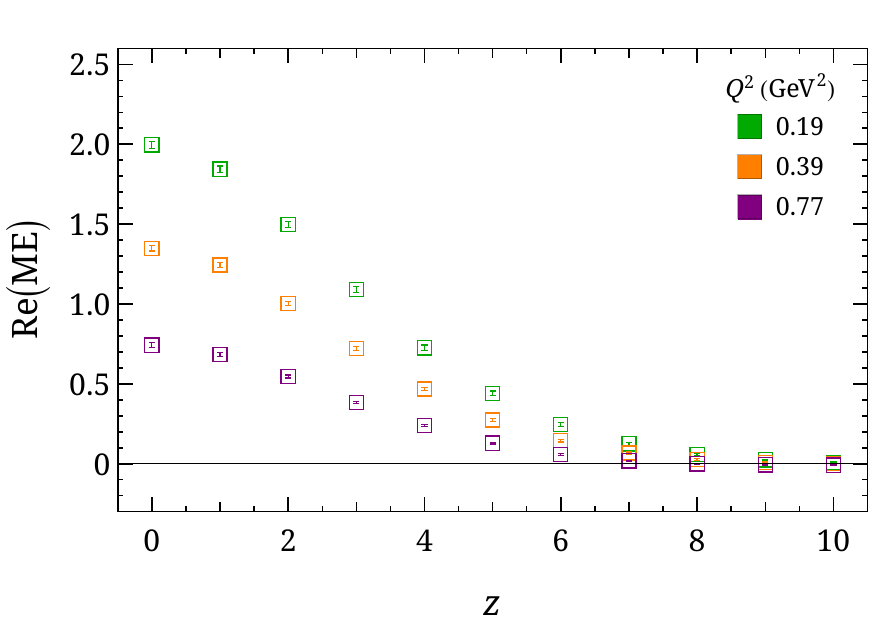}
\includegraphics[width=0.42\textwidth]{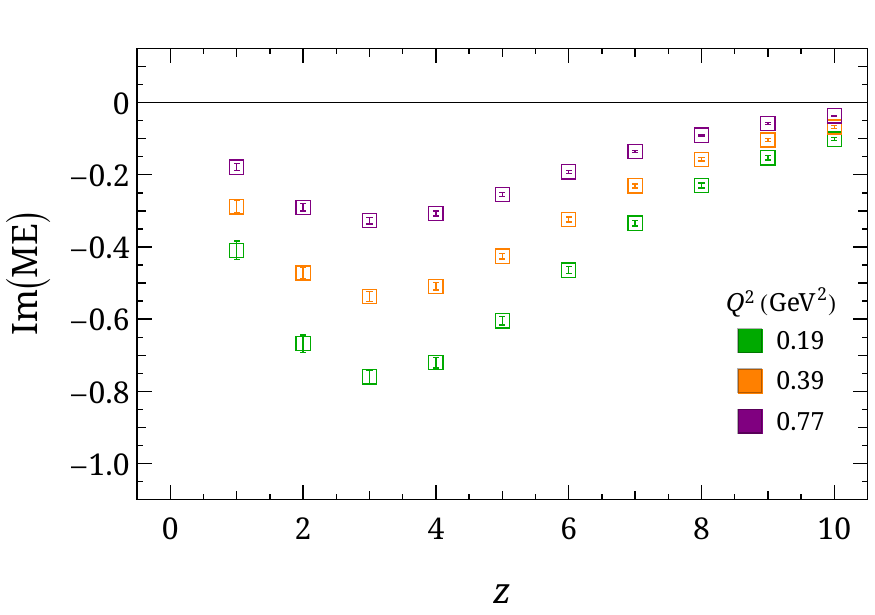}
\caption{
The Wilson-line length displacement $z$-dependence of normalized matrix elements $h_H(P_z,Q^2,z)$ (top) and $h_E(P_z,Q^2,z)$ (bottom) at selected $Q^2\in\{0.19, 0.39, 0.77\}\text{ GeV}^2$. 
\label{fig:H-MEs}}
\end{figure*}

The overdetermined system of linear equations (using multiple $\vec{q}$ and projection operators) allows for solution of $h_H(P_z,Q^2,z)$ and $h_E(P_z,Q^2,z)$, similar to vector form factor calculations with our chosen projection operator and momentum configuration:
The ground-state matrix elements are proportional to 
\begin{widetext}
\begin{equation}
\label{eq:ffunc}
 \frac{  \Tr \left\{ \mathbb{P}_\rho \, [ -i \slashed{P_f} + m_N] \: 
 [h_H(P_z,Q^2,z)\gamma_t +h_E(P_z,Q^2,z)\frac{i\sigma^{t\nu}q_\nu}{2M_N}] 
 \, [ -i \slashed{P_i} + m_N] \, \right\} }{4 \, E_N(\vec{P_f}) E_N(\vec{P_i}^{}\,)}
\end{equation}
\end{widetext}
with $\slashed{p} = i E_N(\vec{p}\,) \gamma_4 + \vec{p} \cdot \vec{\gamma}$. 
We obtain a linear system of equations by using different projection operators $\mathbb{P}_\rho$, momenta $P_f$ and $P_i$ to solve for $H$ and $E$ in coordinate space, $h_{\{H,E\}}(P_z,Q^2,z)$. 
Selected $Q^2$ values of $h_{\{H,E\}}(P_z,Q^2,z)$ normalized by $h_{\{H,E\}}(P_z,Q^2=0,z=0)$ are shown in Fig.~\ref{fig:H-MEs}. 
The real matrix elements decrease quickly to zero due to the large boost momentum used in this calculation.
This helps us to use smaller-displacement data to avoid large contributions from higher-twist effects in the larger-$z$ region. 

\begin{figure*}[tb]
\includegraphics[width=0.32\textwidth]{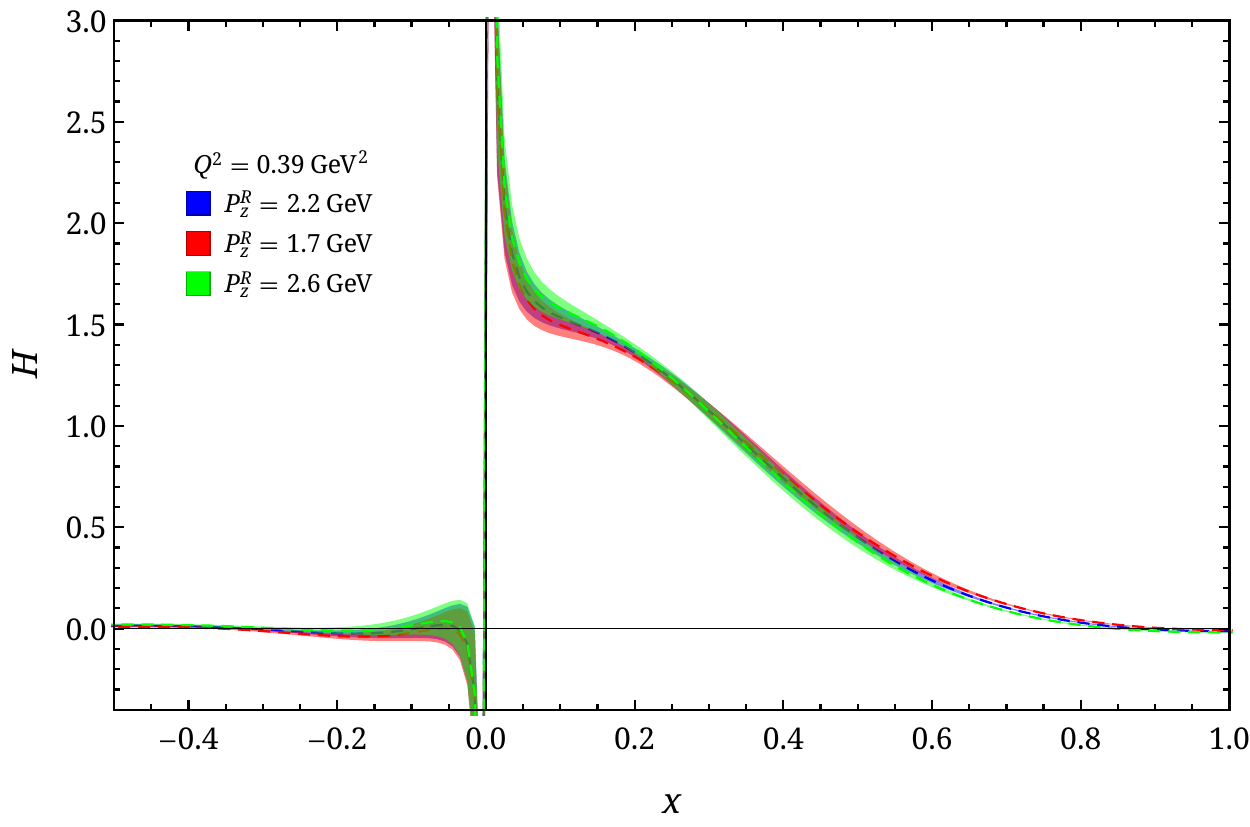}
\includegraphics[width=0.32\textwidth]{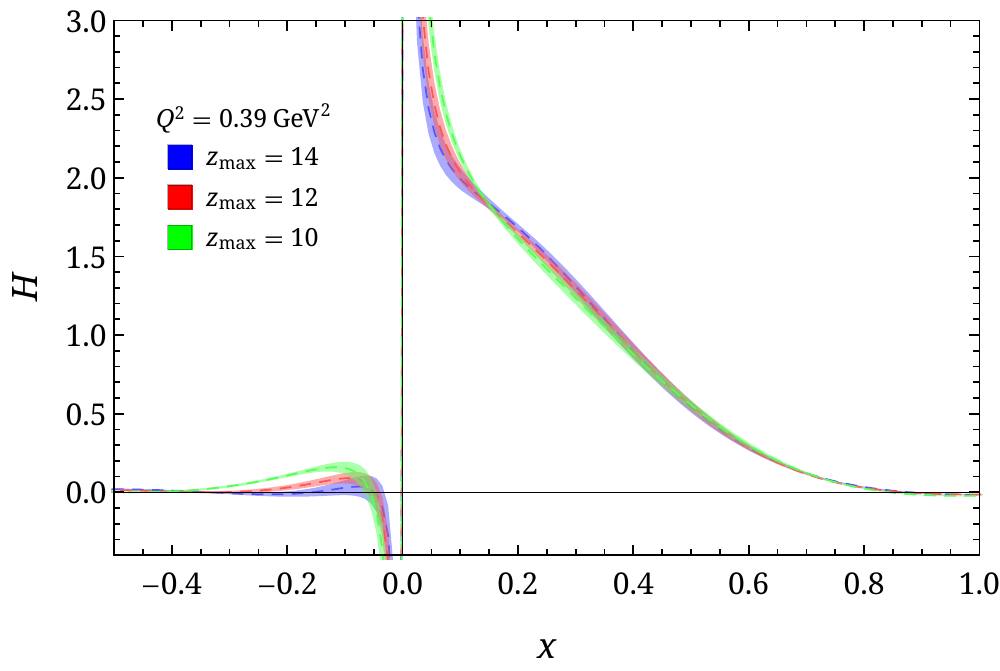}
\includegraphics[width=0.32\textwidth]{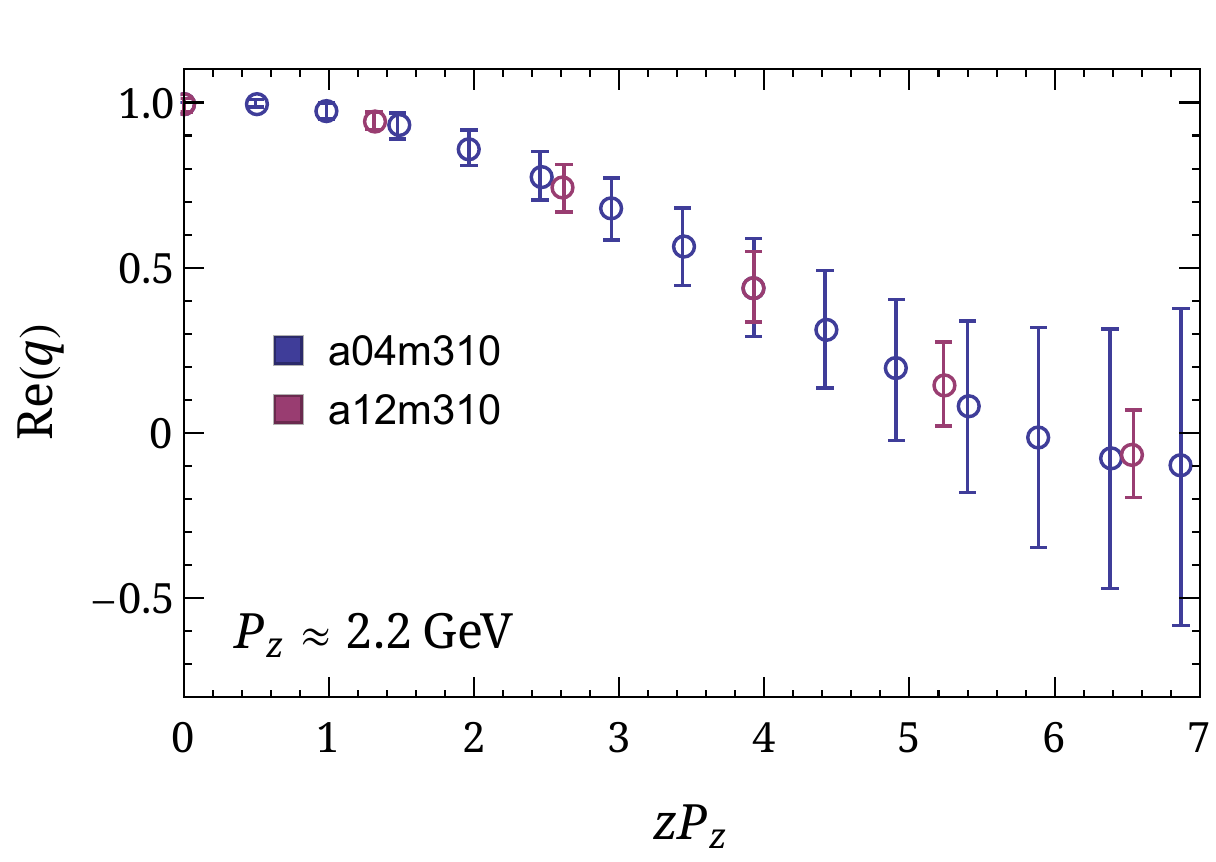}
\caption{\label{fig:systematics}
Examples of the systematics being considered in this calculation. 
The left and middle show examples of post-matching $H$-GPD at $Q^2=0.39$~GeV$^2$ with varying choices of $p_z^R$ and $z_\text{max}$ in the NPR and maximum Wilson-line displacement $z_\text{max}$ used in quasi-GPD.
We found the effects to be mild except in the small-$x$ region. 
The right-hand side shows a comparison of 310-MeV nucleon isovector matrix elements for PDF at lattice spacings of 0.12 and 0.042~fm for around 2.2~GeV nucleon boost momentum. 
}
\end{figure*}

To obtain the quasi-GPDs, we first apply nonperturbative renormalization (NPR) in RI/MOM scheme to the bare matrix elements, using the NPR done in previous work using the same lattice ensembles~\cite{Chen:2018xof,Lin:2018qky,Liu:2018hxv}, itself following the same strategy described in Refs.~\cite{Stewart:2017tvs,Chen:2017mzz}. 
The RI/MOM renormalization constant ${Z}$ is calculated nonperturbatively on the lattice by imposing the following momentum-subtraction condition on the matrix element:
\begin{multline}\label{hRx}
Z(p^R_z, \mu_R) = \\
\left.\frac{\Tr[\mathcal{P} \sum_s \langle ps| \bar\psi_f(\lambda \hat{z}) \gamma_t W(\lambda\hat{z},0) \psi_f(0)|ps\rangle]}
{\Tr[\mathcal{P}  \sum_s \langle ps| \bar\psi_f(\lambda \hat{z})  \gamma_t W(\lambda\hat{z},0) \psi_f(0) |ps\rangle_\text{tree}]} \right|_{\tiny\begin{matrix}p^2=-\mu_R^2 \\ \!\!\!\!p_z=p^R_z\end{matrix}},
\end{multline}
where $\mathcal{P}=\gamma_t-\frac{p_xp_t\gamma_x+p_yp_t\gamma_y}{p_x^2+p_y^2}$. 
On the lattice, $\langle ps|O_{t}(z)|ps\rangle$ is calculated from the amputated Green function of $O_{t}$ with Euclidean external momentum.
In this work, we use the same renormalization scales as used in the previous work~\cite{Chen:2018xof,Lin:2018qky,Liu:2018hxv}:
$\mu_R=3.8$~GeV, 
$p_z^R=2.2$~GeV,  
$\mu^{\overline{\text{MS}}} =3$~GeV.  
The renormalization-scale dependence was studied in Ref.~\cite{Liu:2018uuj}.
We also vary the values of $P_z^R$, and the results are shown in Fig.~\ref{fig:systematics}, focusing on the $\xi=0$ GPDs, where the matching formula is the same as that in the PDFs, as discussed in Ref.~\cite{Liu:2019urm}. 
We normalize all matrix elements by $h_H(P_z, Q^2=0, z=0)$, as in our previous PDF work~\cite{Chen:2018xof,Lin:2018qky,Liu:2018hxv}.
Using matrix-element ratios reduces the lattice systematic error, since in the continuum limit $h_H(P_z, Q^2=0,z=0)$, the vector charge, goes to 1. 

We also perform checks of the $z_\text{max}$ input in the Fourier transformation and lattice-spacing dependence.
The effects are documented in Fig.~\ref{fig:systematics}.

\end{document}